\documentclass[12pt]{article}
\usepackage{amsmath}
\usepackage{subcaption}
\usepackage{graphicx}
\usepackage{graphics}
\usepackage{enumerate}
\usepackage{natbib}
\usepackage{xcolor}
\usepackage{appendix}
\usepackage{url}
\usepackage{multirow}
\usepackage{amsthm}

\usepackage{tablefootnote}
\usepackage{longtable}
\usepackage{threeparttable}
\usepackage{bm}
\usepackage[ruled,vlined]{algorithm2e}
\usepackage{caption}
\usepackage{placeins}
\usepackage{lscape}
\usepackage{authblk}
\usepackage{booktabs}
\usepackage{array}
\usepackage{float}
\usepackage[margin=1.25in]{geometry}
\usepackage{newtxtext}
\usepackage[doublespacing]{setspace}
\newtheorem{theorem}{Theorem}

\newtheorem{lemma}[theorem]{Test}

\newcommand{\blind}{1}

\makeatletter
\newcommand{\pushright}[1]{\ifmeasuring@#1\else\omit\hfill$\displaystyle#1$\fi\ignorespaces}
\newcommand{\pushleft}[1]{\ifmeasuring@#1\else\omit$\displaystyle#1$\hfill\fi\ignorespaces}
\makeatother

\begin{document}

\if1\blind
{
  \title{\bf Evaluating Aggregated Relational Data Models with Simple Diagnostics}

\author[1]{Ian Laga}
\author[1]{Benjamin Vogel}
\author[2]{Jieyun Wang}
\author[2]{Anna Smith}
\author[3]{Owen Ward}
\affil[1]{Department of Mathematical Sciences, Montana State University, Bozeman, MT, USA}
\affil[2]{Dr.
Bing Zhang Department of Statistics, University of Kentucky, Lexington, KY, USA}
\affil[3]{Department of Statistics and Actuarial Science, Simon Fraser University, Burnaby, BC, Canada}
\date{December 21, 2025}
  \maketitle
} \fi

\if0\blind
{
  \bigskip
  \bigskip
  \bigskip
  \begin{center}
    {\LARGE\bf Evaluating Aggregated Relational Data Models with Simple Diagnostics}
\end{center}
  \medskip
} \fi

\bigskip
\begin{abstract}
Aggregated Relational Data (ARD) contain summary information about individual social networks and are widely used to estimate social network characteristics and the size of populations of interest.
Although a variety of ARD estimators exist, practitioners currently lack guidance on how to evaluate whether a selected model adequately fits the data.
We introduce a diagnostic framework for ARD models that provides a systematic, reproducible process for assessing covariate structure, distributional assumptions, and correlation.
The diagnostics are based on point estimates, using either maximum likelihood or maximum a posteriori optimization, which allows quick evaluation without requiring repeated Bayesian model fitting.
Through simulation studies and applications to large ARD datasets, we show that the proposed workflow identifies common sources of model misfit and helps researchers select an appropriate model that adequately explains the data.
\end{abstract}

\noindent%
{\it Keywords: Goodness-of-fit, model assessment, visual inference, size estimation} 
\vfill

\newpage

\section{Introduction}
\label{sec:intro} 

Accurately and efficiently assessing model adequacy is essential for valid inference. Diagnostic tools, both numerical and visual, allow researchers to evaluate whether key assumptions are satisfied and to determine whether the fitted model captures the major features of the data. These evaluations guide decisions about model refinement, including the selection of covariates, the need for alternative distributional assumptions, and whether additional model structure is warranted. These diagnostic needs are especially important, but lacking, in the context of Aggregated Relational Data.

Aggregated Relational Data (ARD) are summary counts of how many people someone knows in specific groups. They are typically collected by asking survey respondents questions of the form ``How many [X]'s do you know?'', where X represents a group, trait, or demographic category of interest. Because ARD leverages the social networks of survey respondents, they provide access to information about populations that are expensive or difficult to sample directly. Researchers frequently use ARD to infer network size (degree), network composition, and characteristics of hidden or hard-to-reach populations.

A wide range of statistical models, both frequentist and Bayesian, have been proposed for ARD and the Network Scale-Up Method (NSUM), an application of ARD to estimate degrees and population sizes. These include models that adjust for non-random mixing between groups, incorporate respondent-level covariates, handle different sampling schemes, or focus on estimating the size of hidden populations. While methodological developments continue to emerge, considerably less attention has been paid to evaluating whether a given model is appropriate for a particular dataset. In practice, analysts often fit a model and either accept the output without systematically examining the adequacy of its assumptions or examine posterior predictive checks relevant to Bayesian models \citep{zheng2006many, laga2023correlated}. Except for leave-one-out cross validation to assess the performance of prevalence estimates for groups with known sizes \citep{park2021segregated, baum2025explaining}, we are not aware of any diagnostics for frequentist ARD models or general-purpose tools applicable to ARD models more broadly.

We address this gap by proposing a general, reproducible diagnostic workflow for ARD/NSUM analyses. Our goal is to provide tools that help both statisticians and applied researchers evaluate model fit, detect common pathologies, and select a suitable modeling strategy before committing to final inference. We organize existing diagnostic ideas from related literatures (e.g., generalized linear model diagnostics) and introduce new diagnostics specifically tailored to ARD structures.

There are two classes of ARD approaches we do not provide diagnostics for. First, we exclude design-based estimators, where inference is driven by sampling design rather than a likelihood model. Because these approaches rely on minimal modeling assumptions, goodness-of-fit diagnostics must instead focus on whether assumptions hold during data collection. An example of this category is the Generalized NSUM \citep{feehan2016generalizing}. Second, we exclude network-based models that use ARD to infer complete network characteristics, such as the models developed by \cite{breza2019social} and \cite{breza2019consistently}. These models are distinct in the literature and require diagnostics geared toward node-level network structure. Instead, we restrict attention to the most commonly used model-based estimators which allow for direct assessment of model adequacy.

The remainder of this paper is organized as follows. Section~\ref{sec:background} provides background on ARD models and introduces a working example. Section~\ref{sec:tools} presents our primary contribution: a comprehensive diagnostic framework demonstrated through this example. %Section~\ref{sec:advice} offers practical guidance for implementation, and
Section~\ref{sec:discussion} concludes with a broad discussion. All proposed diagnostics are implemented in the \texttt{networkscaleup}
R package \citep{networkscaleup, rlang}.

\section{Aggregated Relational Data}
\label{sec:background}

ARD is most commonly used to study social networks and hard-to-reach populations. \cite{bernard1989estimating} first proposed leveraging social network information to estimate the number of people who died in the 1985 Mexico City earthquake. Since its inception, ARD has been collected in many countries to study diverse populations, including people who use drugs \citep{kadushin2006scale, salganik2011assessing}, children who experience choking or injuries \citep{snidero2007use, snidero2012scale}, and groups related to HIV, including female sex workers and men who have sex with men \citep{paniotto2009estimating}. While this paper focuses on traditional ARD designs, several extensions have been proposed, including enriched ARD \citep{feehan2016generalizing} and ARD collected in combination with venue-based sampling \citep{verdery2019estimating}. We provide a brief overview of ARD and the Network Scale-Up Method (NSUM) here and refer readers to \cite{bernard2010counting}, \cite{mccormick2020network}, \cite{laga2021thirty}, and \cite{ward2025bayesian} for comprehensive reviews.

We formalize our notation for the remainder of the manuscript. Given a sample of $n$ respondents from $K$ groups, the ARD form an $n \times K$ matrix $\mathbf{Y}$, where each entry $y_{ik}$ denotes the number of people respondent $i$ reports knowing in group $k$. The basic NSUM estimator assumes
\begin{equation*}
    Y_{ik} \sim \text{Binomial}\left(\exp(\alpha_i), \exp(\beta_k)\right),
\end{equation*}
where $\alpha_i$ is the log-degree of respondent $i$, and $\beta_k$ is the log-prevalence of group $k$ in the corresponding spatial area (typically a city, region, or country) \citep{killworth1998social}. Conceptually, the model assumes that if group $k$ has a prevalence of 1\%, then, on average, 1\% of respondents' social networks belong to group $k$. In practice, however, various biases violate this assumption, motivating the development of more sophisticated models that account for covariates, overdispersion, correlation, and related factors.

Two types of covariates currently exist for ARD: respondent-level covariates and respondent/group-level covariates. Respondent-level covariates are collected in an $n \times p$-dimensional matrix $\bm{Z}$, where the rows include information collected for each respondent which do not vary across groups (e.g., age, gender, level of education, access to the internet). These respondent-level covariates can affect the ARD in one of two ways: \textit{globally}, such that they affect the responses for each group in the same way, or \textit{locally}, such that the relationship between a covariate and the response depends on the group. That is, given a covariate $z_i$, if we assume the expected value of $Y_{ik}$ is given by the relation $E(Y_{ik}) = \mu_{ik}$, the global-covariates are such that $E(Y_{ik}) = \mu_{ik} + \beta z_i$, whereas local-covariates are such that $E(Y_{ik}) = \mu_{ik} + \beta_k z_i$.

On the other hand, a single respondent/group-level covariate is an $n \times K$-dimensional matrix $\bm{X}$, where $x_{ik}$ contains distinctive information about respondent $i$ and group $k$. For example, in an ARD collected in Ukraine, respondents were asked about their level of respect for each group, such that $x_{ik}$ contained respondent $i$'s level of respect towards each group \citep{paniotto2009estimating}. These covariates are relatively rare in the ARD literature, but they provide rich, directly relevant information about the respondent's connections to members of each group. Thus, we generally recommend researchers always account for respondent/group-level covariates when possible.

While group-level covariates can exist in theory (e.g., group sociability measures), they would likely be derived from auxiliary data sources external to ARD surveys. No such covariates are currently available for use with existing ARD.

\iffalse
\begin{equation*}
    E(y_{ik}) \propto \exp\left(\sum_{j=1}^p \beta_p x_{i,p} \right),
\end{equation*}
while the local coefficients model is equivalent to
\begin{equation*}
    E(y_{ik}) \propto \exp\left(\sum_{j=1}^p \beta_{k,} x_{i,p} \right).
\end{equation*}
These two situations are identified by two distinct but similar visualizations.
\fi

\subsection{ARD Models}

Below we summarize the ARD/NSUM model classes considered in this work. While these formulations do not exactly match all existing model-based estimators in the literature, they capture the essential structure of published approaches and are sufficiently general to guide model selection and diagnostic evaluation. The proposed diagnostics focus on the first three models, which are easy to fit, while the final two models, shown in italics, capture additional biases but are impractical to estimate at the exploratory stage. We define notation as they appear.

\noindent\textbf{1. Binomial Model} \hfill \citep{killworth1998social, killworth1998estimation, maltiel2015estimating}

\begin{equation*}
    Y_{ik} \sim \text{Binomial}\left(\exp(\alpha_i), \exp(\beta_k)\right),
\end{equation*}
where \( \alpha_i \) is respondent \( i \)'s log-degree and \( \beta_k \) is the log-prevalence for group \( k \). Existing Binomial ARD models do not incorporate covariates, so we do not include them in our analysis.

\noindent\textbf{2. Poisson Model} \hfill \citep{teo2019estimating}

\begin{equation*}
    Y_{ik} \sim \text{Poisson}\left(\exp\left(\alpha_i + \beta_k + \bm{\beta}^\top \mathbf{z}_i + \gamma_k x_{ik}\right) \right),
\end{equation*}
with additional respondent covariates \( \mathbf{z}_i \), and respondent/group covariates \( x_{ik} \).

\noindent\textbf{3. Negative Binomial Model} \hfill \citep{zheng2006many}

\begin{equation*}
    Y_{ik} \sim \text{Negative-Binomial}\left(
    \mu = \exp\left(\alpha_i + \beta_k + \bm{\beta}^\top \mathbf{z}_i + \gamma_k x_{ik}\right),\; \text{overdispersion} = \omega_k \right)
\end{equation*}
allowing group-specific overdispersion via \( \omega_k \).

\noindent\textbf{4. \textit{Group Correlated Model}}\hfill \citep{laga2023correlated}

\begin{equation*}
\begin{aligned}
Y_{ik} &\sim \text{Poisson}\left(\exp\left(\alpha_i + \beta_k + \bm{\beta}^\top \mathbf{z}_i + \gamma_k x_{ik} + b_{ik}\right)\right), \\
\mathbf{b}_i &\sim \text{MVN}_K(\bm{\mu}, \Sigma),
\end{aligned}
\end{equation*}
introducing correlation across group-level deviations \( b_{ik} \).

\noindent\textbf{5. \textit{Degree Correlated Model}} \hfill \citep{vogel2025accounting}

\begin{equation*}
\begin{aligned}
Y_{ik} &\sim \text{Poisson}\left(\exp\left(\alpha_i + \beta_k + \bm{\beta}^\top \mathbf{z}_i + \gamma_k x_{ik} + b_{ik}\right)\right), \\
(\alpha_i, \mathbf{b}_i) &\sim \text{MVN}_{K+1}(\bm{\mu}, \Sigma),
\end{aligned}
\end{equation*}
allowing correlation between log-degrees \( \alpha_i \) and group-level deviations \( b_{ik} \).

\subsection{Working Example}

For this manuscript, we work through real ARD for each diagnostic. We consider the Ukraine 2008-2009 ARD, which contains data from 9,241 respondents about 15 groups \citep{paniotto2009estimating}. The original ARD contained additional groups, but for demonstration purposes, we match the procedure outlined in \cite{laga2023correlated}, where probe groups are subset based on leave-one-out accuracy.

The Ukraine data contains eight respondent-level covariates, with one continuous (Age) and six binary (Gender, Ukrainian, Employed, Access to Internet, Secondary Education, Vocational Education, and Academic Education). Age is transformed to be between zero and one for visualization, though this is not necessary for modeling. The groups include those related to demographic information (e.g., males 20-30, kids), events (e.g., died in 2007, divorced in 2007), names (e.g., Pavlo), and activities (e.g., intravenous drug users, men who have sex with men). The Ukraine data also contained a respondent/group-level covariate where each respondent is asked about their level-of-respect towards members of each group. Though omitted for our proposed diagnostics, level-of-respect should be included in final models.

We apply our proposed diagnostic process to additional simulated ARD in the Supplementary Information.

% To fully test the proposed methodology with ARD of sufficient size and complexity, we use the McCarty 1998-1999 \citep{mccarty2001comparing} and the Ukraine 2008-2009 ARD \citep{paniotto2009estimating}. The McCarty data we consider contains ARD from 500 respondents about 21 groups, while the Ukraine data contains ARD from 9,241 respondents about 15 groups. The results from the proposed diagnostic procedure for each ARD are shown below.

\subsection{Diagnostic Structure}
\label{sec:diagnostics}

\begin{figure}[!t]
    \centering
    \includegraphics[width=1\textwidth]{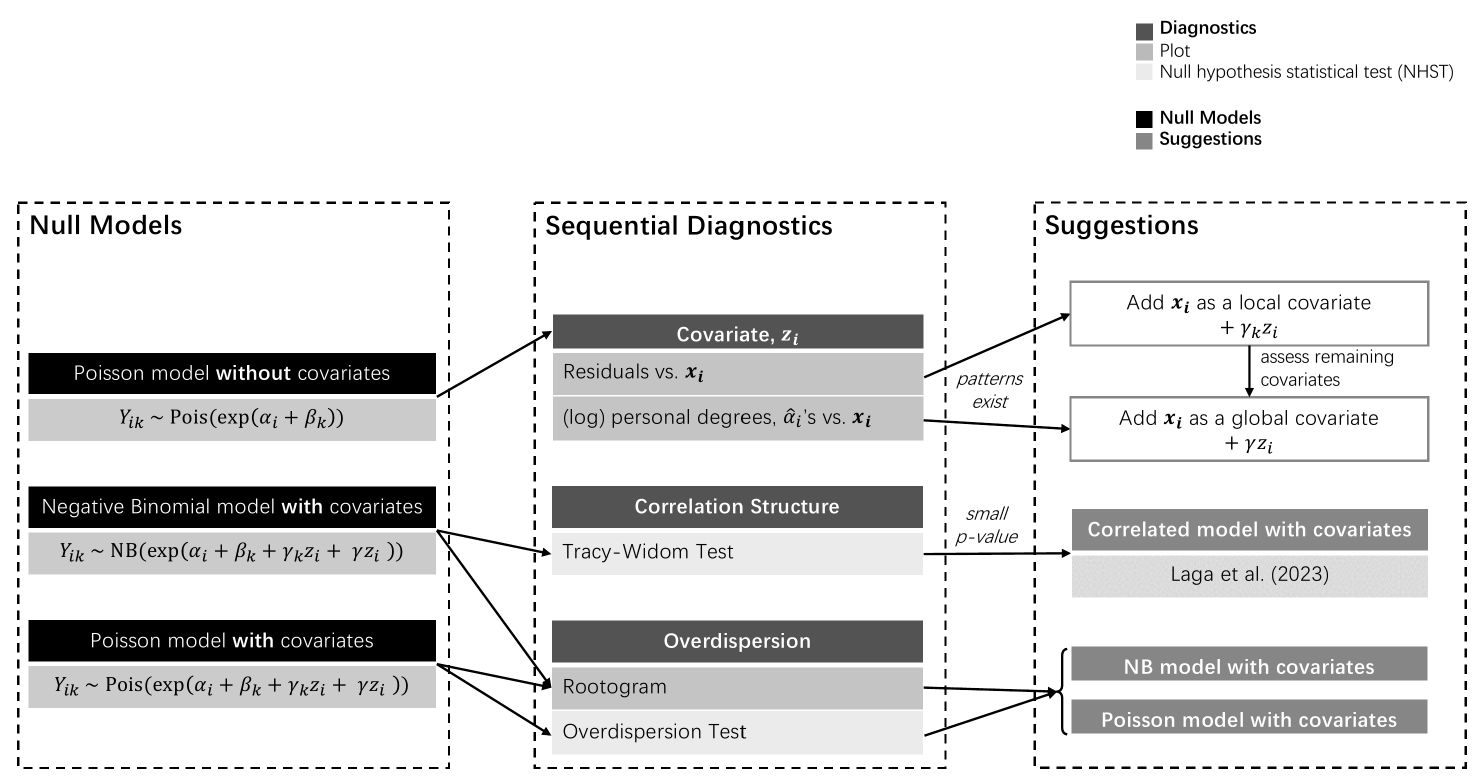}
    \caption{Proposed diagnostic structure for ARD models. A maximum of three null models are fit, and all diagnostics depend on these null models.}
    \label{fig:workflow}
\end{figure}

While more flexible ARD models are often closer to the true data-generating process, the most complex specifications are computationally expensive to estimate. We therefore recommend using diagnostics to identify which model features are necessary rather than defaulting to the most complex model. The diagnostic framework we propose evaluates three key aspects: covariate structure, correlation, and distributional assumptions. Among these, covariates are typically most informative and interpretable, while correlation and distributional diagnostics help determine whether added model complexity is justified by the data. The diagnostics help determine which departures from the simplest models are most influential and which modeling components are most important to include. In Section \ref{sec:tools}, we detail our recommended diagnostic process, outlined in Figure~\ref{fig:workflow} here. The procedure begins by fitting a simple Poisson ARD model without covariates, proceeds to identify relevant covariates, and concludes by using covariate-adjusted models to assess whether the correlation and distributional assumptions are satisfied or whether a more complex model is required.

Fitting multiple models, especially Bayesian ones, can be computationally expensive, so all diagnostics in this paper use point estimates. We take a frequentist approach and use maximum likelihood estimates (MLE), though many of the same models have Bayesian versions with researcher-specified priors. Our diagnostic framework extends naturally to these Bayesian settings by replacing the MLE with maximum a posteriori (MAP) estimates. Since MAP estimation produces point estimates which maximize the posterior density, the diagnostics follow the same procedure as when using the MLE, but now account for changes in the prior distributions. This allows practitioners to apply our diagnostic workflow regardless of whether they ultimately pursue frequentist or Bayesian inference. We use $\hat{\bm{\theta}}$ to denote either MLE or MAP estimates, depending on the modeling approach.

We propose diagnostics based on the models where parameter estimates can be obtained quickly using standard optimization tools (for example, \texttt{glmmTMB} for MLE \citep{glmmTMBR, glmmTMBJSS} and \texttt{Stan} for MAP \citep{stan}). For example, rather than fitting both correlated and uncorrelated models to compare fit, diagnostics from an uncorrelated model reveal whether correlation structure is needed. This avoids the computational expense of fitting all candidate models. This approach is computationally efficient even for large surveys. In our experiments, obtaining the MLE for a negative binomial ARD model on the Ukraine data required approximately 11 minutes to estimate parameters and residuals, while fitting a group correlated model using MCMC took roughly one week. This demonstrates the substantial reduction in computational effort when performing early diagnostics with point estimates based on simple models rather than full posterior inference.

%Assuming the covariate structure is correctly specified, we find that unmodeled correlation manifests as additional variance, while distributional misspecification (e.g., fitting Poisson when the data exhibit negative binomial overdispersion) does not create spurious residual correlation. Moreover, existing models that account for correlation assume Poisson-distributed counts with additional noise, implicitly approximating a negative binomial distribution \citep{laga2023correlated}. Despite having specified the correlation structure, distributional diagnostics remain valuable by quantifying the severity of overdispersion arising from both the correlation model itself and any residual unaccounted noise. By examining distributional assumptions for each group independently, these diagnostics reveal which groups exhibit greater overdispersion and allow comparison against a priori expectations.

\section{Diagnostic Tools}
\label{sec:tools}

We present a suite of diagnostic tools for assessing model adequacy in ARD models. These tools address three key aspects of model specification: identifying appropriate covariate structures through residual and coefficient plots, detecting correlation patterns across groups using an eigenvalue-based test, and evaluating distributional assumptions through rootograms and overdispersion metrics. Together, these diagnostics provide a systematic framework for model validation, allowing practitioners to determine whether local or global covariates are needed, test for group correlation structure, and assess whether a Poisson or negative binomial distribution better captures the data-generating process.

We recommend first explaining as much variation as possible through covariates, so that residual noise is captured by meaningful, interpretable variables that are straightforward to measure and collect. As shown in Section 5 of the Supplementary Information, deviating from this sequence can lead to misleading conclusions. Missing covariates often appear as additional noise, resulting in a negative binomial rather than a Poisson likelihood, or as spurious correlation, leading to an unnecessary correlation structure. However, in our experiments, the diagnostics reliably identify the correct covariate structure even when the distributional or correlation assumptions are misspecified. These findings are informally supported by the work of \cite{gourieroux1984pseudo}, who showed that Poisson regression coefficients remain consistent under distributional misspecification when the conditional mean is correctly specified.

After identifying the covariate structure, researchers should assess both correlation and distributional assumptions. Existing correlation models assume Poisson counts with additional noise, implicitly approximating a negative binomial distribution \citep{laga2023correlated}. The diagnostics therefore help determine whether correlation is strong enough to require explicit modeling or whether a negative binomial model adequately captures the overdispersion without added complexity. When correlation is weak or absent, the distributional diagnostics help distinguish whether a Poisson or negative binomial model is more appropriate, i.e., whether significant overdispersion remains after adjusting for covariates.

We use both Pearson residuals and randomized quantile residuals for the diagnostics \citep{mccullagh1989generalized, dunn1996randomized}. In our experiments, both residual types led to the same conclusions, so we do not recommend one over the other. We do not use raw residuals because for discrete count data like ARD, they exhibit non-normality, heteroscedasticity, and limited distinct values, making them unsuitable for diagnostics \citep{dunn1996randomized}. We use ``residuals'' in the manuscript to emphasize that both types may be used for all diagnostics except for those related to correlation structure, but all visualizations in this manuscript are produced with randomized quantile residuals.

\subsection{Covariates}

To identify relevant covariates, we examine residual patterns in two ways. First, we plot the residuals against their respective covariates, faceted by group $k$, to identify potential local covariates. When covariates are uncorrelated (or weakly correlated) with the responses, these plots will appear relatively flat. Second, we plot the estimated $\hat{\alpha}_i$'s (personal degrees, on the log scale) against their respective covariates to identify potential global covariates. As before, flat patterns suggest weak or no association, moderate to strong linear patterns suggest that the covariate should be included in our model, while curvature can indicate polynomial or non-linear relationships.

Since the global coefficient visualization often highlights both global and local covariates, after identifying local covariates from the faceted residual plots, we examine the global plots only for the remaining covariates. Note that defining the covariate structure is only a starting point. It identifies the structure needed for the remaining diagnostics, but it does not address issues such as multicollinearity among predictors.

\iffalse
\begin{figure}[!t]
    \centering
    \begin{subfigure}{0.49\textwidth}
        \includegraphics[width=\textwidth]{Diagnostic_Figures/ukraine_local_cov.pdf}
        \caption{}
        \label{fig:ukraine_cov_local}
    \end{subfigure}
    \begin{subfigure}{0.49\textwidth}
        \includegraphics[width=\textwidth]{Diagnostic_Figures/ukraine_global_cov.pdf}
        \caption{}
        \label{fig:ukraine_cov_global}
    \end{subfigure}
\caption{Global (a) and local (b) covariate plots for the Ukraine ARD.}
\label{fig:ukraine_cov}
\end{figure}
\fi

\begin{figure}[!t]
    \centering
    \includegraphics[width=1\textwidth]{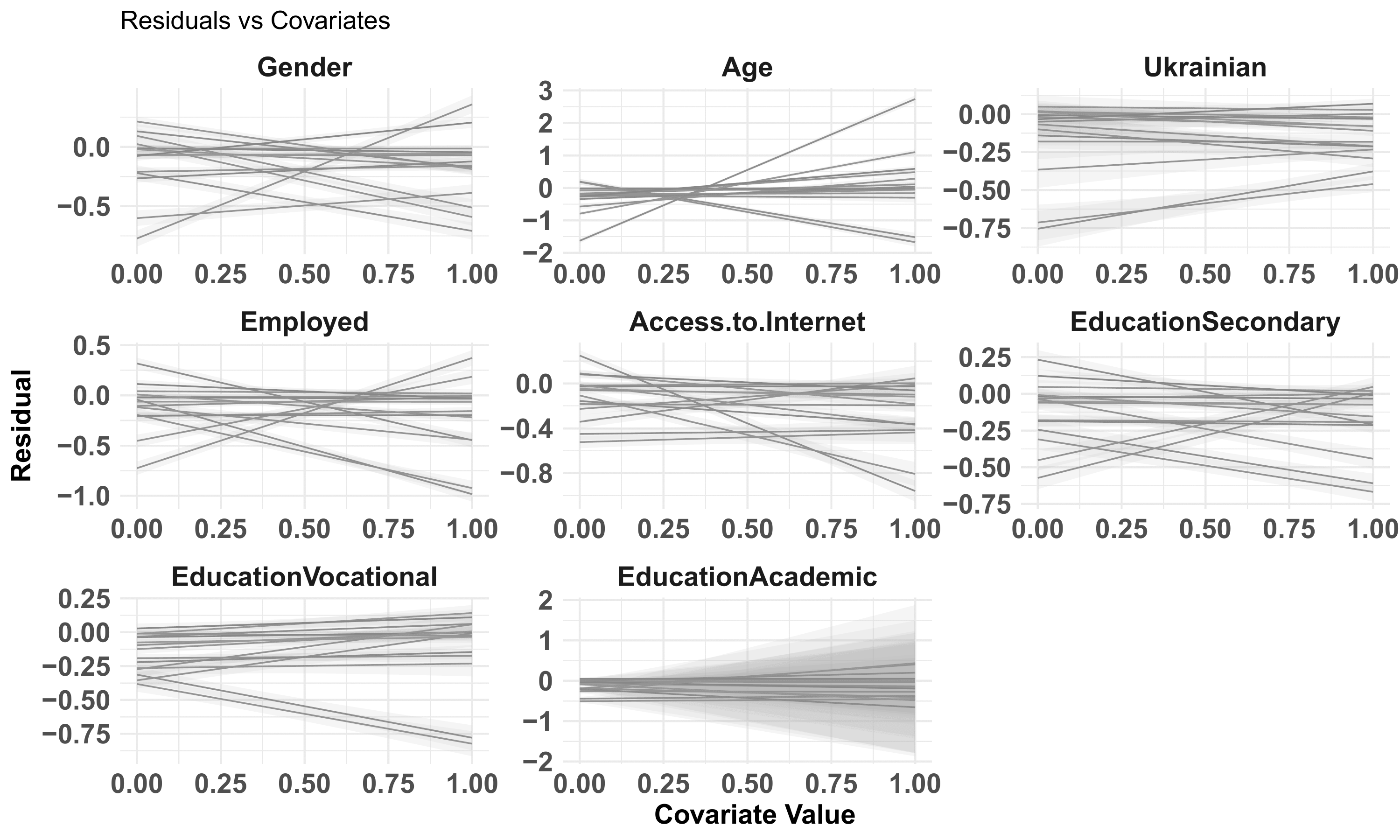}
    \caption{Local covariate plots for the Ukraine ARD.}
    \label{fig:ukraine_cov_local}
\end{figure}

\begin{figure}[!t]
    \centering
    \includegraphics[width=1\textwidth]{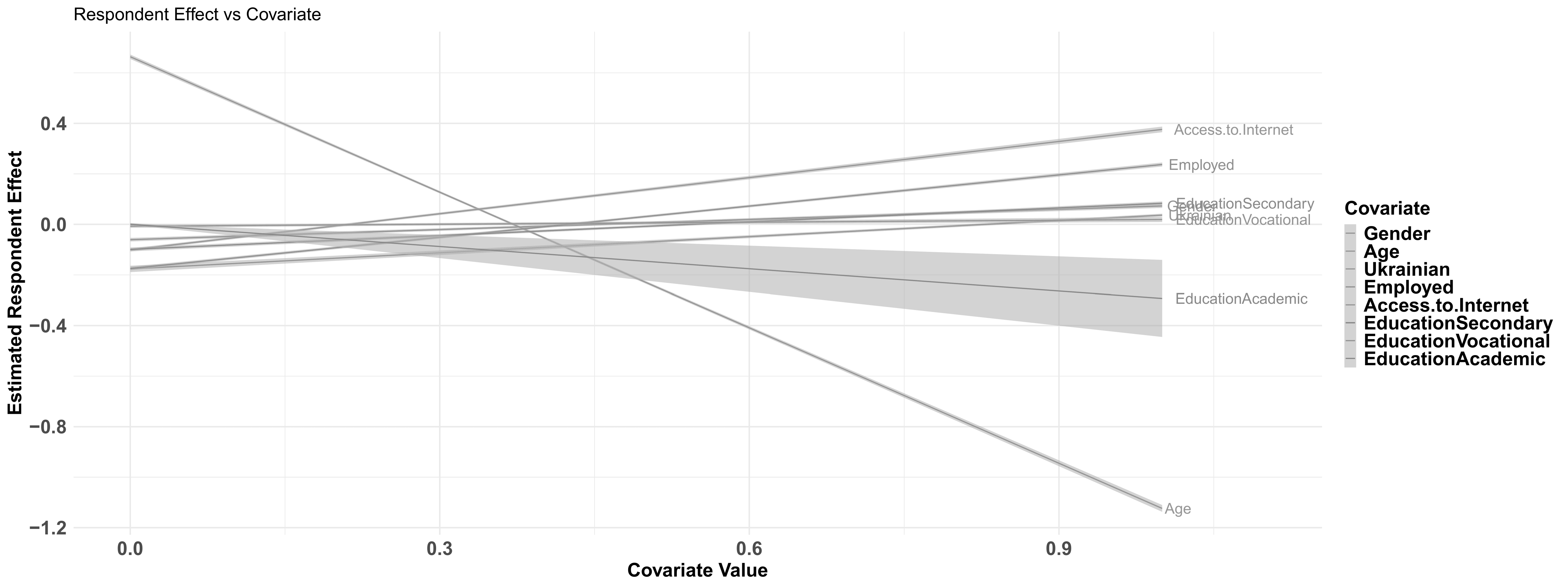}
    \caption{Global covariate plots for the Ukraine ARD.}
    \label{fig:ukraine_cov_global}
\end{figure}

\paragraph{Application to Ukraine ARD.}

We first examine residuals vs. covariates in Figure~\ref{fig:ukraine_cov_local} to identify local covariates, which include Gender, Age, Employed, Access to Internet, and Secondary Education. We then examine the $\hat{\alpha}_i$'s vs. covariates in Figure~\ref{fig:ukraine_cov_global} for the remaining covariates to identify global effects, which include Ukrainian, Vocational Education, and Academic Education. While all covariates are plotted here, we focus on those not identified as local. We note that other researchers may prefer to swap some of these local and global covariates depending on their comfort with complex covariate interactions and desire for parsimonious models.

\subsection{Correlation structure}

We analyze the residual correlation structure by examining the eigenvalues of the sample covariance matrix of the randomized quantile residual matrix $\bm{R}$. Under the null hypothesis of no group correlation, the largest eigenvalue of the sample covariance matrix of $\bm{R}$ should follow predictions from random matrix theory.

\cite{marvcenko1967distribution} study the convergence behavior of eigenvalues of $\bm{R}^T\bm{R}$, and \cite{tracy1994level} and \cite{tracy1996orthogonal} show that the largest eigenvalue $\lambda_1$, after appropriate centering and scaling, converges in distribution to the Tracy–Widom distribution:
\begin{equation*}
    \frac{\lambda_1 - \mu_n}{\sigma_n} \xrightarrow{d} \mathrm{TW}_1,
\end{equation*}
where
\begin{align*}
    \mu_n &= \left(\sqrt{n - 1} + \sqrt{K}\right)^2, \\
    \sigma_n &= \left(\sqrt{n - 1} + \sqrt{K}\right)
                \left(\frac{1}{\sqrt{n - 1}} + \frac{1}{\sqrt{K}}\right)^{1/3},
\end{align*}
and $\mathrm{TW}_1$ denotes the Tracy–Widom distribution for the Gaussian Orthogonal Ensemble \citep{johnstone2001distribution}. \cite{johnstone2001distribution} show that this convergence holds if $n, K \to \infty$ with $n/K \to \gamma \in (0, \infty)$, and \cite{karoui2003largest} extend this result to the cases $n/K \to \infty$ and $n/K \to 0$. For ARD, $K$ remains constant, violating the necessary conditions. However, the theory is useful as an approximation for small sample sizes.

For finite samples, \cite{ma2012accuracy} and \cite{johnstone2006high} recommend improved centering and scaling constants based on second-order accuracy:
\begin{align*}
    \mu_n &= \left(\sqrt{n - \tfrac{1}{2}} + \sqrt{K - \tfrac{1}{2}}\right)^2, \\
    \sigma_n &= \left(\sqrt{n - \tfrac{1}{2}} + \sqrt{K - \tfrac{1}{2}}\right)
               \left(\frac{1}{\sqrt{n - \tfrac{1}{2}}} + \frac{1}{\sqrt{K - \tfrac{1}{2}}}\right)^{1/3}.
\end{align*}

\iffalse
In these results, the authors do not account for the loss of degrees of freedom that arises from estimating parameters when computing $\bm{R}$. In our setting, we estimate $n$ respondent-specific parameters $\alpha_i$ and $K$ group-specific parameters $\beta_k$, which imposes constraints on the residuals. Motivated by this observation, we propose a slightly larger penalty of $n - 1$ and $K - 1$ in the centering and scaling terms:
\begin{align*}
    \mu_n &= \left(\sqrt{n - 1} + \sqrt{K - 1}\right)^2, \\
    \sigma_n &= \left(\sqrt{n - 1} + \sqrt{K - 1}\right)
               \left(\frac{1}{\sqrt{n - 1}} + \frac{1}{\sqrt{K - 1}}\right)^{1/3}.
\end{align*}
Empirically, all centering and scaling adjustments produce conservative results, meaning they are unlikely to reject the null hypothesis when it is true. Conservative tests are preferable in this setting since we do not want to fit correlated models unless correlation is clearly present. We find that our proposed adjustment provides more accurate type I error rates while remaining conservative, and all options are equivalent as $n, K \to \infty$. Users who prefer a more conservative test may still choose the original adjustments, all of which are implemented in the accompanying software.
\fi

\cite{johnstone2006high} propose a hypothesis test for the largest eigenvalue that we adapt to the ARD context. As they note, testing the largest eigenvalue is equivalent to testing whether the covariance matrix is equal to the identity.

\begin{lemma}[Tracy–Widom Test for Group Correlation]
    Let $\bm{R}$ be an $n \times K$ matrix of randomized quantile residuals, where $n$ is the number of respondents and $K$ is the number of groups. Define the sample covariance matrix
    \begin{equation*}
        \bm{S} = \frac{1}{n-1} \bm{R}^T \bm{R},
    \end{equation*}
    and let $\lambda_1$ be the largest eigenvalue of $\bm{S}$.
    
    Under the null hypothesis
    \begin{equation*}
        H_0: \text{the columns of } \bm{R} \text{ are uncorrelated},
    \end{equation*}
    the residuals are asymptotically independent across groups.
    
    The test statistic is
    \begin{equation*}
        T = \frac{\lambda_1 - n\mu_n}{n\sigma_n},
    \end{equation*}
    for $\mu_n$ and $\sigma_n$ defined above, and we reject $H_0$ at level $\alpha$ if $T > q_{1-\alpha}$, where $q_{1-\alpha}$ is the $(1 - \alpha)$ quantile of the Tracy–Widom distribution $\mathrm{TW}_1$.
    
    Large values of $T$ indicate that the largest eigenvalue is larger than expected under random noise, providing evidence of residual correlation across groups.
\end{lemma}

This Tracy–Widom test is valid because when model parameters are consistently estimated under the true model, the randomized quantile residuals $R_{i,k}$ are asymptotically standard normal \citep{dunn1996randomized}. Although finite samples introduce nonzero correlations among residuals due to shared parameter estimates, these correlations vanish asymptotically. We note again that the sample covariance matrix is computed using $n - 1$, while the centering and scaling constants are defined using $n$. This adjustment is based on empirical findings and leads to improved type I error control in small samples.

The same reasoning applies when a negative binomial model is fit to Poisson data. Under a true Poisson data-generating process, the negative binomial model approximates the Poisson distribution as the overdispersion parameter grows, yielding an $\bm{R}$ matrix that behaves almost identically. The converse is not true, and randomized quantile residuals from a Poisson model fit to negative binomial ARD do not follow a standard normal distribution. For this reason, we recommend implementing Test 1 using a fitted negative binomial model.

We evaluated the type I error rate of the permutation test under small-sample conditions and under distributional misspecification. The results (reported in Section 2 of the Supplementary Information) show that the test is slightly conservative across settings, particularly when a negative binomial model is fit to Poisson-generated data. This conservatism is desirable in our setting because we do not recommend fitting correlated models unless correlation is clearly present. Overall, the type I error was well controlled and never exceeded 0.05 across 100 simulations, except when fitting a Poisson model to data generated from a negative binomial distribution, where the type I error was 1. These findings indicate that Test 1 provides reliable inference in practice when the assumed distribution is not severely misspecified.

\begin{figure}[!t]
    \centering
        \includegraphics[width=\textwidth]{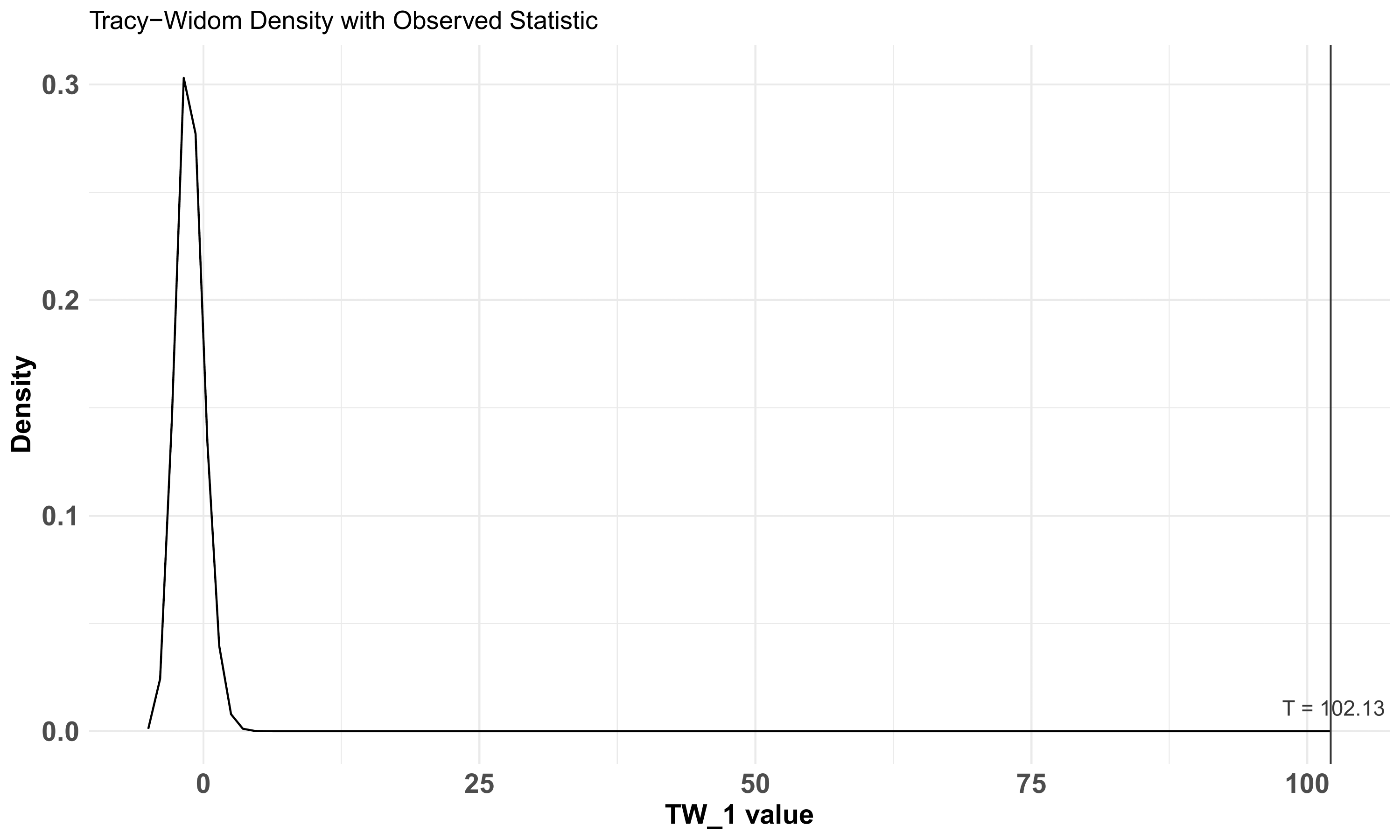}
        \caption{Tracy-Widom correlation test for the Ukraine ARD.}
        \label{fig:ukraine_tw}
\end{figure}

\paragraph{Application to Ukraine ARD.}

The Tracy-Widom test in Figure~\ref{fig:ukraine_tw} provides strong evidence of group correlation ($T = 102.13$), consistent with the findings in \cite{laga2023correlated}. The observed test statistic falls far into the tail of the Tracy-Widom distribution, indicating that the largest eigenvalue of the residual correlation matrix is substantially larger than expected under independence.

\subsection{Distributional Tests}

\subsubsection{Rootogram}

\begin{figure}[ht]
    \centering
    \begin{subfigure}[b]{0.8\textwidth}
        \includegraphics[width=\textwidth]{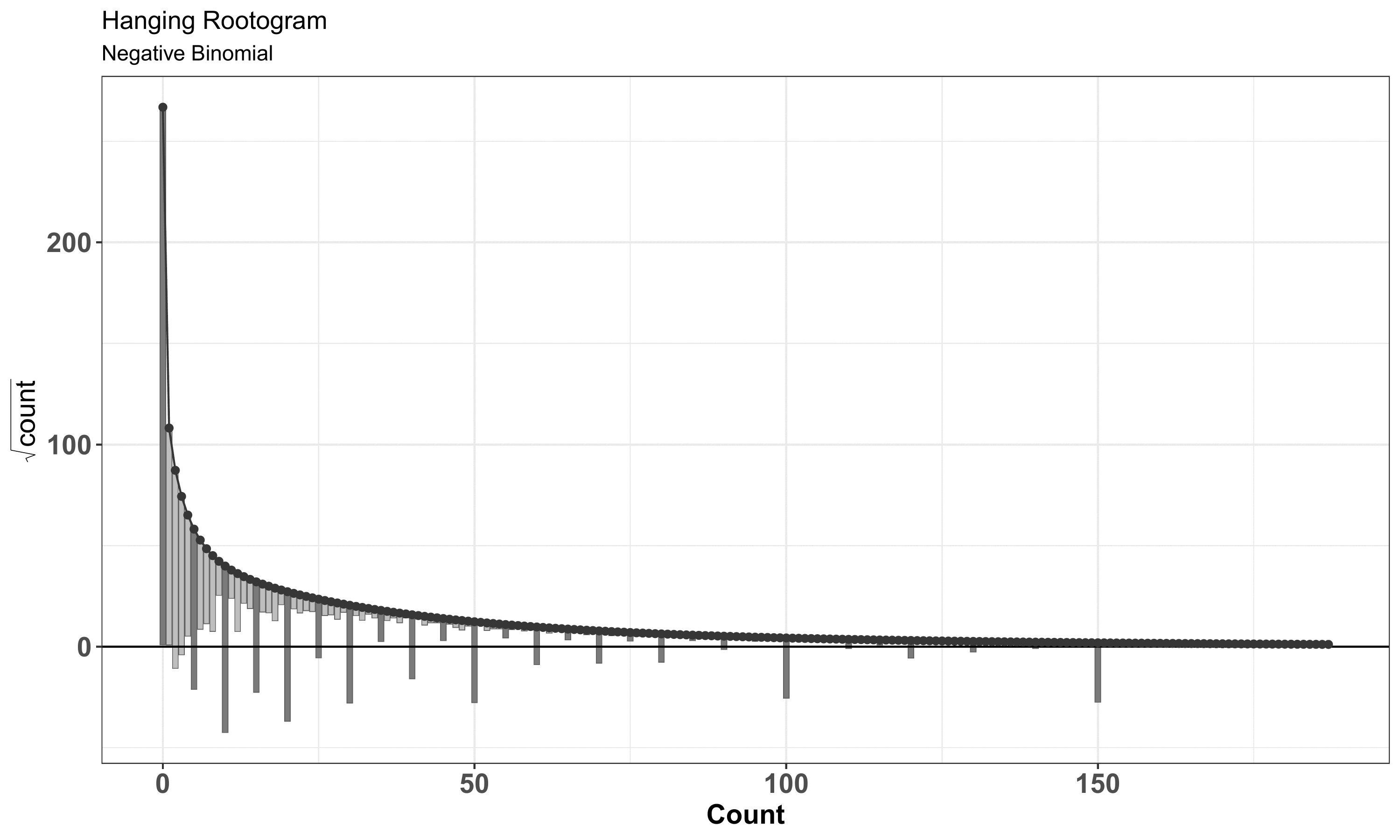}
    \end{subfigure}

    \begin{subfigure}[b]{1\textwidth}
        \includegraphics[width=\textwidth]{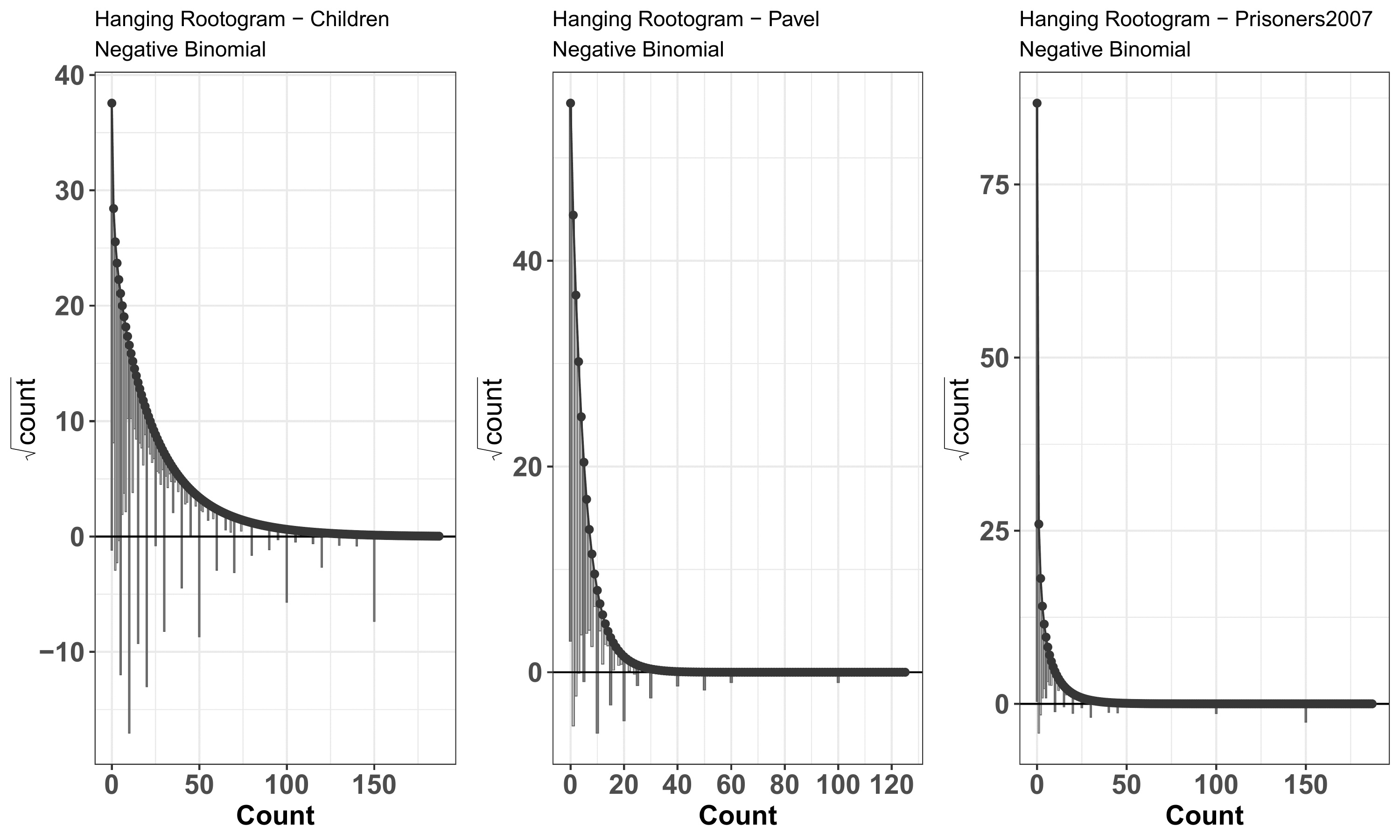}
    \end{subfigure}

    \caption{Overall and group rootograms for the Ukraine ARD. The groups correspond to ``children aged 10-13", ``men named Pavel'', and ``men who were in prison in 2007". Multiple of five are shown in darker gray.}
    \label{fig:ukraine_root}
\end{figure}

We first propose evaluating the appropriate distribution using the rootogram. The rootogram plots the discrepancy between empirical and expected frequencies and can be used to evaluate general goodness of fit of a probability model. Formally, given $\hat{\bm{\theta}}$, the rootogram plots some form of $\sqrt{\text{obs}_j}$ vs $\sqrt{\text{exp}_j}$, where
\begin{align*}
    \text{obs}_j &= \sum_{m=1} I(y_m = j)\\
    \text{exp}_j &= \sum_{m=1} f(j; \hat{\bm{\theta}_m}),
\end{align*}
where $j = 0, 1, 2, \ldots$ represents possible integer outcomes and $f(\cdot; \cdot)$ is the density of the response distribution \citep{kleiber2016visualizing}. Specifically, ``hanging'' rootograms plot bars from $\text{exp}_j$ to $\text{exp}_j - \text{obs}_j$ with an overlaid curve of expected frequencies $\text{exp}_j$, ``standing'' rootograms plot bars with height $\text{obs}_j$ with an overlaid curve of expected frequencies, and ``suspended'' rootograms plot bars with height $\text{exp}_j - \text{obs}_j$ with an overlaid curve of expected frequencies. All visualization lead to identical conclusions. For the Ukraine data, we create hanging rootograms in Figure~\ref{fig:ukraine_root}. Observations where the gray bars do not touch the x-axis indicate poor fit, with bars above the x-axis indicating overestimation and those that hang below corresponding to underestimation.

In addition to examining the rootogram across all observations, group-level rootograms can be constructed to evaluate model fit within each group. By plotting the discrepancy between observed and expected frequencies for each group, these rootograms explore how the assumed response distribution behaves across groups. Comparing group-level rootograms to the overall rootogram allows one to assess whether misfit arises uniformly across the data or is concentrated in specific groups. Together, overall and group-level rootograms provide a complementary approach for diagnosing the adequacy of the fitted model both globally and within subgroups.

\paragraph{Application to Ukraine ARD.}

The overall rootogram in Figure~\ref{fig:ukraine_root} (top) suggests that the negative binomial model provides a reasonable fit to the observed ARD, with discrepancies primarily attributable to rounding toward multiples of five (shown in darker gray). However, the group-specific rootograms (bottom) reveal substantial heterogeneity across groups, with the negative binomial model struggling to capture the complexity of most groups.

\subsubsection{Overdispersion test}

Next, we propose studying the choice of distribution via dispersion metrics. One can evaluate the acceptability of the Poisson distribution by studying the predictive performance of the dispersion index test-statistic
\begin{equation*}
    D_k = \frac{1}{\bar{y}}\sum_{i=1}^n (y_{ik} - \hat{\lambda}_{ik})^2 / \hat{\lambda}_{ik}.
\end{equation*}
Under the null distribution, $D_k \sim \chi^2_{n - p}$, where $p$ is the total number of covariates. Based on the dispersion index, we can measure how overdispersed the data is within each group with respect to the Poisson distribution. The overdispersion test can be thought of as a numeric analogue to the rootogram. The normal distribution version of the overdispersion test implemented in the \texttt{dispersiontest} function in the \texttt{AER} R package may also be used and produces almost identical p-values \citep{kleiberAER, cameron1990regression}.

\begin{figure}
\centering
    \includegraphics[width=0.9\textwidth]{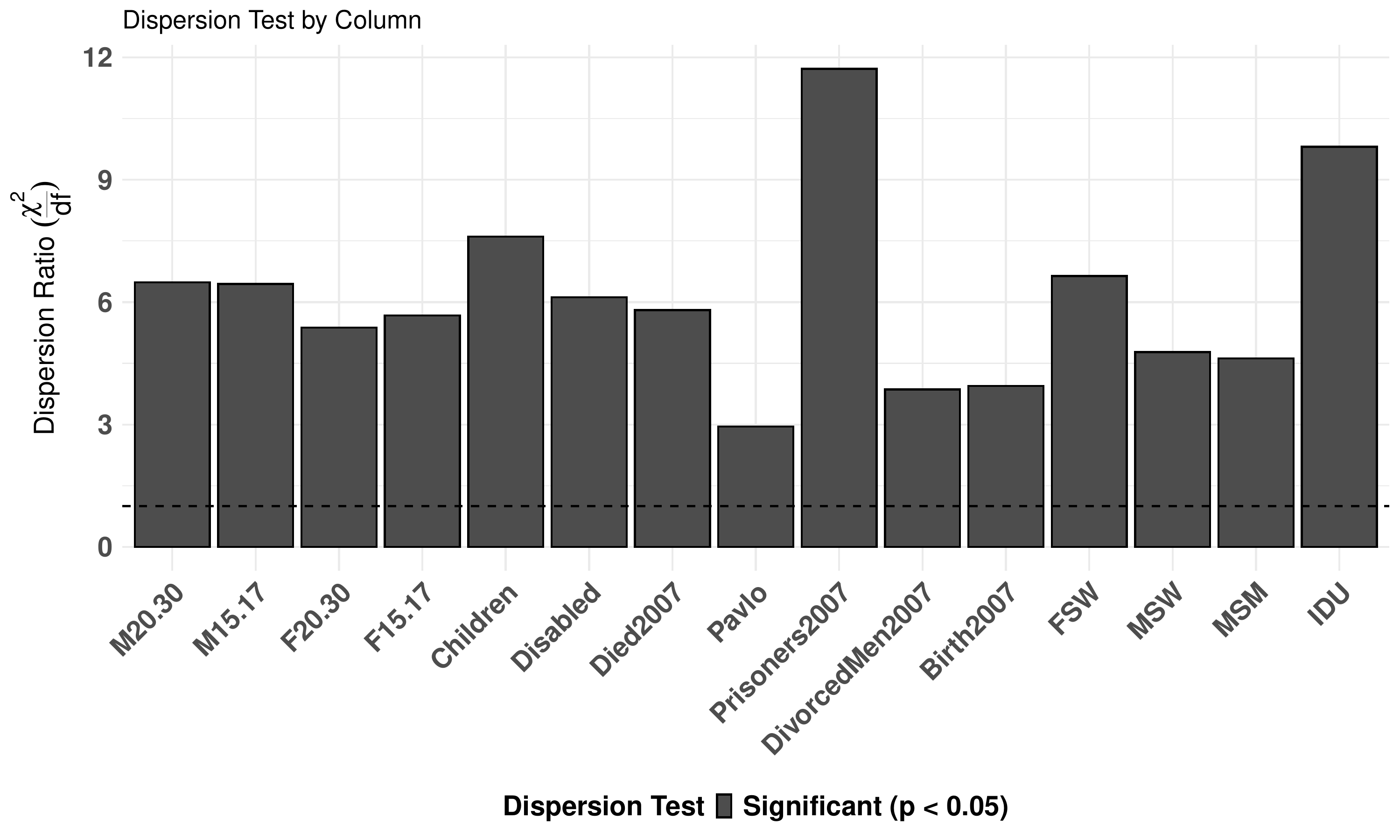}
\caption{Dispersion metric for Ukraine Poisson fit.}
\label{fig:ukraine_disp}
\end{figure}

\paragraph{Application to Ukraine ARD.}

Similar to the group rootograms, the estimated dispersion ratios for each group for a Poisson model in Figure~\ref{fig:ukraine_disp} suggest that some groups are better approximated by a Poisson likelihood (e.g., Pavlo) while other groups exhibit large overdispersion, necessitating a negative binomial likelihood.

%\subsection{Size estimates}

%For completeness, we briefly review the leave-one-out prevalence estimates commonly used in ARD studies. Let $k$ index the groups for which $\beta_k$ is known (often called probe groups or known populations). Leave-one-out cross-validation proceeds by iterating over these groups, treating $\beta_k$ as unknown, estimating it from the model, and then comparing the estimated value to the true prevalence.

It is unclear how these leave-one-out estimates should be used beyond providing a general sense of how accurately the model recovers known prevalences. Although some authors have suggested discarding probe groups whose leave-one-out estimates perform poorly \citep{habecker2015improving}, there is little evidence that doing so improves prevalence estimates for the target populations. We therefore recommend interpreting these results with caution.

\FloatBarrier % Just for temporary spacing
\subsection{Working Example Summary}

Across our diagnostic analyses for the Ukraine ARD, we find that both local and global covariates contribute meaningful information, with local effects including Gender, Age, Employment, Internet Access, and Secondary Education, and global effects including Ukrainian, Vocational Education, and Academic Education. The Tracy-Widom test provides strong evidence of group correlation, indicating substantial residual structure beyond independence. Model fit assessments show that a negative binomial model generally captures the overall ARD distribution, though group-specific analyses reveal heterogeneity, with certain groups (e.g., Prisoners) deviating from model expectations. Dispersion estimates further highlight that while some groups are well-approximated by a Poisson likelihood, others exhibit substantial overdispersion, justifying the use of a negative binomial model. Based on these results, and given the remaining lack of fit in the negative binomial model, we recommend using a correlated or degree-correlated model to obtain final \citep[e.g.,][]{laga2023correlated}, reliable interpretations.

\FloatBarrier

\FloatBarrier

\section{Discussion}
\label{sec:discussion}

In this manuscript we provide a suite of diagnostics, both visual and numeric, to help practitioners choose an appropriate ARD model. These diagnostics are not intended to give binary answers. Instead, they allow researchers to judge the extent to which a model aligns with the data and to identify where departures occur. Because social networks are complex, it is ultimately the researcher's responsibility to determine when a model provides a good-enough representation for the scientific question at hand.

The diagnostic tools are general and can be adapted or extended to suit specific analytical goals. For instance, the covariate plots can easily be modified to examine interactions by faceting on a second covariate. Similar modifications can be made for the distributional and correlation diagnostics, allowing researchers to explore patterns that are particular to their application or dataset.

While we propose diagnostics to help choose covariate, correlation, and distributional structure, researchers have traditionally relied on leave-one-out procedures to evaluate population size estimates against known group sizes. However, it remains unclear how these leave-one-out estimates should be used beyond providing a general sense of model performance. Although some authors have suggested discarding probe groups with poor leave-one-out estimates \citep{habecker2015improving}, there is little evidence that doing so improves prevalence estimates for target populations. We therefore recommend interpreting these results with caution.

As new ARD models continue to be developed, we encourage authors to provide fast and accessible diagnostics that help assess whether their proposed methods are appropriate for a given dataset. Doing so will allow analysts to compare models efficiently and transparently, rather than selecting a model arbitrarily or relying solely on computationally intensive estimation and model checking techniques. Broad adoption of diagnostic tools will support more rigorous model evaluation and help practitioners match their data to the methods that are most suitable for their goals.

A limitation of the proposed diagnostics is that they depend on point estimates and ignore any measure of uncertainty. While this is an intentional choice to reduce computational efforts, diagnostics may differ when using full posteriors.

Despite extensive efforts, we were unable to identify any visualization or diagnostic based solely on an uncorrelated model that reliably detects degree correlation in the observed data. Our findings indicate that the correlation structure across groups must first be accounted for before any residual dependence can be meaningfully attributed to degree. Because group and degree correlations are highly related, unaccounted for group correlation appears to manifest as degree correlation. A natural idea is to examine the relationship between the residual matrix and the estimated log-degrees. However, when the degrees are estimated under an uncorrelated model, the estimates are themselves correlated with the residuals, making such diagnostics uninformative. We explored multiple alternatives, including parametric bootstrap, nonparametric bootstrap, and permutation approaches, but none successfully removed this induced dependence. Instead, when group correlation is present, practitioners should either fit the degree-correlated model directly or rely on posterior predictive checks based on the correlated model.

There are also modifications of traditional ARD models that we do not consider. For example, \cite{diaz2026robust} propose robust versions of NSUM estimators designed to reduce the influence of outliers and related reporting errors. Although these methods build on the model-based estimators discussed above, it is not straightforward to adapt our diagnostics to these robust settings. Our diagnostics rely heavily on residual behavior, while robust estimators intentionally downweight or ignore influential observations, causing their residuals to behave differently and limiting their interpretability. As additional extensions of model-based ARD estimators are developed, it will be important to study how diagnostic tools can be modified so that they remain informative and compatible with these robust methods.

A final direction for future work is the development of diagnostics that address the sampling side of ARD. The diagnostics introduced here focus on model adequacy and do not evaluate whether respondents understood the questions, how they rounded or heaped their answers, or whether certain groups were systematically misreported or confused with others. These issues can introduce bias before modeling even begins, and they may vary across surveys, populations, and modes of data collection. Tools that detect such patterns, quantify their effect, or suggest improvements to survey design would provide a valuable complement to model-based diagnostics and strengthen the reliability of inferences drawn from ARD.

\bibliography{ard_bib}
\bibliographystyle{apalike}

\clearpage

\appendix

% Redefine figure and table naming for the appendix
\renewcommand{\figurename}{Supplementary Figure}
\renewcommand{\tablename}{Supplementary Table}
\renewcommand{\thefigure}{\arabic{figure}}
\renewcommand{\thetable}{\arabic{table}}
\setcounter{figure}{0}
\setcounter{table}{0}

\begin{center}
    {\LARGE\bf Supplementary Information for ``Evaluating Aggregated Relational Data Models with Simple Diagnostics''}
\end{center}
  \medskip

\section{Overview}

This document provides supplementary materials for the main manuscript, including detailed simulation studies and additional methodological validation.

In Section 2, we evaluate the small sample size type I error rates of the Tracy-Widom test for group correlation across various sample sizes ($n \in \{100, 500, 1000\}$), number of groups ($K \in \{10, 20\}$), distributional assumptions (Poisson and negative binomial), and correction factors. The results demonstrate that the test maintains appropriate type I error control when the fitted model is at least as complex as the true data-generating model.

In Section 3, we apply our diagnostic workflow to four simulated datasets that span the range of scenarios encountered in practice. The simulations vary in their distributional family (Poisson vs.\ negative binomial), covariate structure (0-4 active covariates with varying local and global effects), and correlation structure (none, group correlation, or degree correlation). For each simulation, we demonstrate how the covariate, correlation, and distribution diagnostics guide model selection when applied in the correct order.

Finally, in Section 4, we use the data from Simulation 1 to demonstrate the consequences of performing diagnostics in the wrong order. Specifically, we show how checking for correlation and distributional issues before accounting for covariate effects can lead to incorrect conclusions, as missing covariates can manifest as apparent correlation or overdispersion.

\section{Tracy-Widom Test Calibration}

We evaluate the small sample size type I error rate for the Tracy-Widom test via a simulation study. We simulated and fit 100 data sets for all combinations of $n \in \{100, 500, 1000\}$, $K \in \{10, 20\}$, and either no correction factor or a correction factor of 1/2. For each replication, we simulated data from a Poisson and negative binomial distribution and fit Poisson and negative binomial models to each dataset.

We conducted the group correlation Tracy-Widom test across all model combinations, with type I error rates presented in Supplementary Table~\ref{tab:error_rate}. The results demonstrate that when the assumed model is at least as complex as the true data-generating model (i.e., assuming Poisson or negative binomial for truly Poisson data, or assuming negative binomial for truly negative binomial data), the type I error rate remains below the nominal 0.05 level. Applying a correction factor of 1/2 yields higher type I error rates compared to using no correction and was always closer to the nominal level.

As shown in Supplementary Section~\ref{sec:sim4}, the test maintains sufficient power to detect true correlation even when a negative binomial distribution is assumed for Poisson-generated data. These findings suggest that assuming a negative binomial distribution when performing the Tracy-Widom test for group correlation is both safer and more advantageous, as this approach errs on the side of parsimony by making it harder to incorrectly reject the null hypothesis of no correlation.

\begin{table}[!h]
\centering
\caption{Type I error rate of the Tracy-Widom test for all corrections.}
\label{tab:error_rate}
\resizebox{\textwidth}{!}{%
\begin{tabular}[t]{lllrrrrrr}
\toprule
\multicolumn{3}{c}{ } & \multicolumn{3}{c}{K = 10} & \multicolumn{3}{c}{K = 20} \\
\cmidrule(l{3pt}r{3pt}){4-6} \cmidrule(l{3pt}r{3pt}){7-9}
Data Distribution & Model Distribution & correction & n=100 & n=500 & n=1000 & n=100 & n=500 & n=1000\\
\midrule
 &  & none & 0.02 & 0.01 & 0.01 & 0.03 & 0.00 & 0.00\\
 & \multirow{-2}{*}{\raggedright\arraybackslash Poisson} & 1/2 & 0.02 & 0.01 & 0.01 & 0.05 & 0.00 & 0.00\\
 &  & none & 0.00 & 0.00 & 0.00 & 0.00 & 0.00 & 0.00\\
\multirow{-4}{*}[0.5\dimexpr\aboverulesep+\belowrulesep+\cmidrulewidth]{\raggedright\arraybackslash Poisson} & \multirow{-2}{*}{\raggedright\arraybackslash Negative Binomial} & 1/2 & 0.00 & 0.00 & 0.00 & 0.00 & 0.00 & 0.00\\
\cmidrule{1-9}
 &  & none & 1.00 & 1.00 & 1.00 & 1.00 & 1.00 & 1.00\\
 & \multirow{-2}{*}{\raggedright\arraybackslash Poisson} & 1/2 & 1.00 & 1.00 & 1.00 & 1.00 & 1.00 & 1.00\\
 &  & none & 0.00 & 0.00 & 0.00 & 0.00 & 0.00 & 0.00\\
\multirow{-4}{*}[0.5\dimexpr\aboverulesep+\belowrulesep+\cmidrulewidth]{\raggedright\arraybackslash Negative Binomial} & \multirow{-2}{*}{\raggedright\arraybackslash Negative Binomial} & 1/2 & 0.00 & 0.00 & 0.00 & 0.01 & 0.00 & 0.00\\
\bottomrule
\end{tabular}
}
\end{table}

\clearpage

\section{Simulation study}
\label{sec:simulations}

We apply our diagnostic workflow to four simulated datasets to demonstrate its performance under different data-generating scenarios. Each simulation consists of aggregate relational data (ARD) with $n = 500$ respondents and $K = 20$ groups.

\textbf{Simulation 1} generates data from a Poisson distribution with six potential covariates, of which three are active (two local and one global), and no correlation structure.

\textbf{Simulation 2} generates data from a negative binomial distribution with no covariates and group correlation.

\textbf{Simulation 3} generates data from a negative binomial distribution with six potential covariates, of which three are active (one local and two global), and degree correlation.

\textbf{Simulation 4} generates data from a Poisson distribution with six potential covariates, of which four are active (two local and two global), and group correlation.

These simulations span the range of scenarios encountered in practice: different distributional families (Poisson vs.\ negative binomial), varying covariate structures, and different types of correlation (none, group, or degree).

\clearpage

% Simulation 1
\subsection{Simulation 1}

\textbf{Covariate Structure:} The covariate structure can be identified using the diagnostic plots in Supplementary Figure~\ref{fig:sim1_cov}. These plots indicate that X3 and X6 function as local covariates, while X5 acts as a global covariate. Although X6 also shows a nonzero slope in the global plot, its effect in the local plots make it clear that it should be classified as a local covariate.\\

\noindent \textbf{Correlation Structure:} The correlation structure is evaluated using the Tracy-Widom test, with results shown in Supplementary Figure~\ref{fig:sim1_corr}. The observed test statistic is small relative to the null distribution for both the Poisson and negative binomial models, indicating no evidence of residual correlation structure.\\

\begin{figure}[H]
    \centering
    \includegraphics[width=0.6\textwidth]{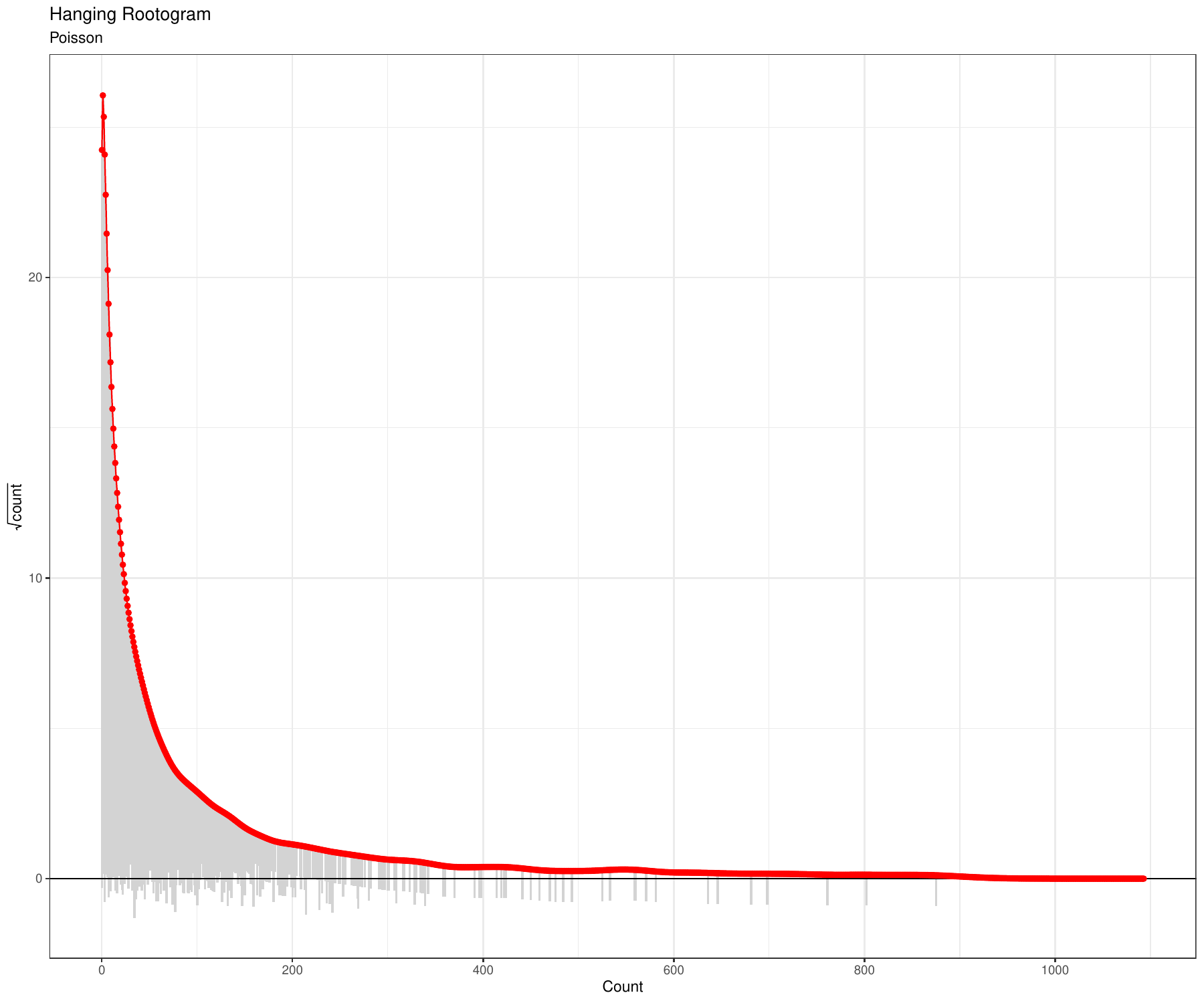}
    
    \includegraphics[width=0.6\textwidth]{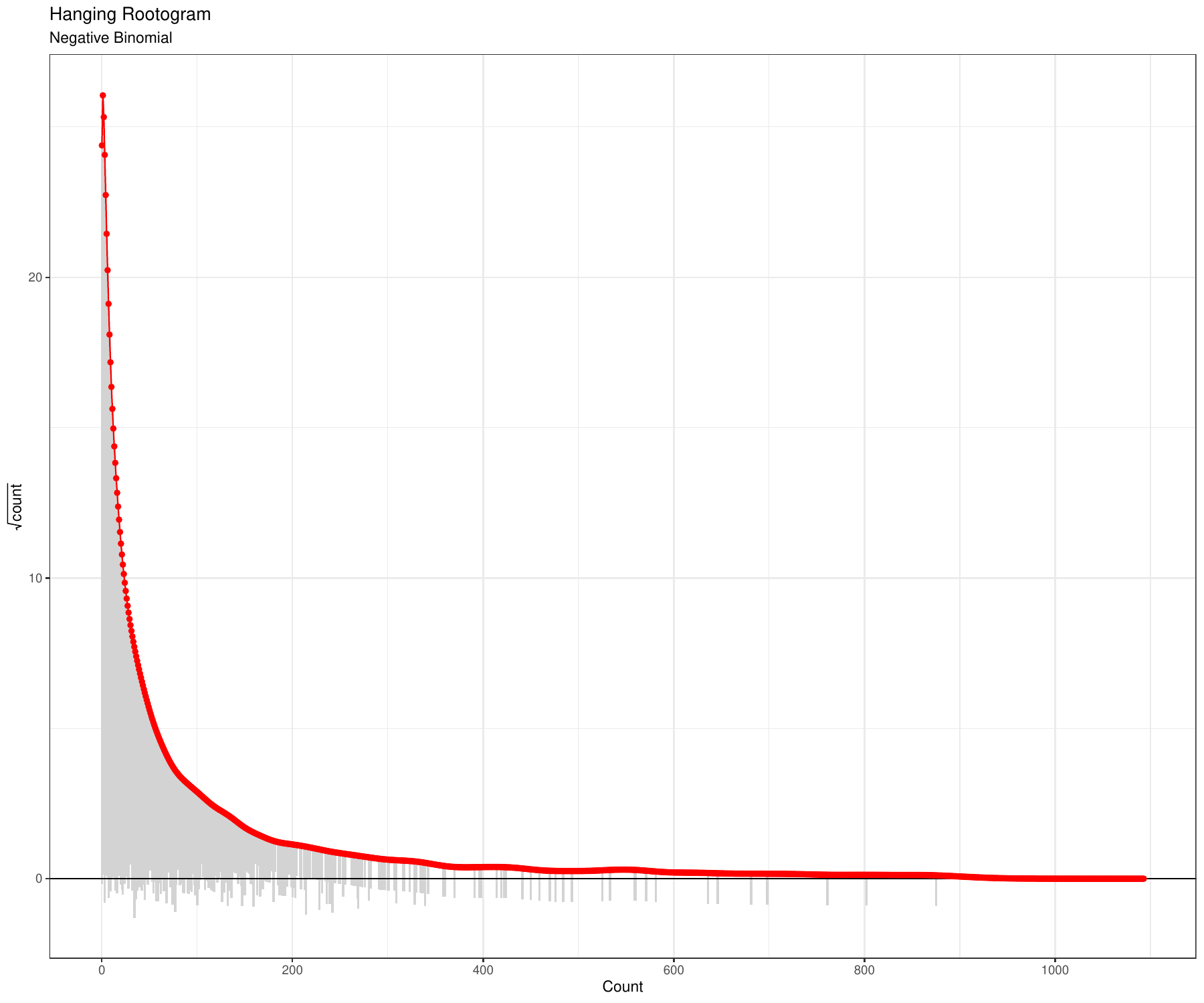}
    \caption{(Simulation 1) Overall rootogram diagnostics for Poisson (top) and negative binomial (bottom) models}
    \label{fig:sim1_root_all}
\end{figure}

\begin{figure}[H]
    \centering
    \includegraphics[width=\textwidth]{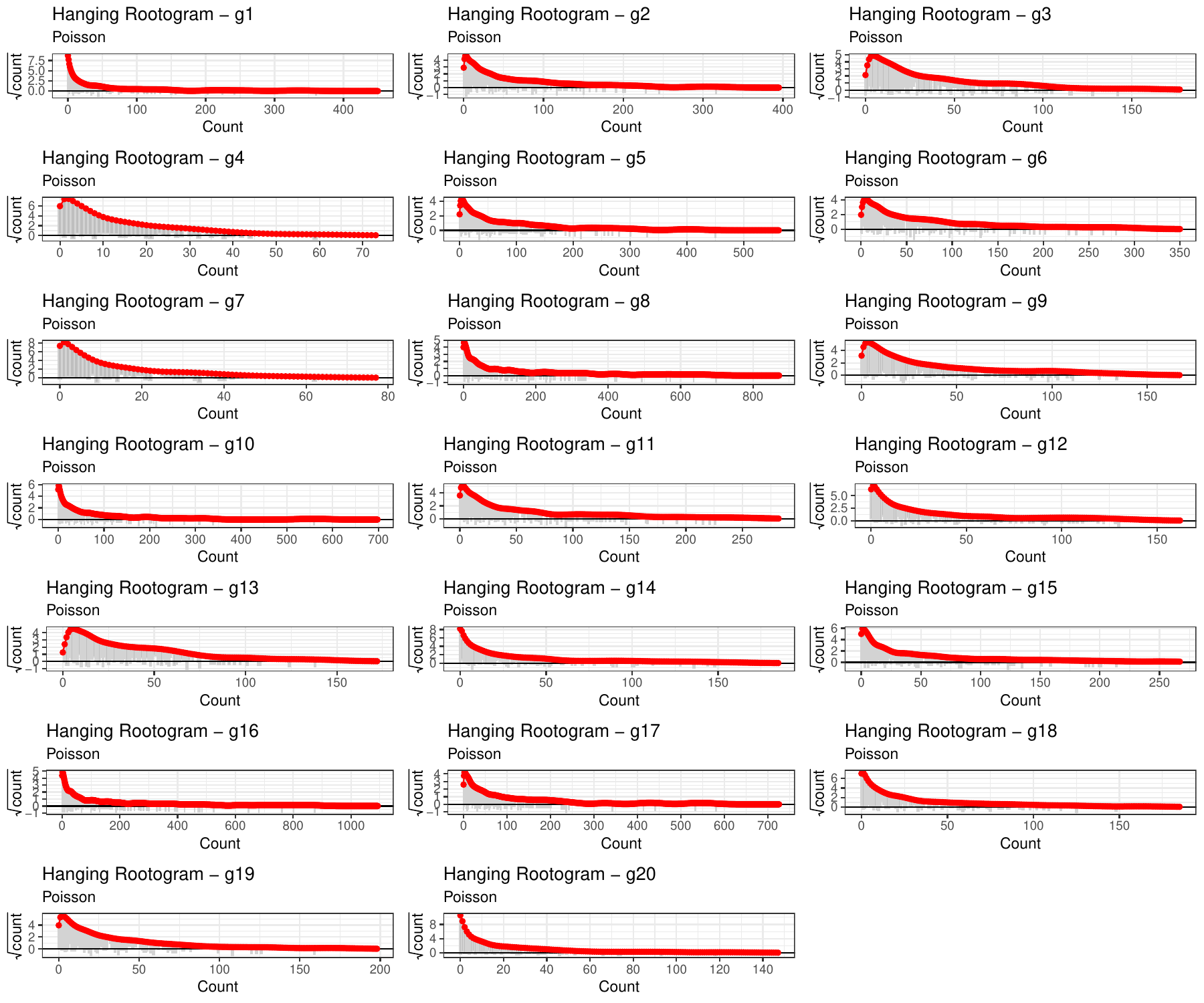}
    \caption{(Simulation 1) Poisson rootogram diagnostics by group}
    \label{fig:sim1_pois_root_group}
\end{figure}

\begin{figure}[H]
    \centering
    \includegraphics[width=\textwidth]{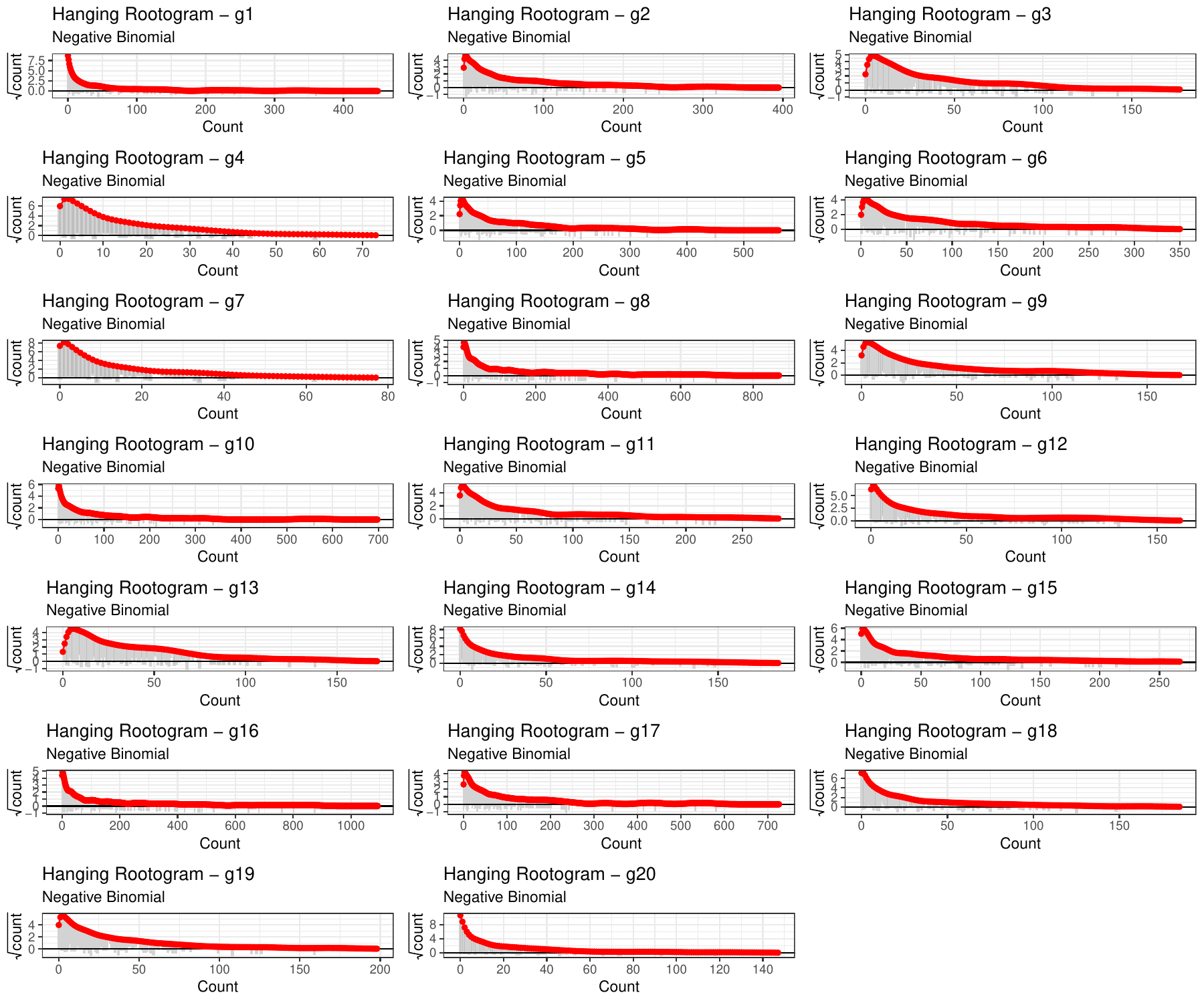}
    \caption{(Simulation 1) Negative binomial rootogram diagnostics by group}
    \label{fig:sim1_nb_root_group}
\end{figure}

\begin{figure}[H]
    \centering
    \includegraphics[width=0.7\textwidth]{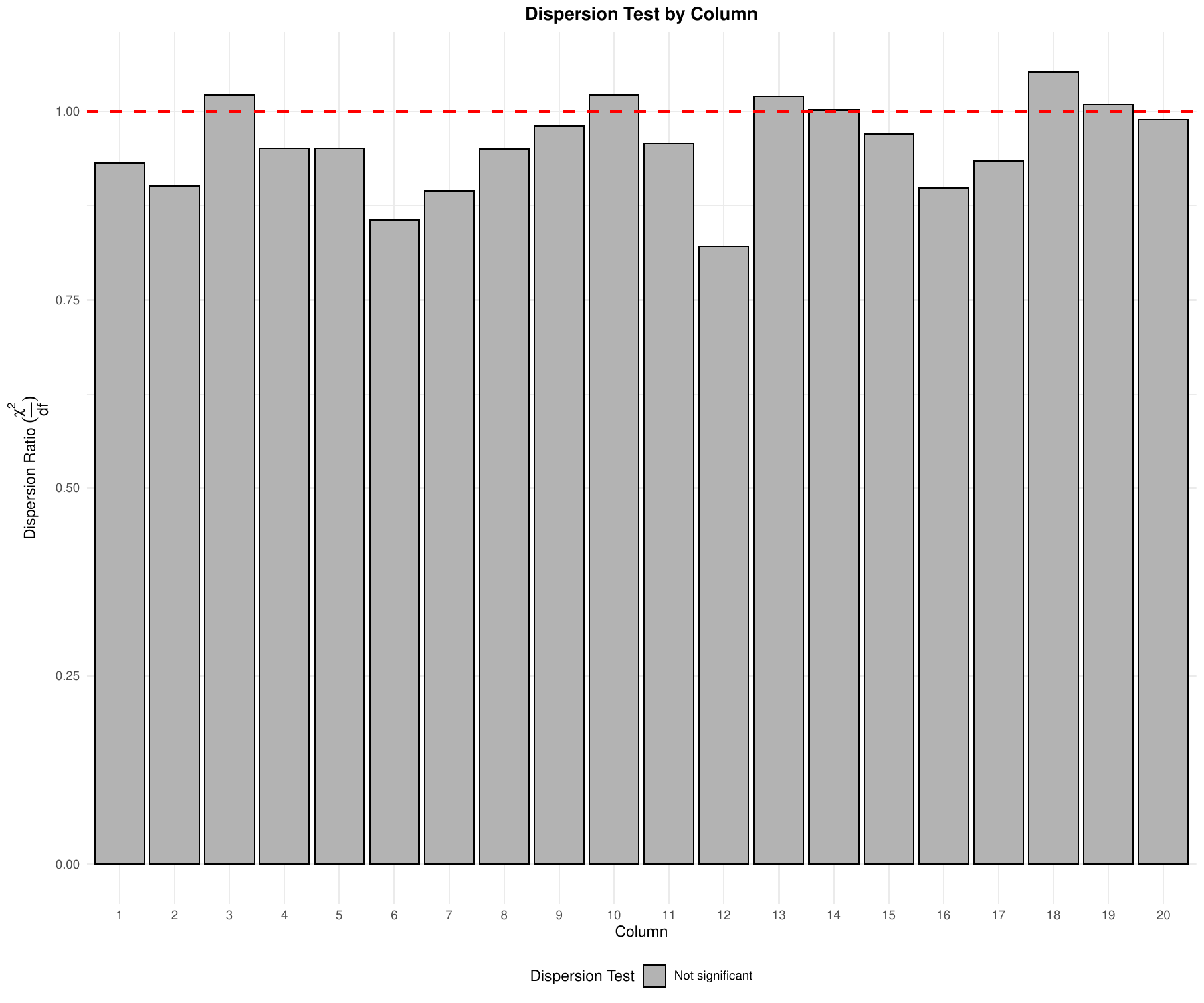}
    \caption{(Simulation 1) Dispersion diagnostics}
    \label{fig:sim1_disp}
\end{figure}

\noindent  \textbf{Distribution:} The distributional fit is assessed through rootograms (Supplementary Figures~\ref{fig:sim1_root_all}, \ref{fig:sim1_pois_root_group}, and \ref{fig:sim1_nb_root_group}) and dispersion diagnostics (Supplementary Figure~\ref{fig:sim1_disp}). The overall rootogram shows good agreement between observed and expected frequencies, with group-specific rootograms confirming this pattern holds across all groups. Combined with the dispersion metrics showing values near 1, these results indicate that the Poisson distribution provides an appropriate fit for all groups.

\begin{figure}[H]
    \centering
    \begin{subfigure}[b]{0.48\textwidth}
        \centering
        \includegraphics[width=\textwidth]{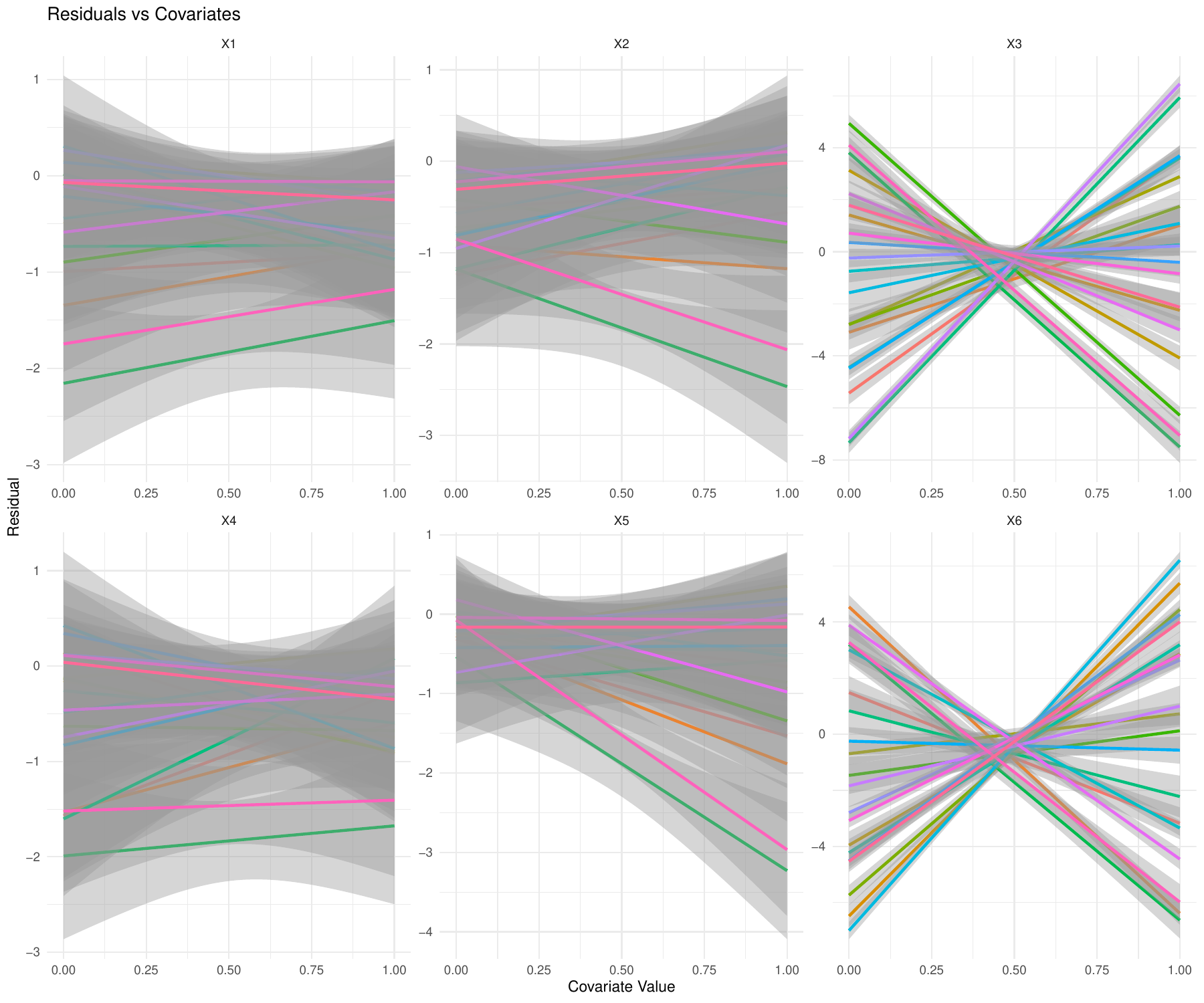}
        \caption{Local covariate effects}
        \label{fig:sim1_local_cov}
    \end{subfigure}
    \hfill
    \begin{subfigure}[b]{0.48\textwidth}
        \centering
        \includegraphics[width=\textwidth]{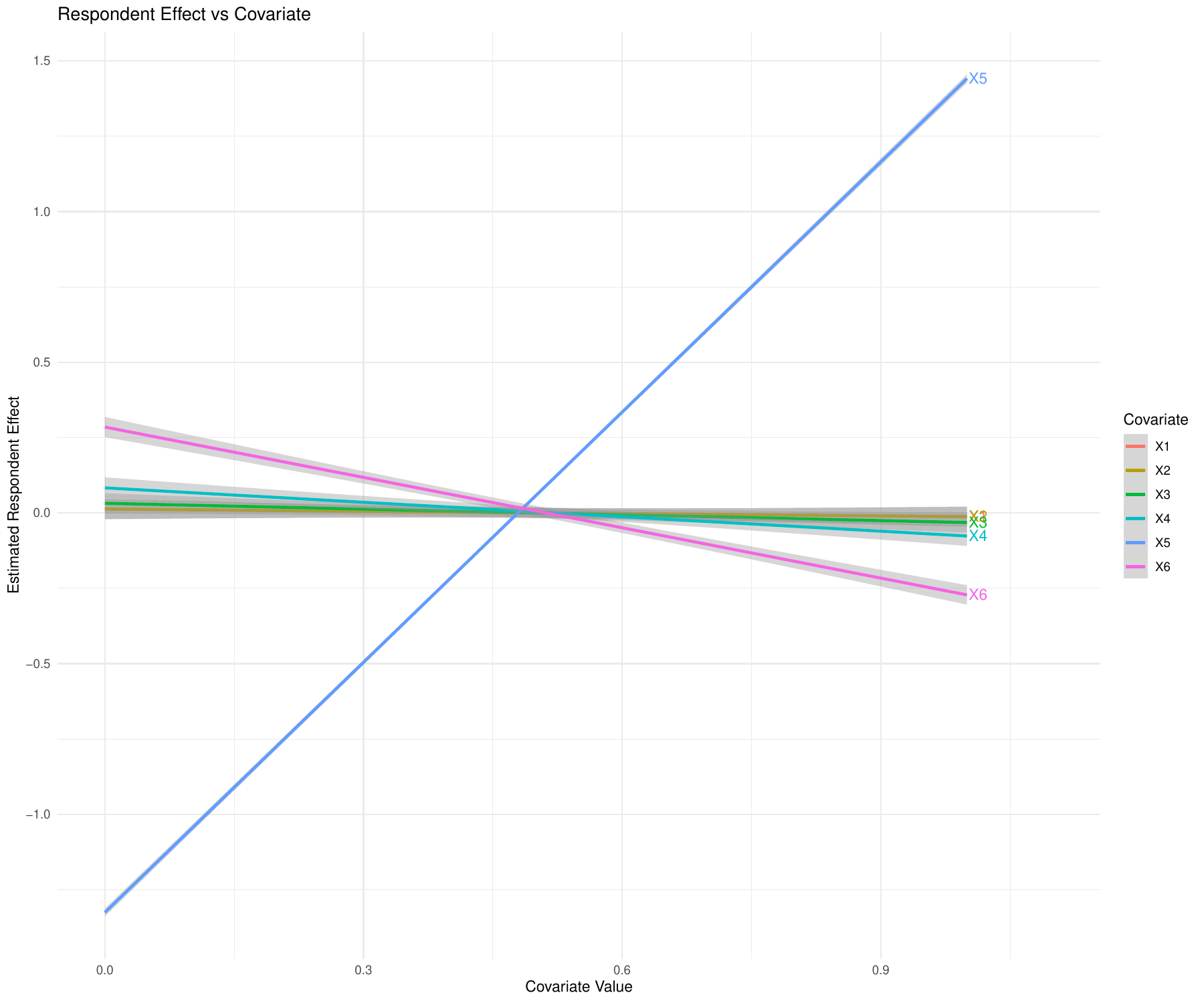}
        \caption{Global covariate effects}
        \label{fig:sim1_global_cov}
    \end{subfigure}
    \caption{(Simulation 1) Covariate selection diagnostics}
    \label{fig:sim1_cov}
\end{figure}

\begin{figure}[H]
    \centering
    \begin{subfigure}[b]{0.48\textwidth}
        \centering
        \includegraphics[width=\textwidth]{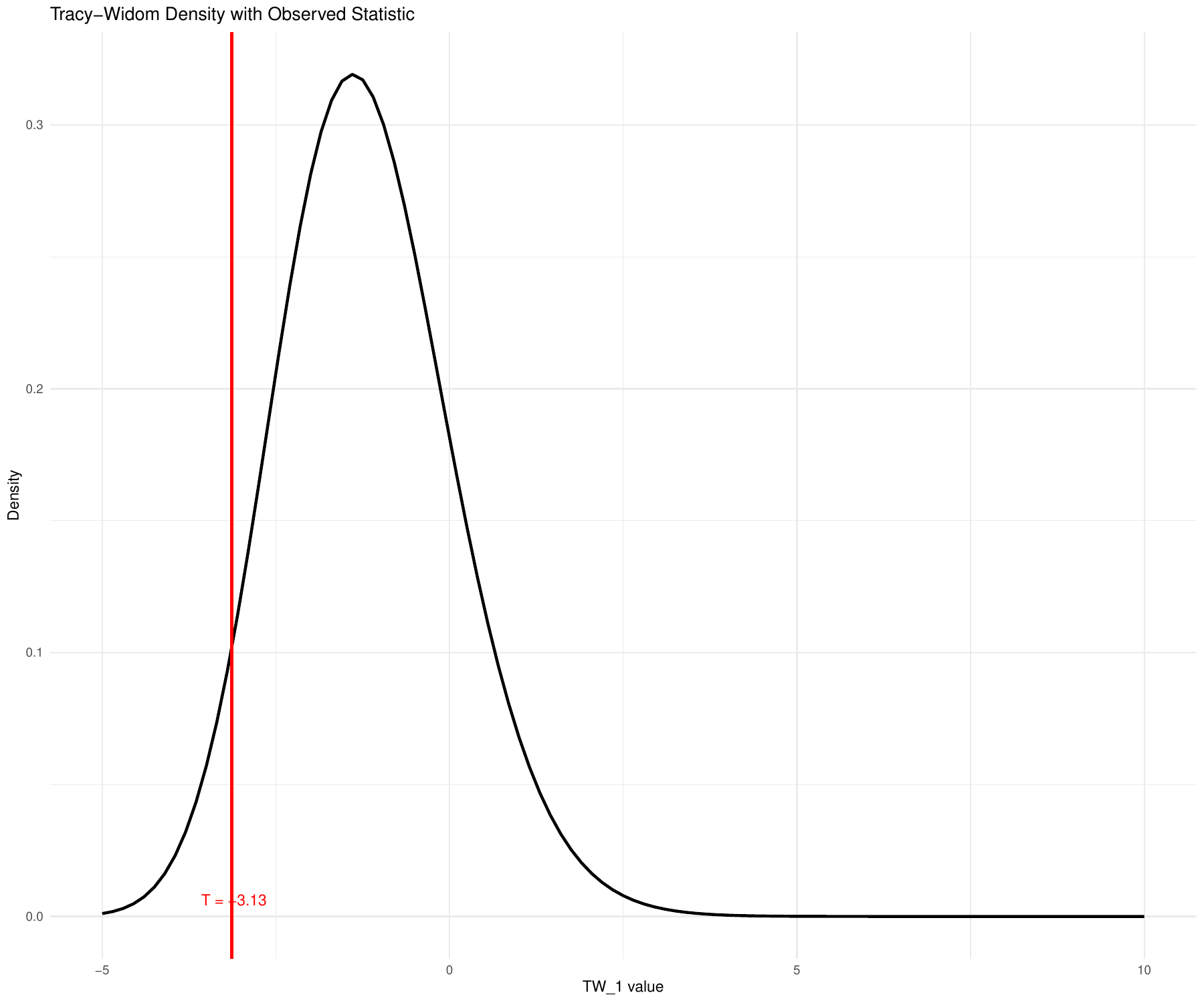}
        \caption{Poisson model}
        \label{fig:sim1_corr_pois}
    \end{subfigure}
    \hfill
    \begin{subfigure}[b]{0.48\textwidth}
        \centering
        \includegraphics[width=\textwidth]{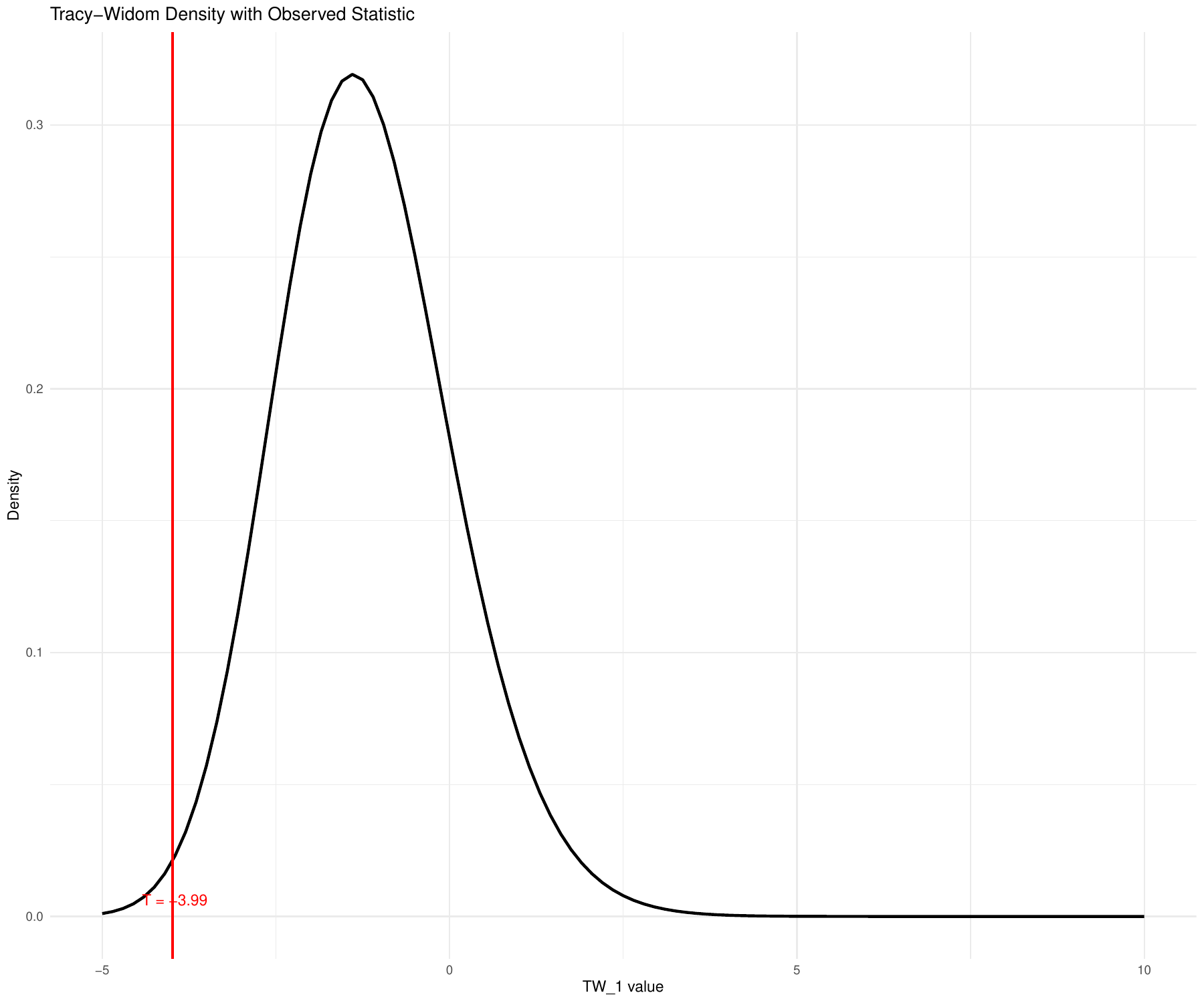}
        \caption{Negative binomial model}
        \label{fig:sim1_corr_nb}
    \end{subfigure}
    \caption{(Simulation 1) Group correlation diagnostics}
    \label{fig:sim1_corr}
\end{figure}

\clearpage

% Simulation 2
\subsection{Simulation 2}

\textbf{Covariate Structure:} The covariate structure can be identified using the diagnostic plots in Supplementary~\ref{fig:sim2_cov}. There are no obvious non-zero slopes, where all effects have small ranges and large standard errors, indicating the absence of any covariates.\\

\noindent \textbf{Correlation Structure:} The correlation structure is evaluated using the Tracy-Widom test, with results shown in Supplementary~\ref{fig:sim2_corr}. The observed test statistic is very large relative to the null distribution for both the Poisson and negative binomial models, indicating the clear presence of residual group correlation structure.\\

\begin{figure}[H]
    \centering
    \includegraphics[width=0.6\textwidth]{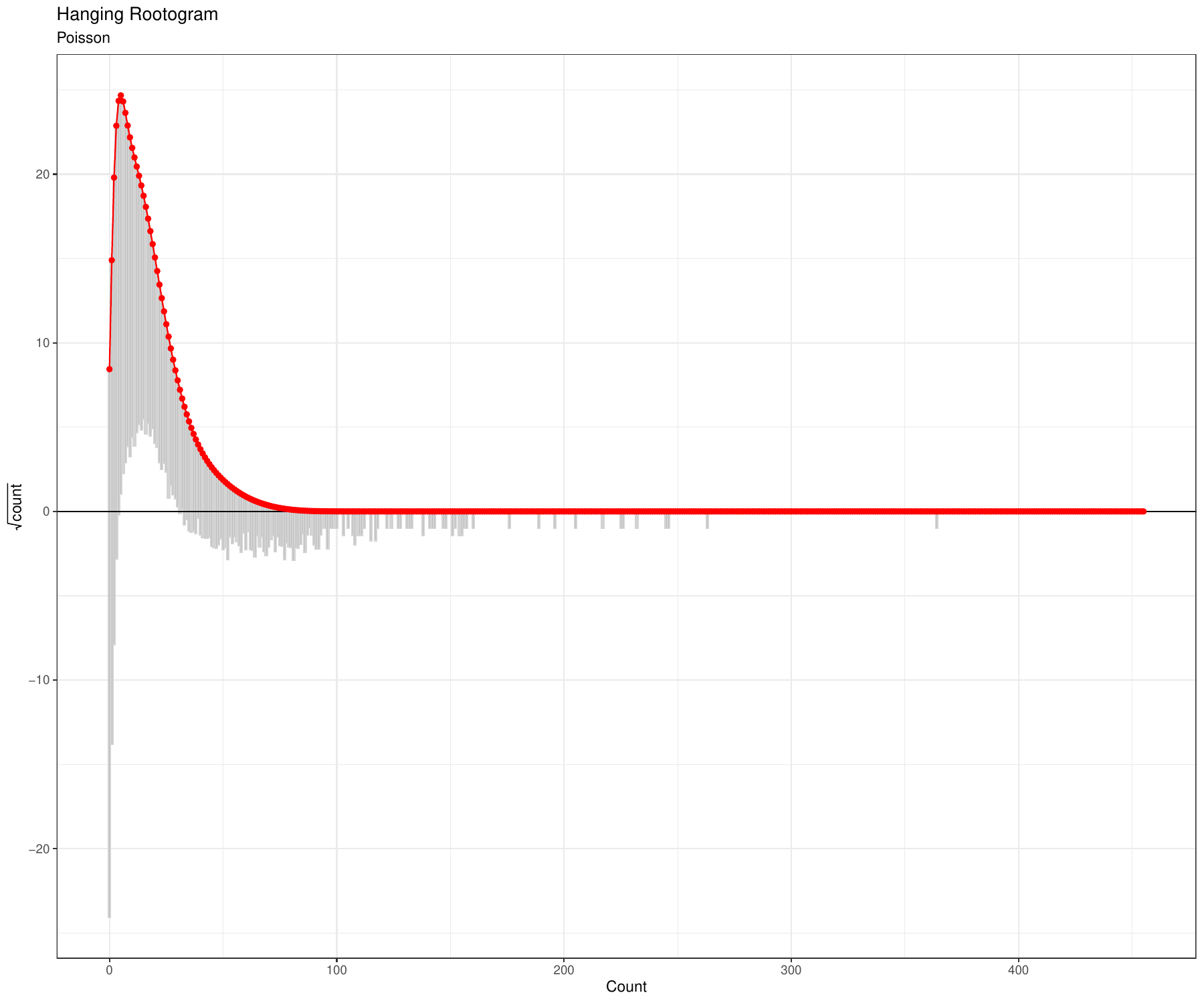}
    
    \includegraphics[width=0.6\textwidth]{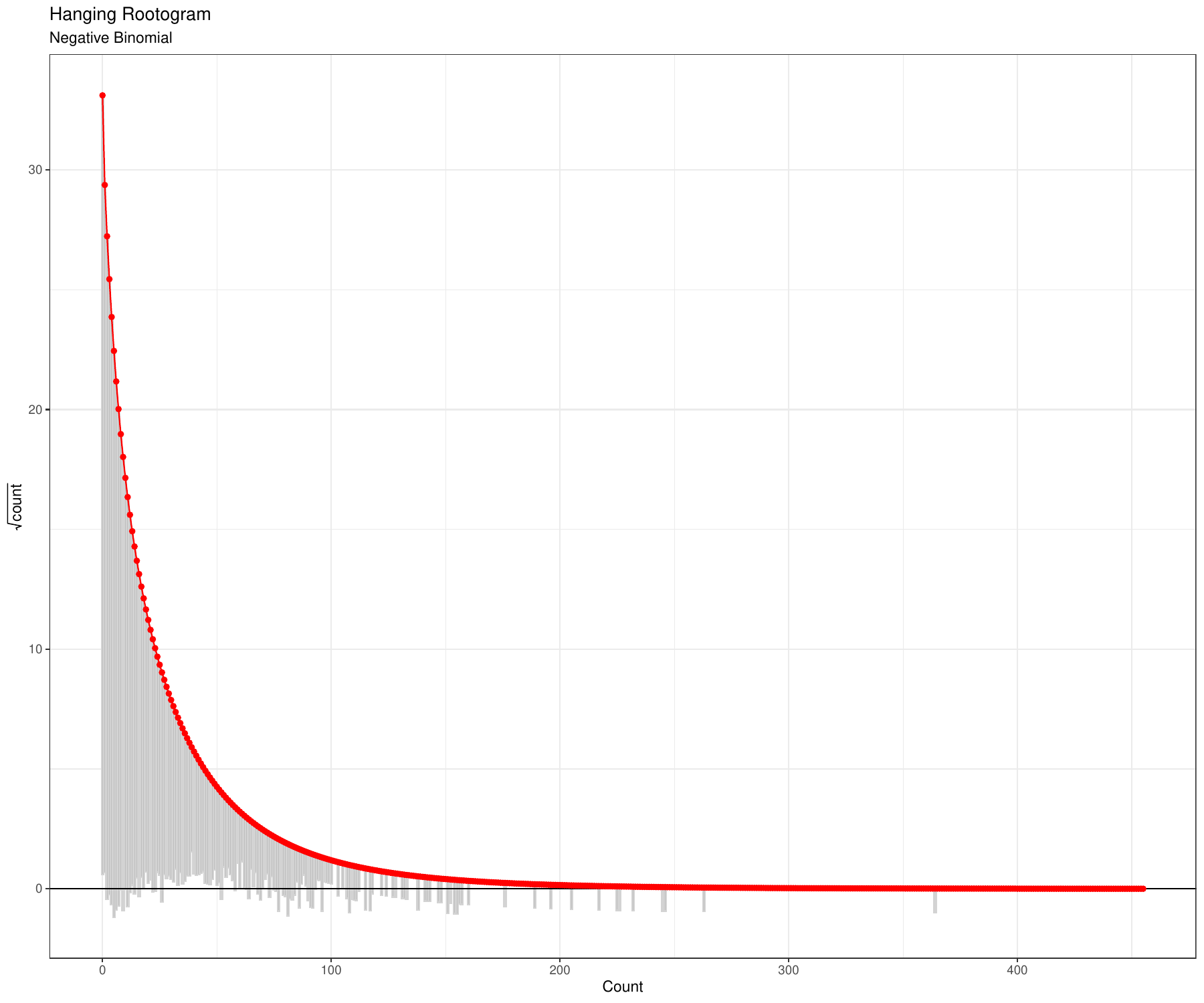}
    \caption{(Simulation 2) Overall rootogram diagnostics for Poisson (top) and negative binomial (bottom) models}
    \label{fig:sim2_root_all}
\end{figure}

\begin{figure}[H]
    \centering
    \includegraphics[width=\textwidth]{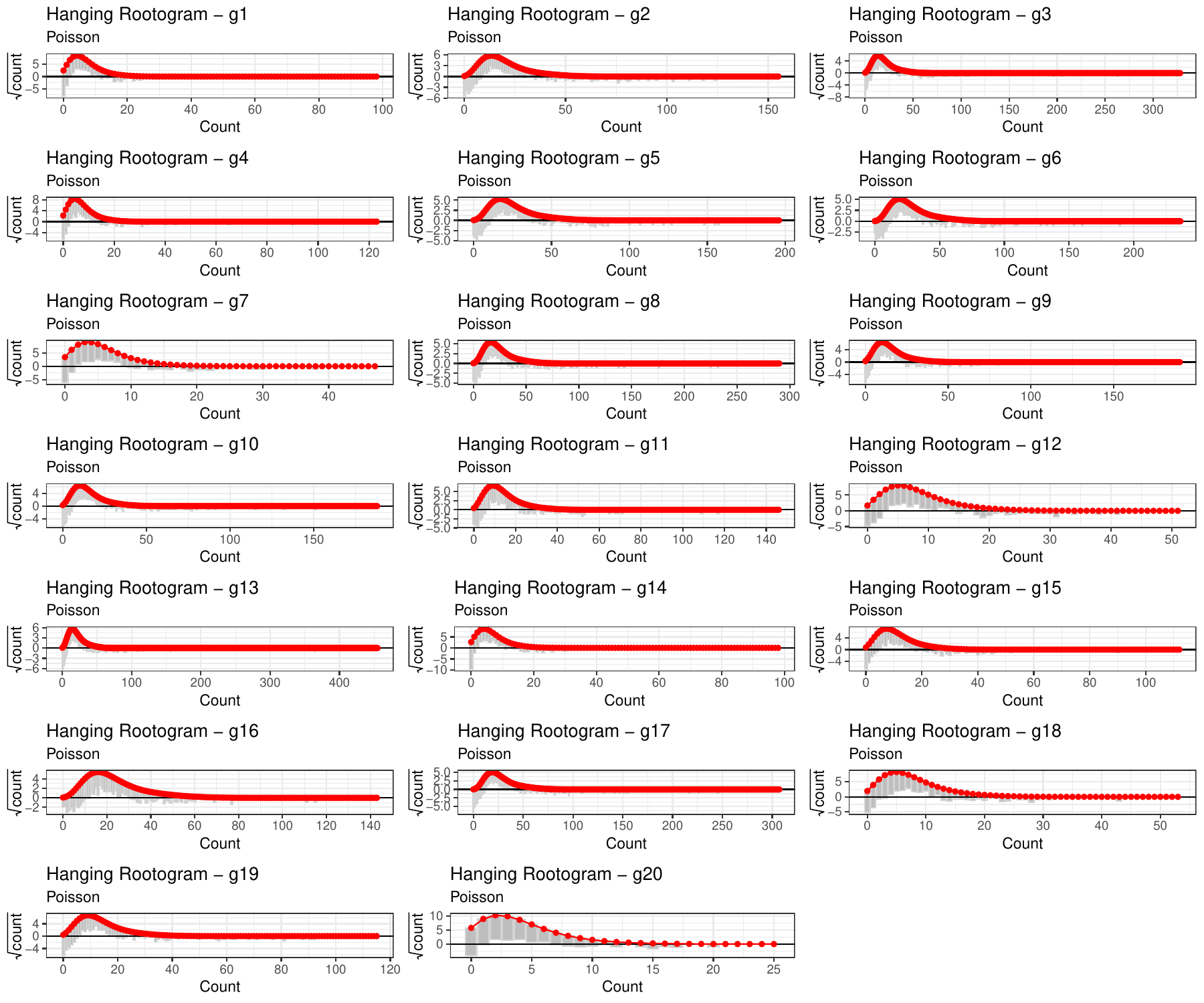}
    \caption{(Simulation 2) Poisson rootogram diagnostics by group}
    \label{fig:sim2_pois_root_group}
\end{figure}

\begin{figure}[H]
    \centering
    \includegraphics[width=\textwidth]{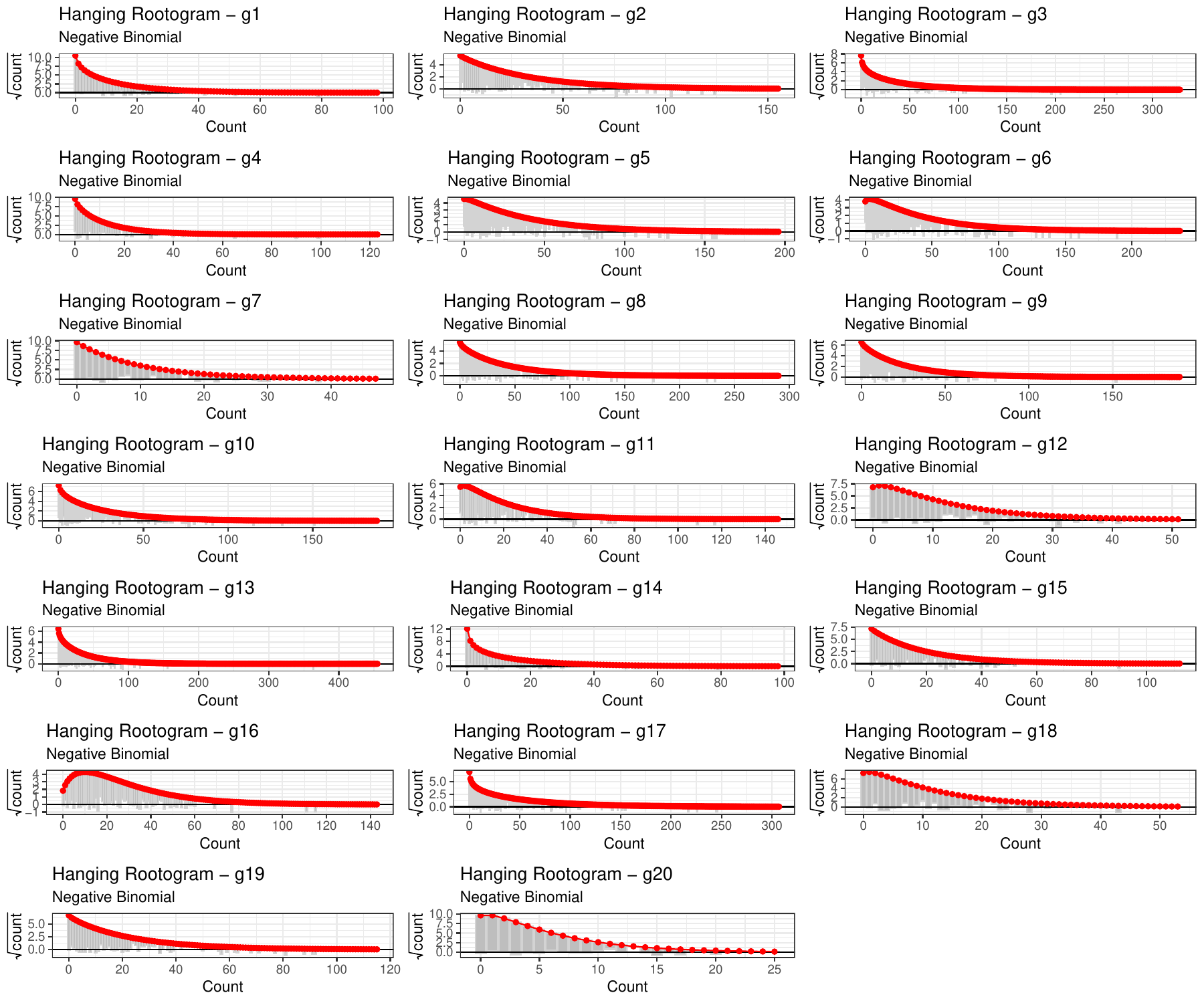}
    \caption{(Simulation 2) Negative binomial rootogram diagnostics by group}
    \label{fig:sim2_nb_root_group}
\end{figure}

\begin{figure}[H]
    \centering
    \includegraphics[width=0.7\textwidth]{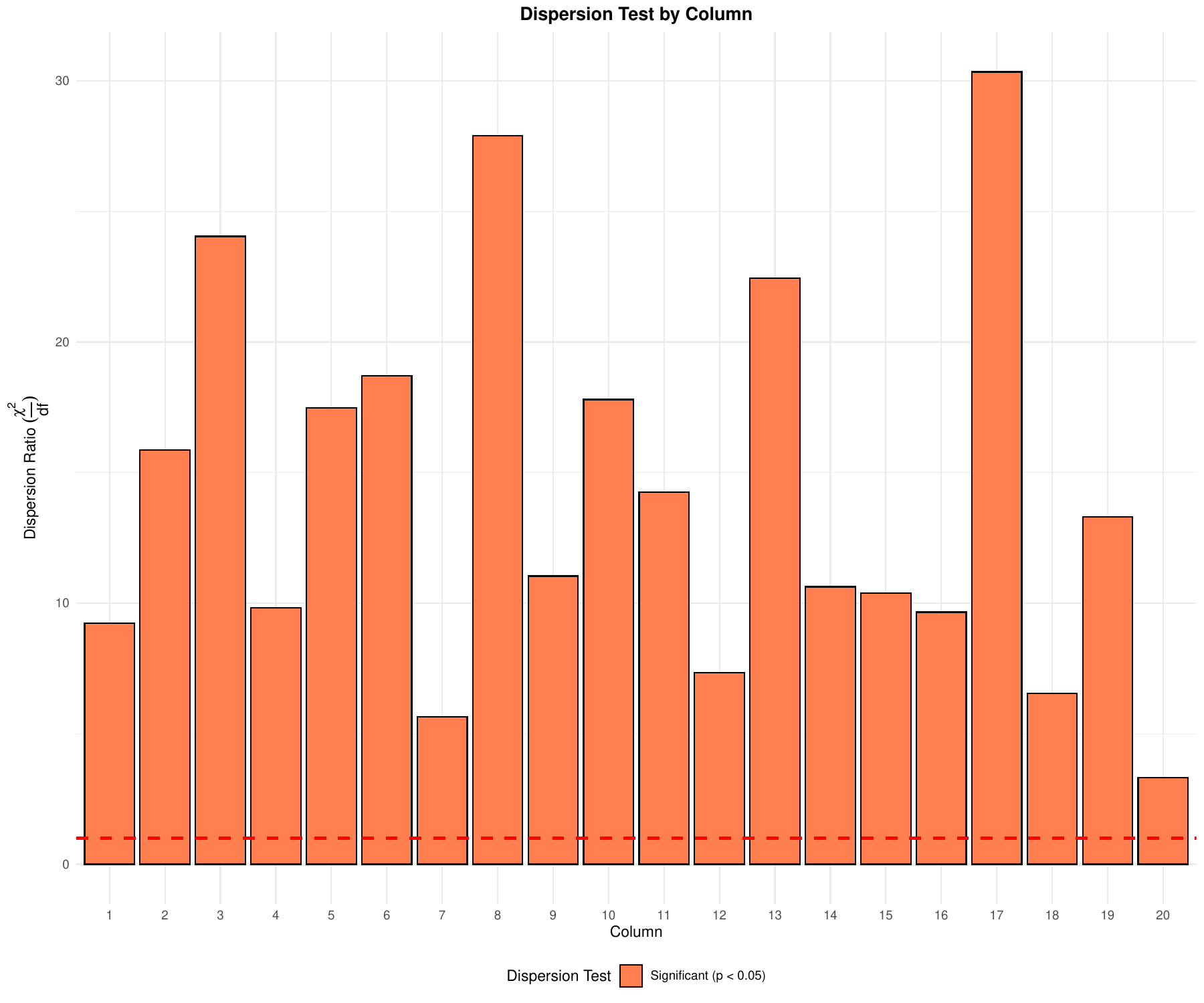}
    \caption{(Simulation 2) Dispersion diagnostics}
    \label{fig:sim2_disp}
\end{figure}

\noindent \textbf{Distribution:} The distributional fit is assessed through rootograms (Supplementary Figures~\ref{fig:sim2_root_all}, \ref{fig:sim2_pois_root_group}, and \ref{fig:sim2_nb_root_group}) and dispersion diagnostics (Supplementary Figure~\ref{fig:sim2_disp}). The overall rootogram shows good agreement between observed and expected frequencies for the negative binomial distribution, but bad agreement for the Poisson distribution. This result is consistent across each group. Combined with the dispersion metrics showing values substantially larger than 1, these results indicate that the negative binomial distribution provides an appropriate fit for all groups. Overall, the results suggest that the negative binomial distribution is able to adequately account for the true correlation structure as well.

\begin{figure}[H]
    \centering
    \begin{subfigure}[b]{0.48\textwidth}
        \centering
        \includegraphics[width=\textwidth]{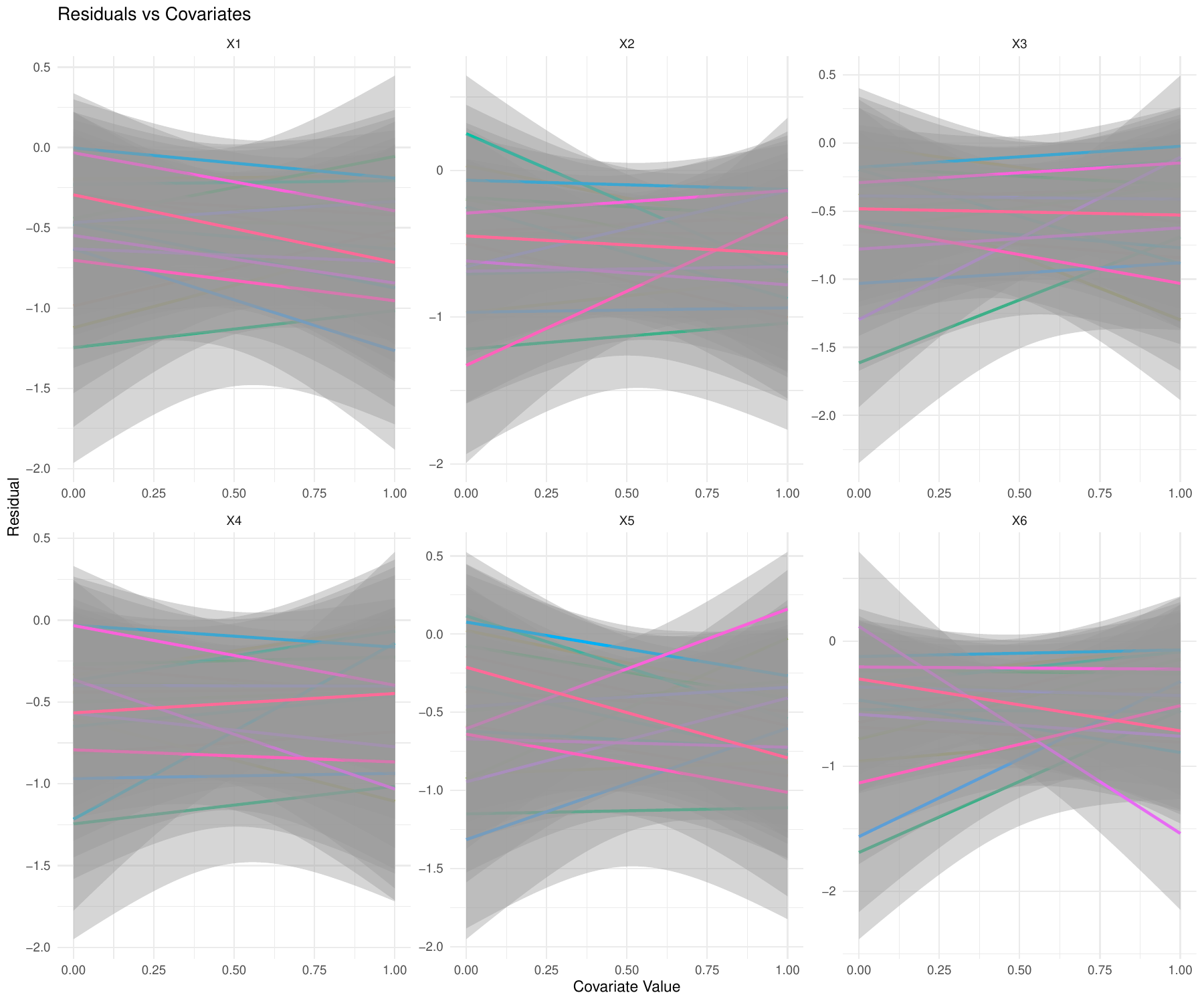}
        \caption{Local covariate effects}
        \label{fig:sim2_local_cov}
    \end{subfigure}
    \hfill
    \begin{subfigure}[b]{0.48\textwidth}
        \centering
        \includegraphics[width=\textwidth]{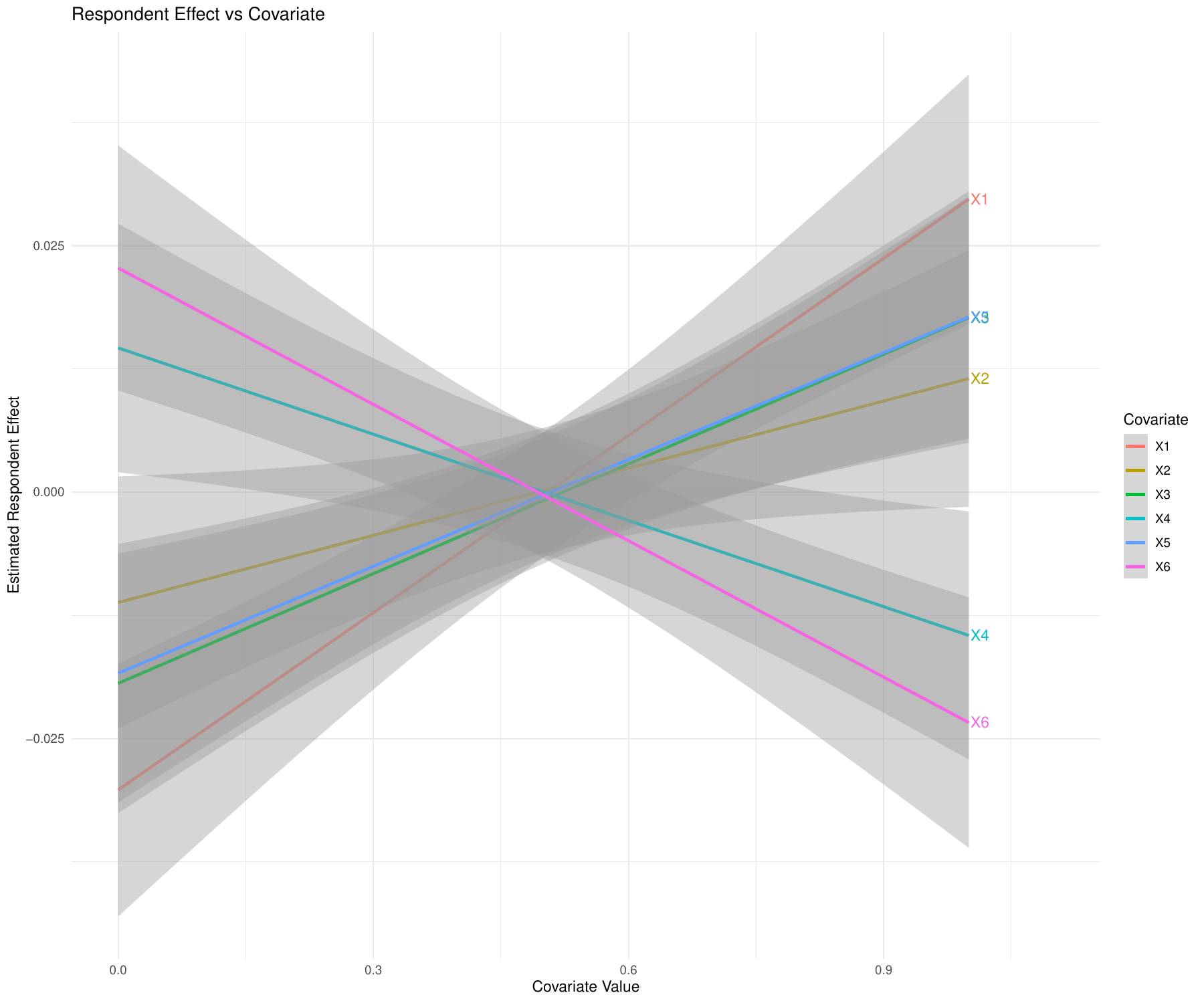}
        \caption{Global covariate effects}
        \label{fig:sim2_global_cov}
    \end{subfigure}
    \caption{(Simulation 2) Covariate selection diagnostics}
    \label{fig:sim2_cov}
\end{figure}

\begin{figure}[H]
    \centering
    \begin{subfigure}[b]{0.48\textwidth}
        \centering
        \includegraphics[width=\textwidth]{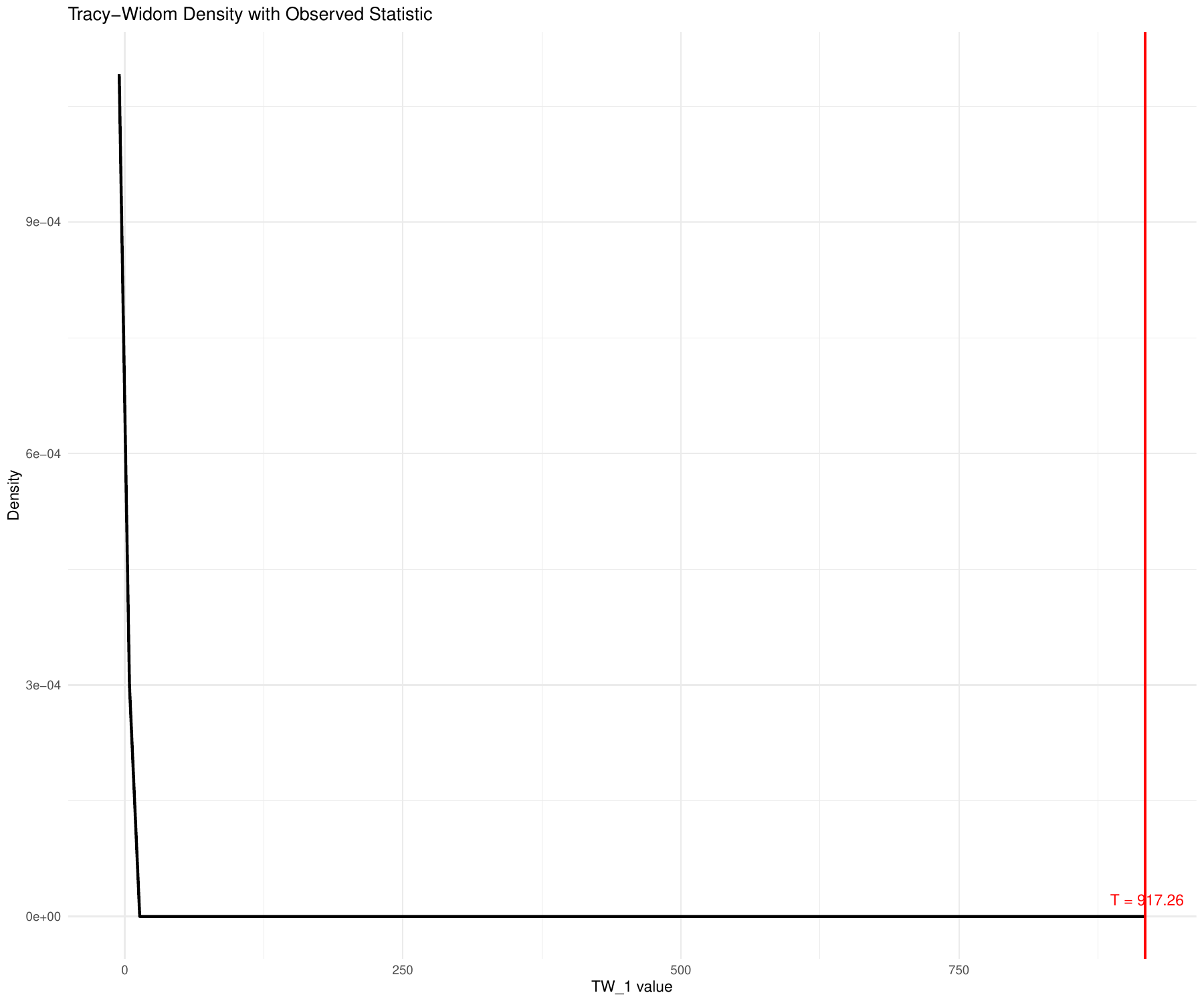}
        \caption{Poisson model}
        \label{fig:sim2_corr_pois}
    \end{subfigure}
    \hfill
    \begin{subfigure}[b]{0.48\textwidth}
        \centering
        \includegraphics[width=\textwidth]{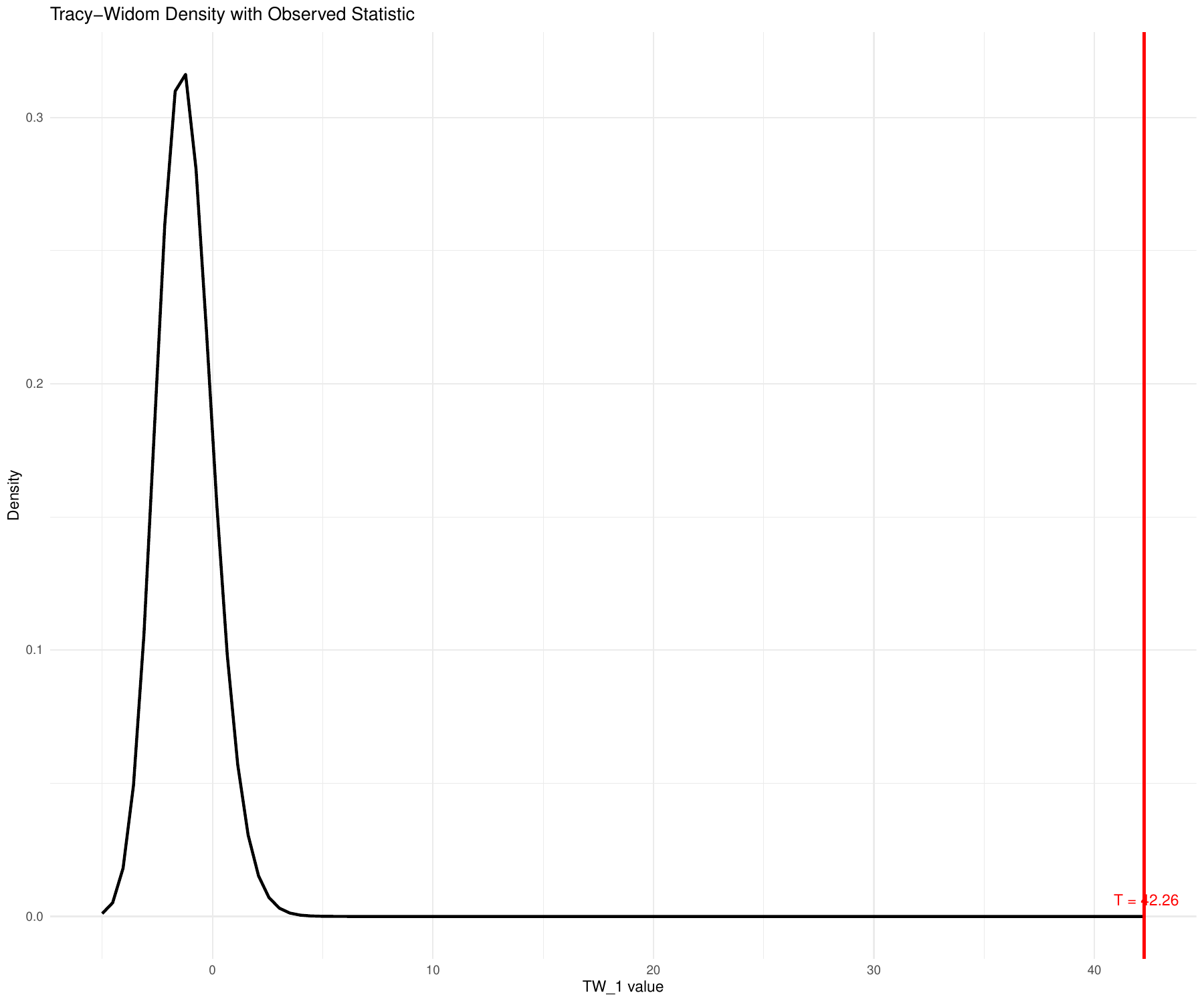}
        \caption{Negative binomial model}
        \label{fig:sim2_corr_nb}
    \end{subfigure}
    \caption{(Simulation 2) Group correlation diagnostics}
    \label{fig:sim2_corr}
\end{figure}

\clearpage

% Simulation 3
\subsection{Simulation 3}

\textbf{Covariate Structure:} The covariate structure can be identified using the diagnostic plots in Supplementary~\ref{fig:sim3_cov}. These plots indicate that X1 is as local covariates, while X5, X6, and potentially X2 and X1 act as a global covariate. While the true model has only X5 and X6 as global covariates, depending on the researchers desire for parsimony, they could include X2 and X1 as additional global covariates and perform additional diagnostics on a full model.\\

\noindent  \textbf{Correlation Structure:} The correlation structure is evaluated using the Tracy-Widom test, with results shown in Supplementary~\ref{fig:sim3_corr}. The observed test statistic is large relative to the null distribution for both the Poisson and negative binomial models, indicating the clear presence of residual group correlation structure. While the true model also has degree correlation, we are unable to determine the exact correlation structure from these simple diagnostics and additional diagnostics from a full model would be required.\\

\begin{figure}[H]
    \centering
    \includegraphics[width=0.6\textwidth]{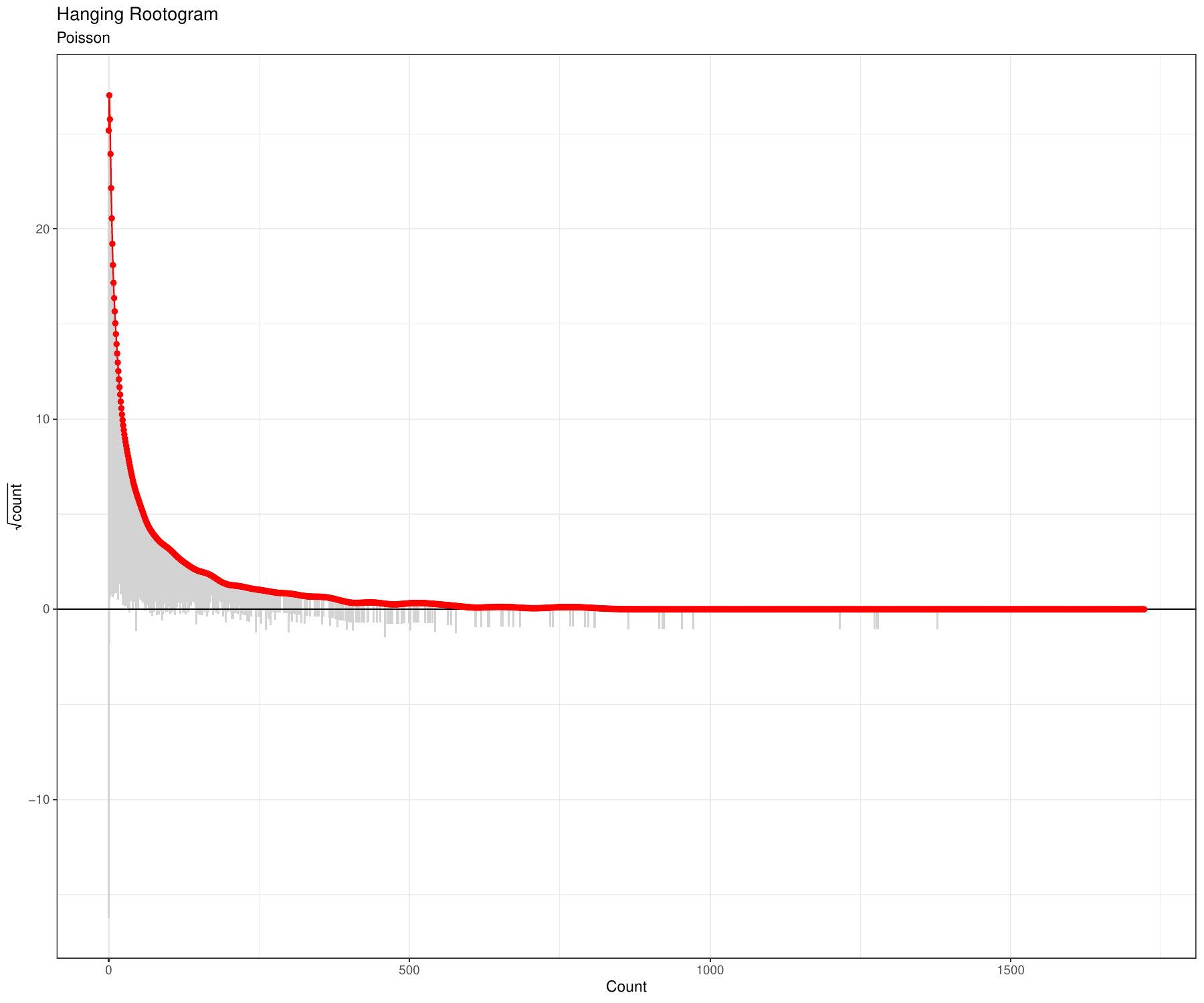}
    
    \includegraphics[width=0.6\textwidth]{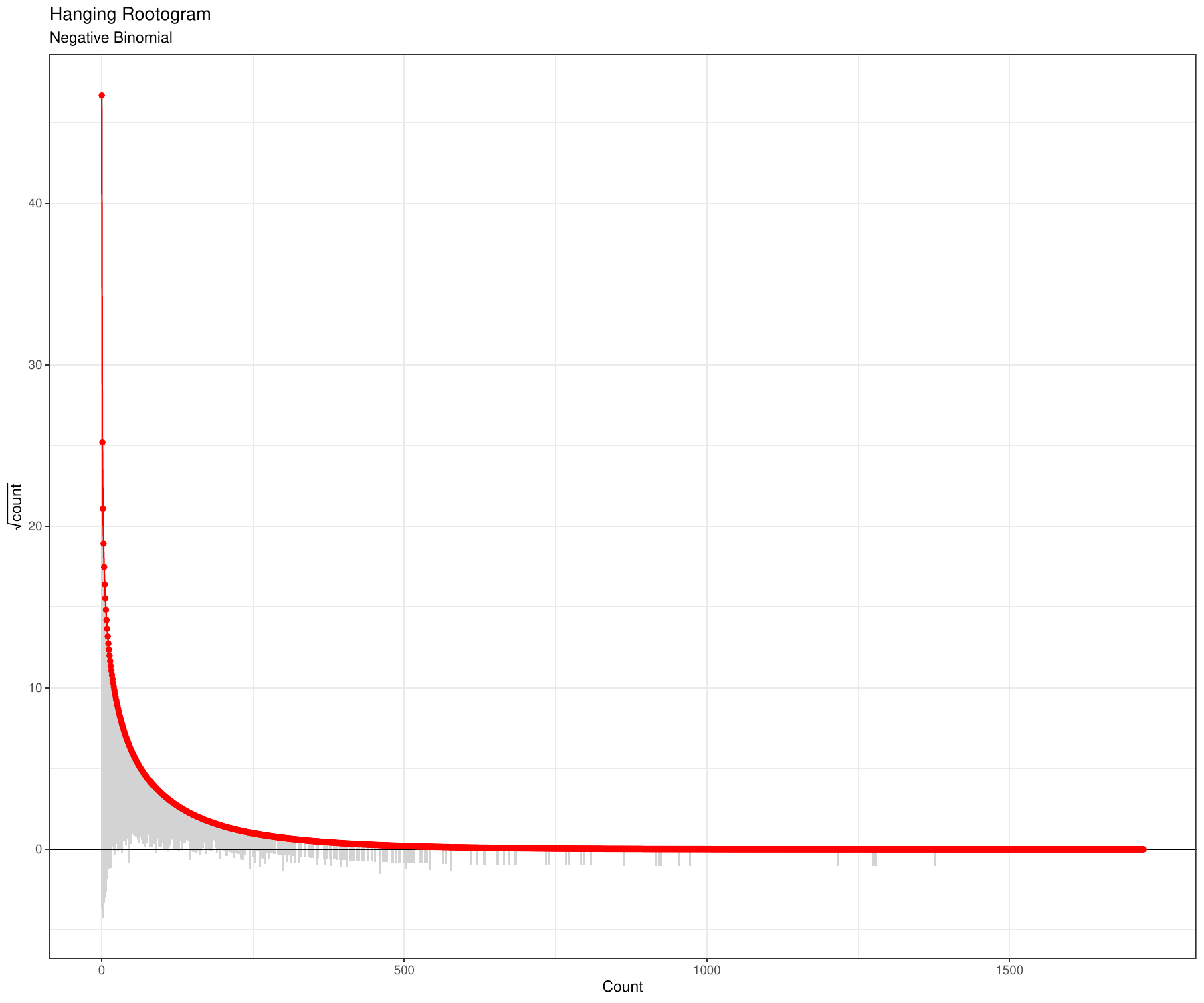}
    \caption{(Simulation 3) Overall rootogram diagnostics for Poisson (top) and negative binomial (bottom) models}
    \label{fig:sim3_root_all}
\end{figure}

\begin{figure}[H]
    \centering
    \includegraphics[width=\textwidth]{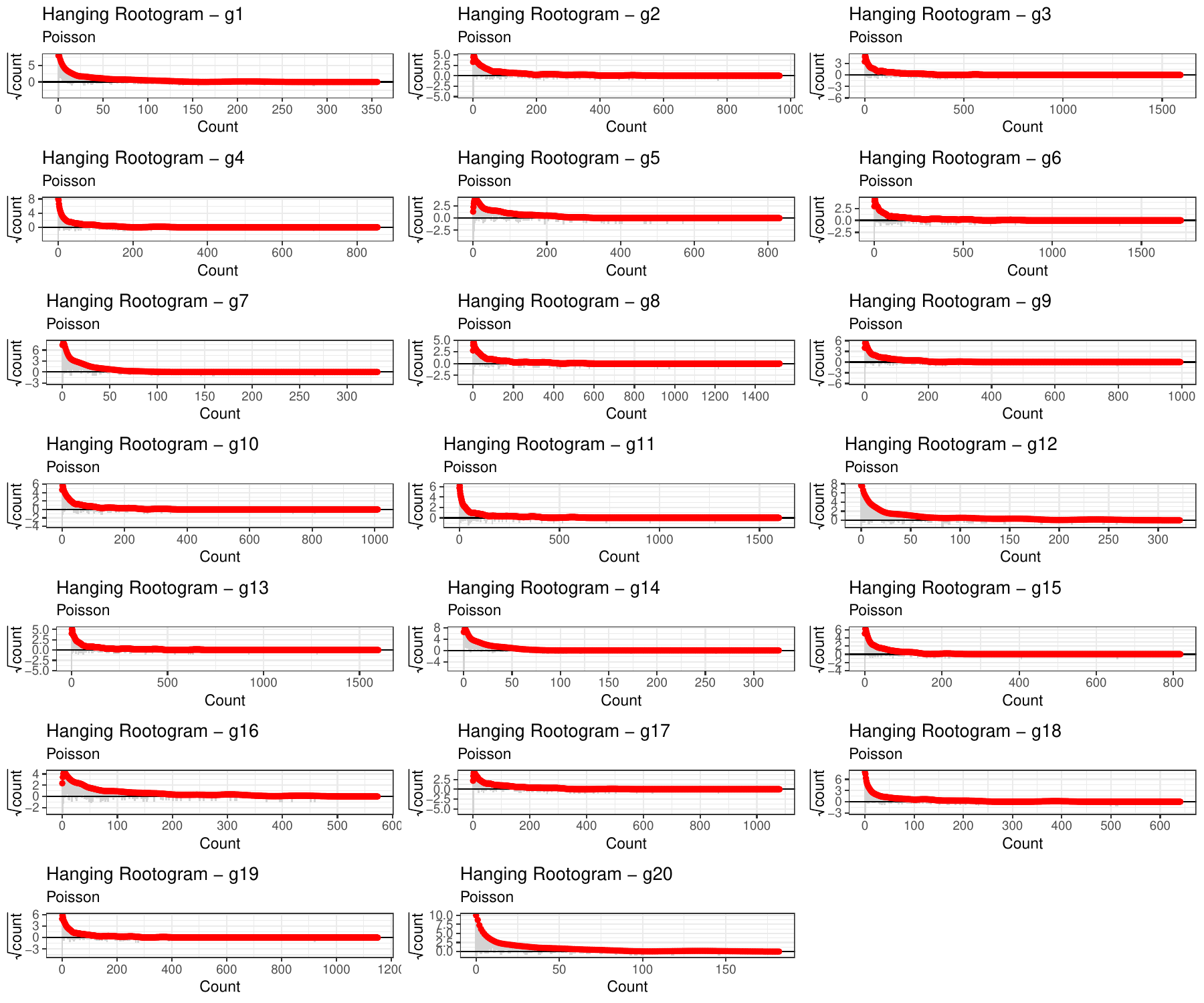}
    \caption{(Simulation 3) Poisson rootogram diagnostics by group}
    \label{fig:sim3_pois_root_group}
\end{figure}

\begin{figure}[H]
    \centering
    \includegraphics[width=\textwidth]{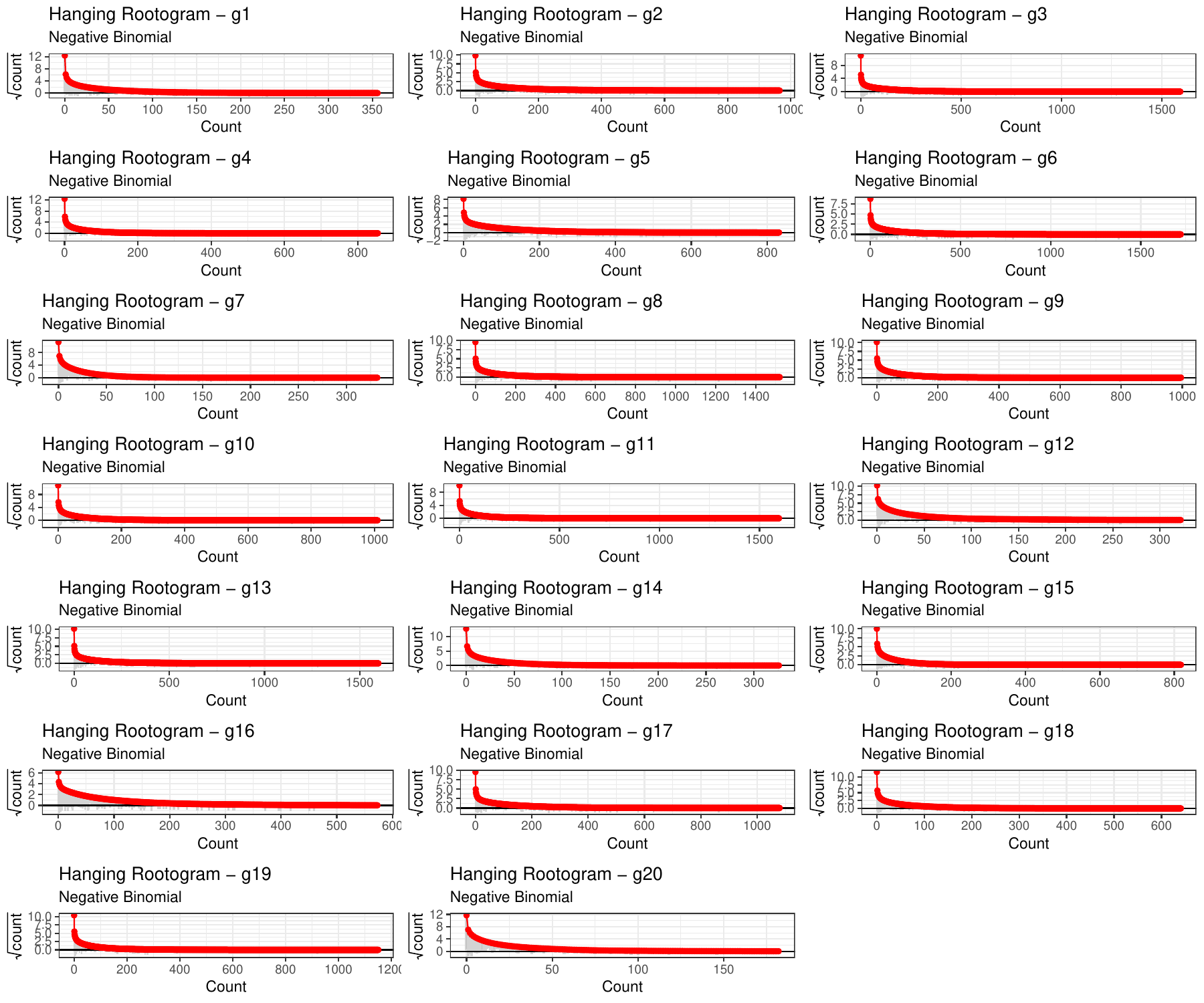}
    \caption{(Simulation 3) Negative binomial rootogram diagnostics by group}
    \label{fig:sim3_nb_root_group}
\end{figure}

\begin{figure}[H]
    \centering
    \includegraphics[width=0.7\textwidth]{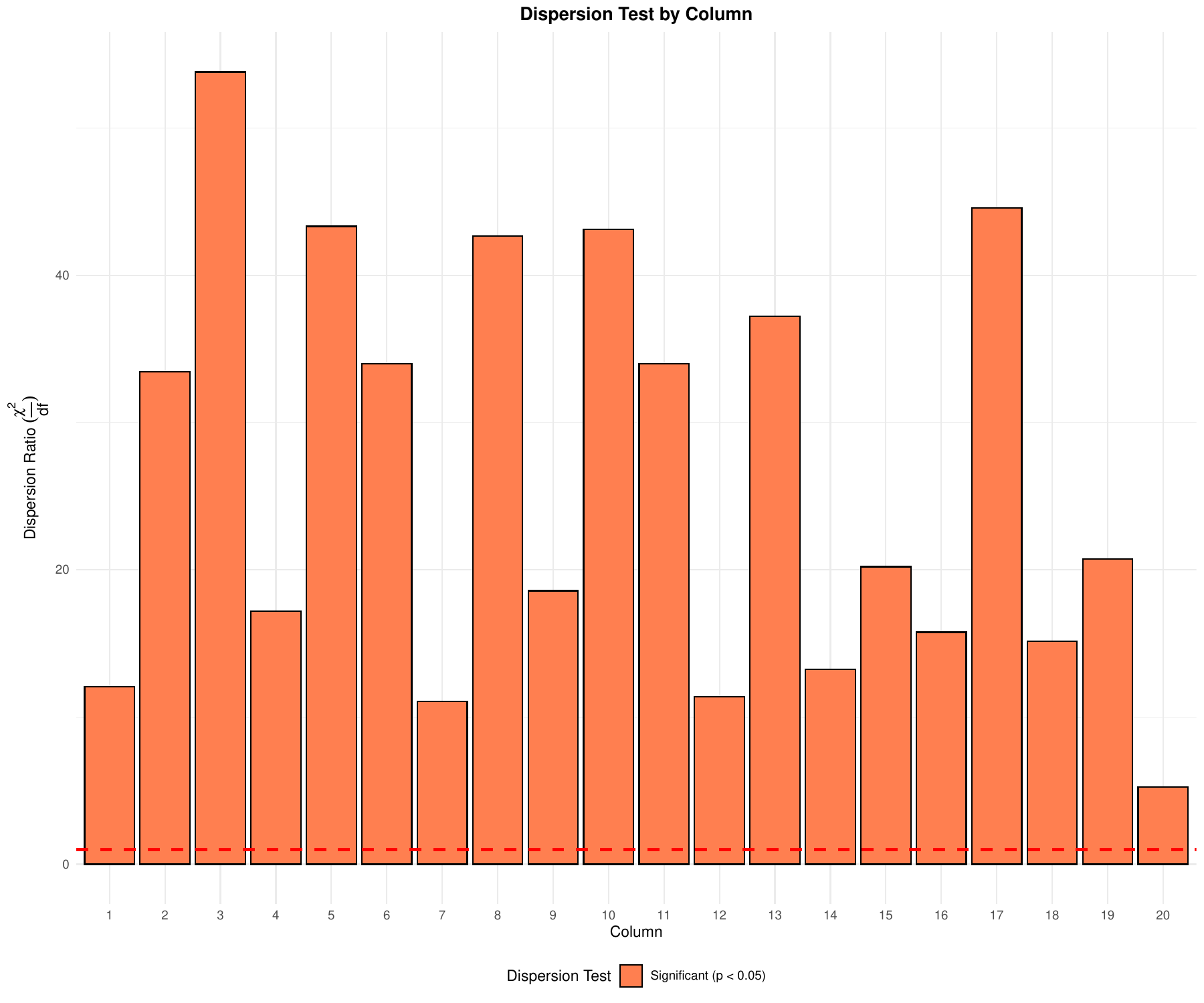}
    \caption{(Simulation 3) Dispersion diagnostics}
    \label{fig:sim3_disp}
\end{figure}

\noindent  \textbf{Distribution:} The distributional fit is assessed through rootograms (Supplementary Figures~\ref{fig:sim3_root_all}, \ref{fig:sim3_pois_root_group}, and \ref{fig:sim3_nb_root_group}) and dispersion diagnostics (Supplementary Figure~\ref{fig:sim3_disp}). The overall rootograms, group rootograms, and dispersion metrics suggest that the neither the Poisson nor negative binomial distribution are sufficiently complex to explain the observed ARD, and therefore a full model would require a correlation structure based on the results above.

\begin{figure}[H]
    \centering
    \begin{subfigure}[b]{0.48\textwidth}
        \centering
        \includegraphics[width=\textwidth]{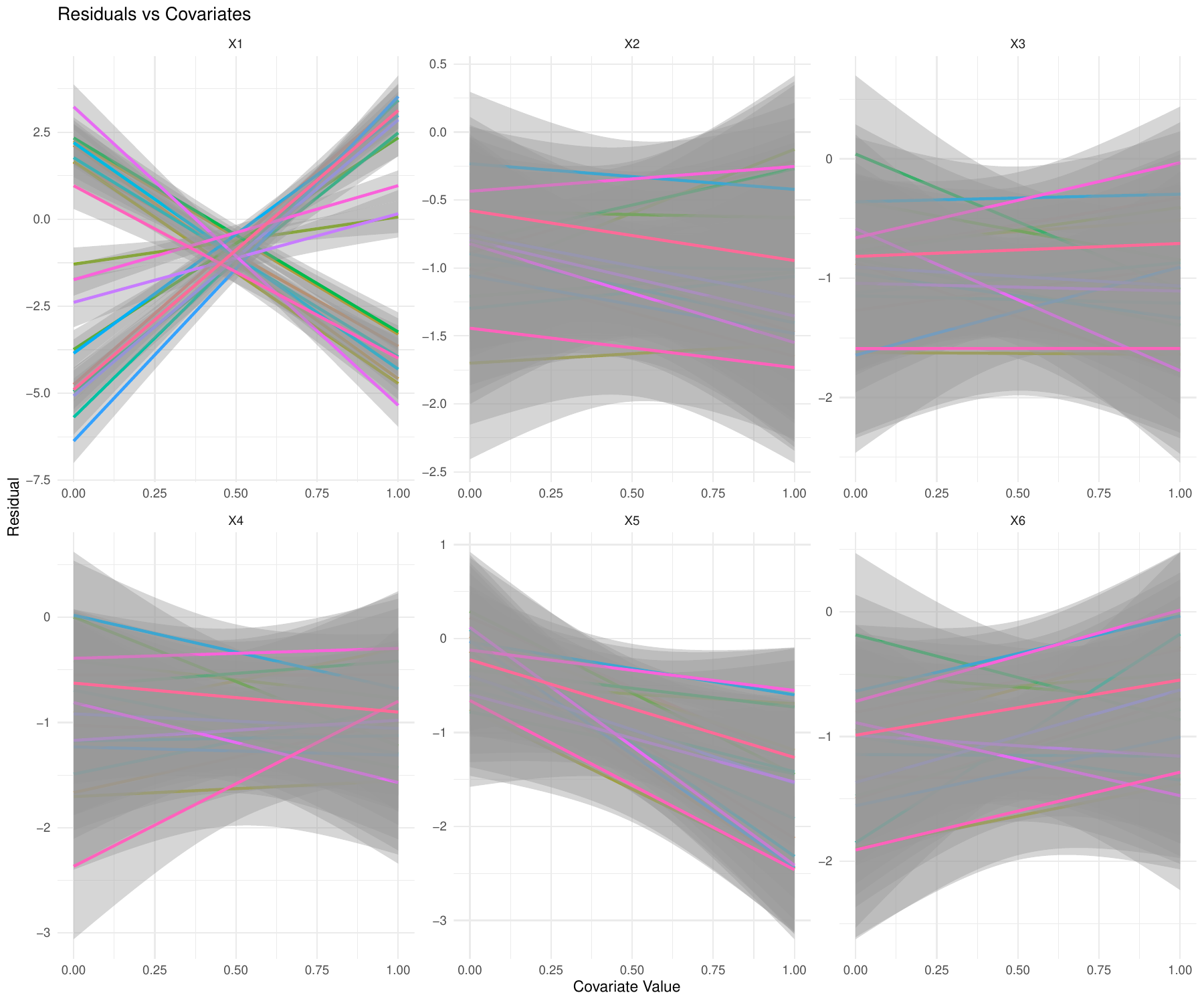}
        \caption{Local covariate effects}
        \label{fig:sim3_local_cov}
    \end{subfigure}
    \hfill
    \begin{subfigure}[b]{0.48\textwidth}
        \centering
        \includegraphics[width=\textwidth]{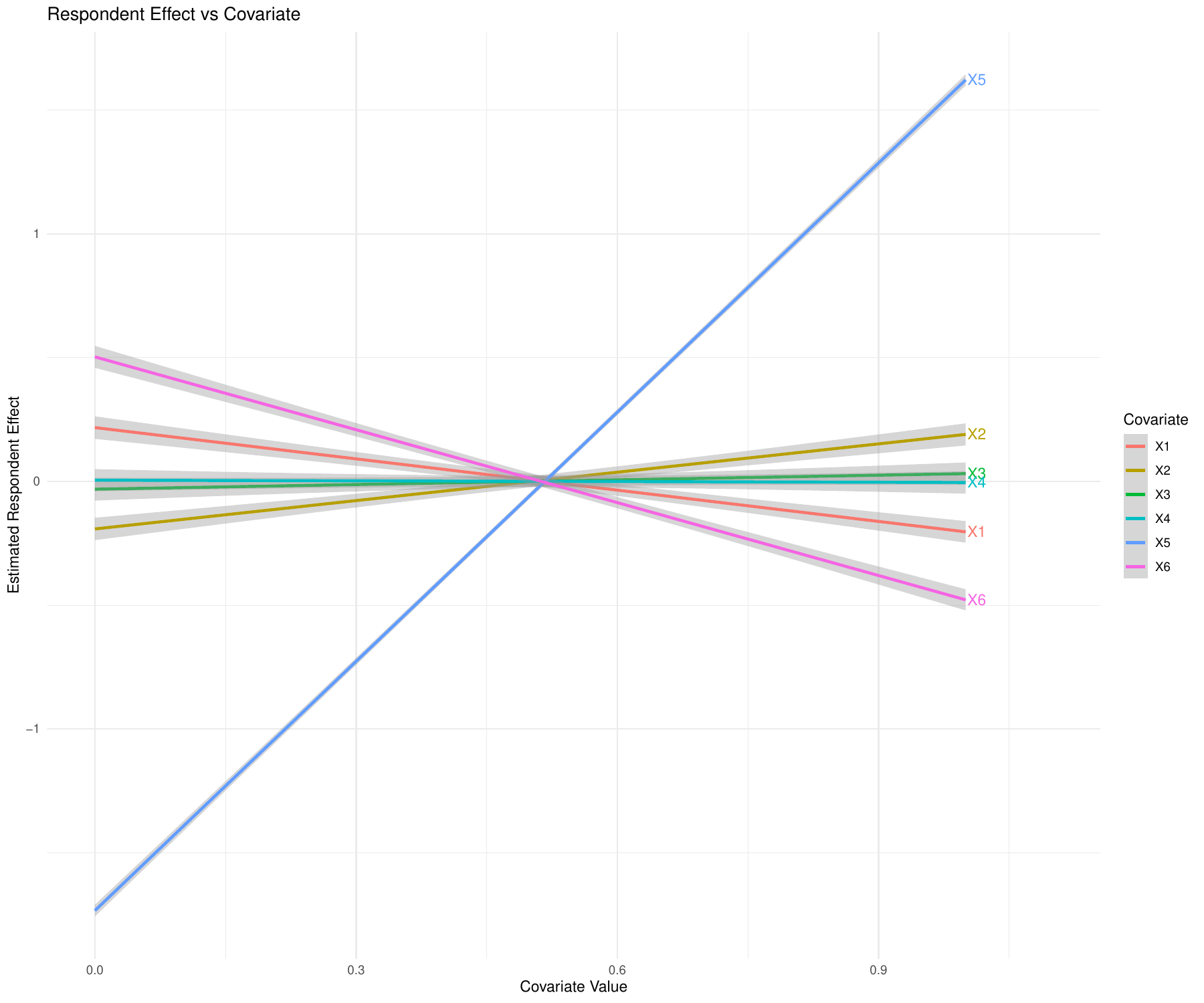}
        \caption{Global covariate effects}
        \label{fig:sim3_global_cov}
    \end{subfigure}
    \caption{(Simulation 3) Covariate selection diagnostics}
    \label{fig:sim3_cov}
\end{figure}

\begin{figure}[H]
    \centering
    \begin{subfigure}[b]{0.48\textwidth}
        \centering
        \includegraphics[width=\textwidth]{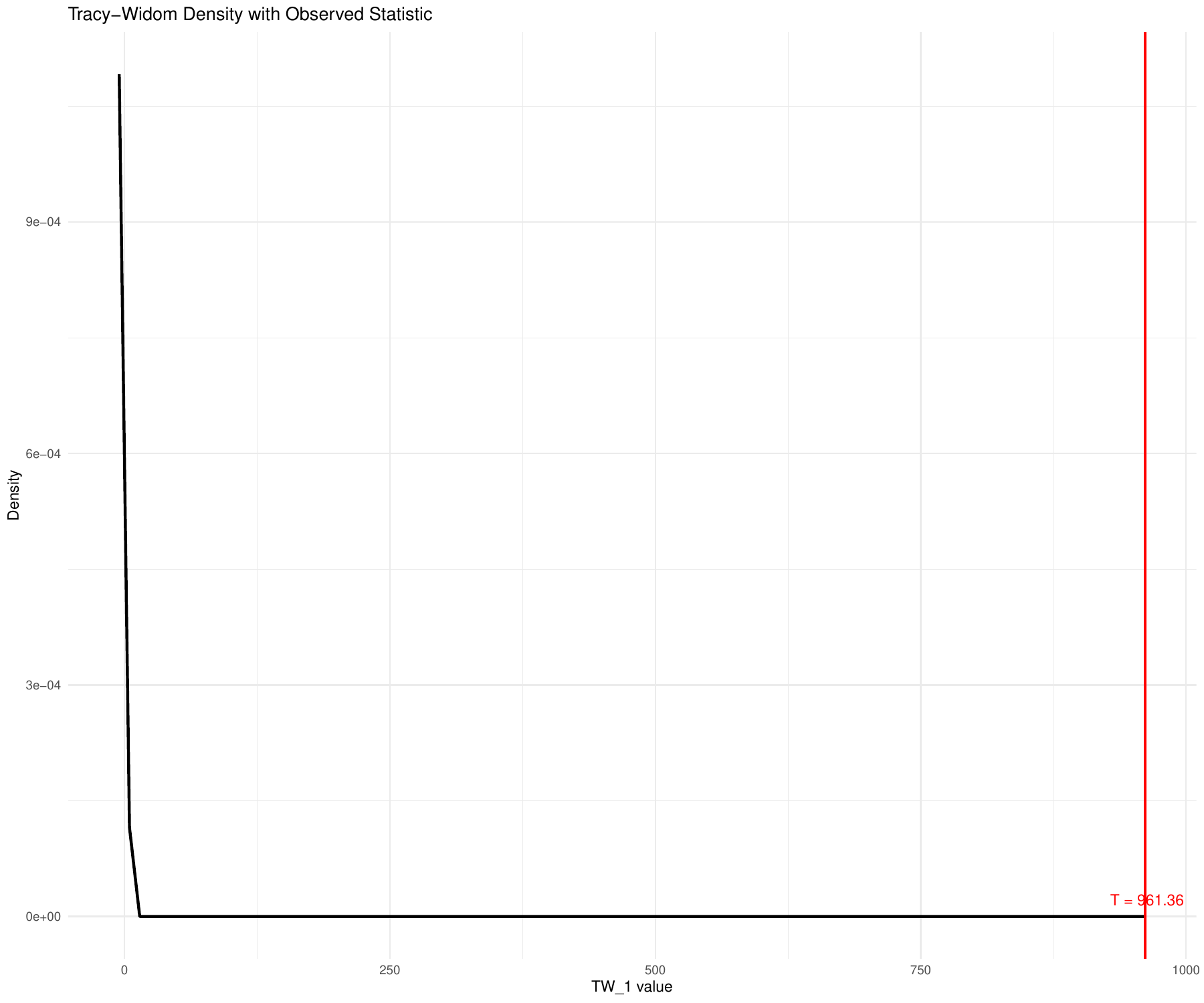}
        \caption{Poisson model}
        \label{fig:sim3_corr_pois}
    \end{subfigure}
    \hfill
    \begin{subfigure}[b]{0.48\textwidth}
        \centering
        \includegraphics[width=\textwidth]{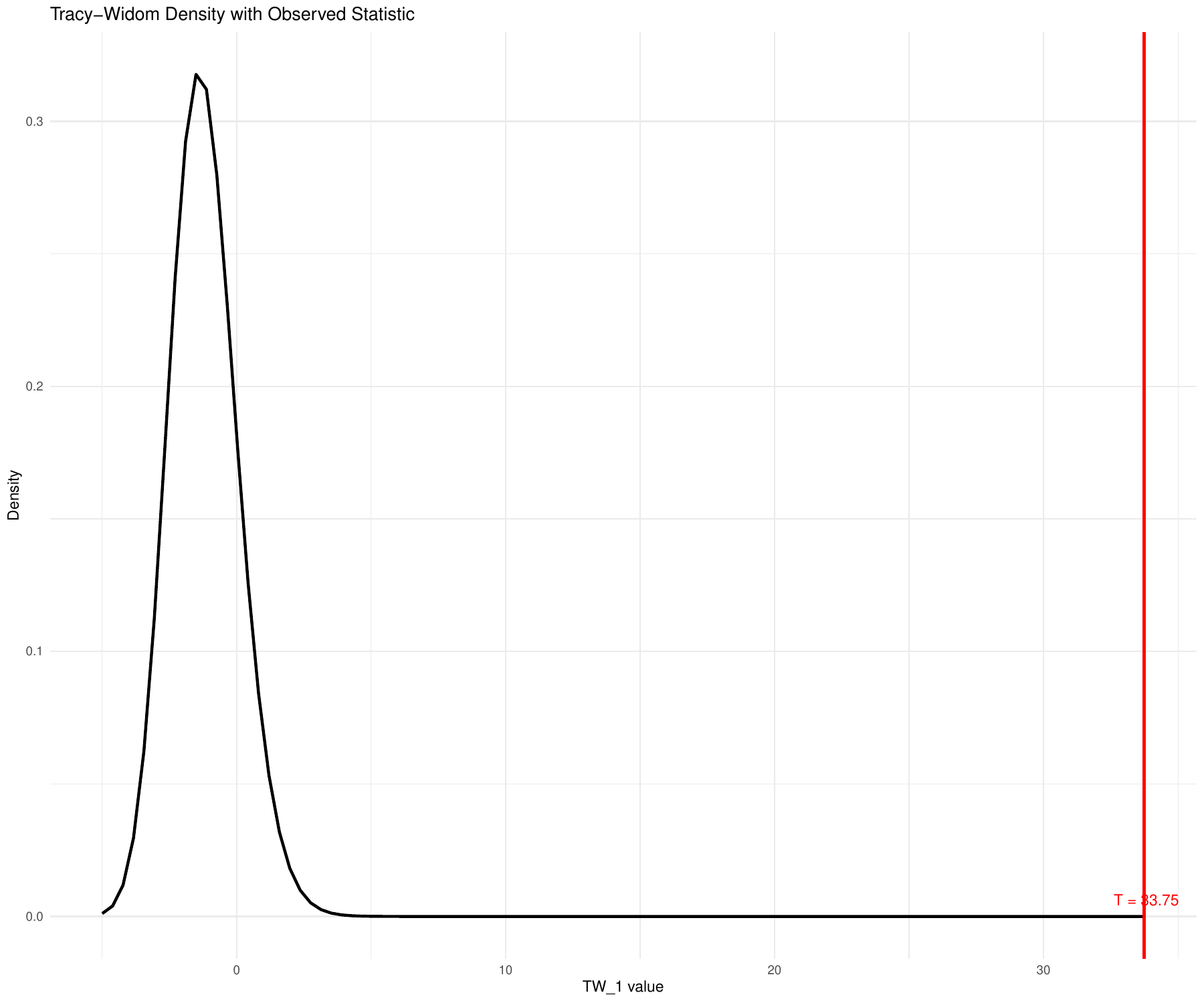}
        \caption{Negative binomial model}
        \label{fig:sim3_corr_nb}
    \end{subfigure}
    \caption{(Simulation 3) Group correlation diagnostics}
    \label{fig:sim3_corr}
\end{figure}

\clearpage

% Simulation 4
\subsection{Simulation 4}
\label{sec:sim4}

\textbf{Covariate Structure:} The covariate structure can be identified using the diagnostic plots in Supplementary~\ref{fig:sim4_cov}. These plots indicate that X1 and X5 function as local covariates, while X4 and X6 act as global covariates.\\

\noindent  \textbf{Correlation Structure:} The correlation structure is evaluated using the Tracy-Widom test, with results shown in Supplementary~\ref{fig:sim4_corr}. The observed test statistic is large relative to the null distribution for both the Poisson and negative binomial models, indicating the clear presence of residual group correlation structure.\\

\begin{figure}[H]
    \centering
    \includegraphics[width=0.6\textwidth]{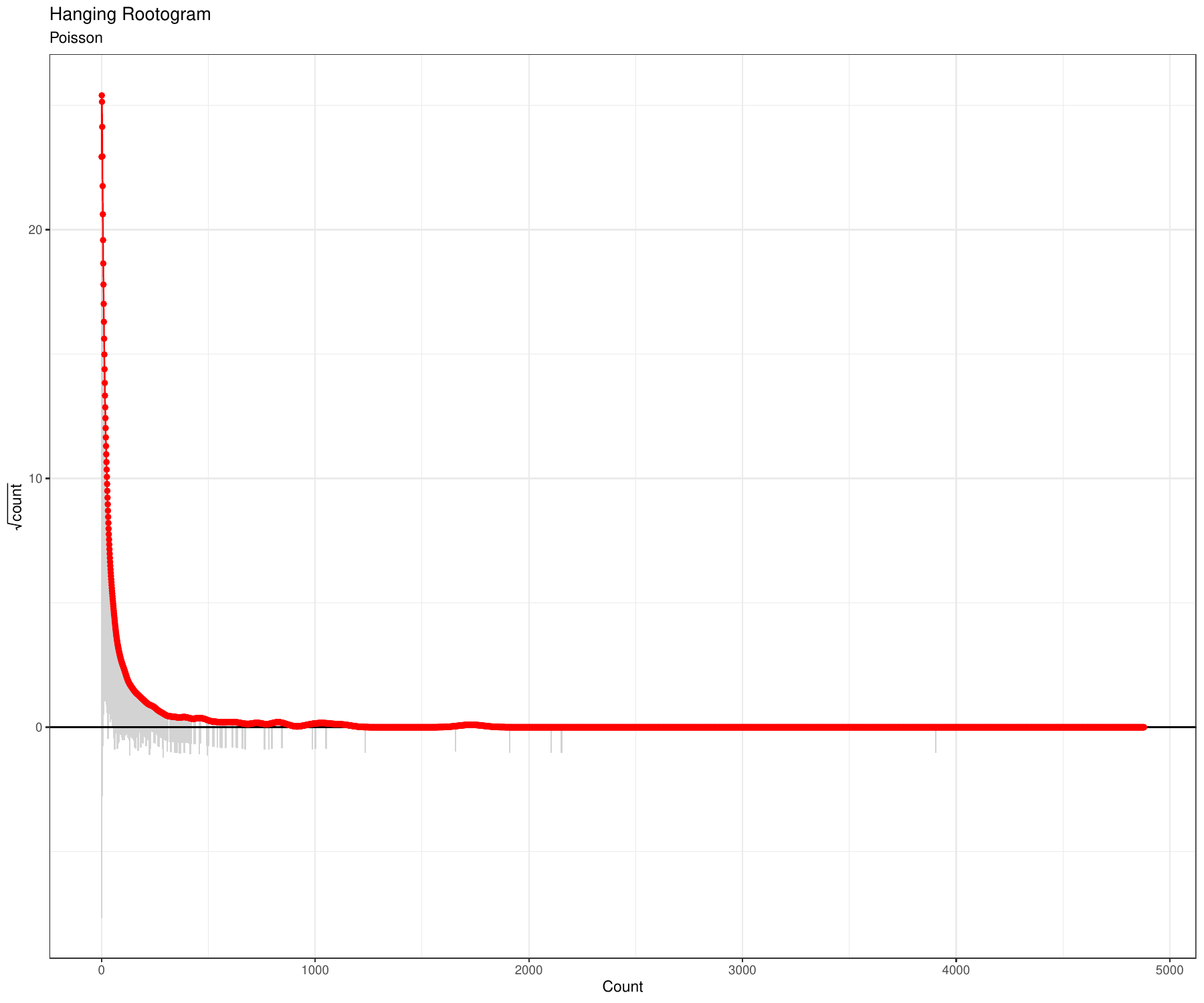}
    
    \includegraphics[width=0.6\textwidth]{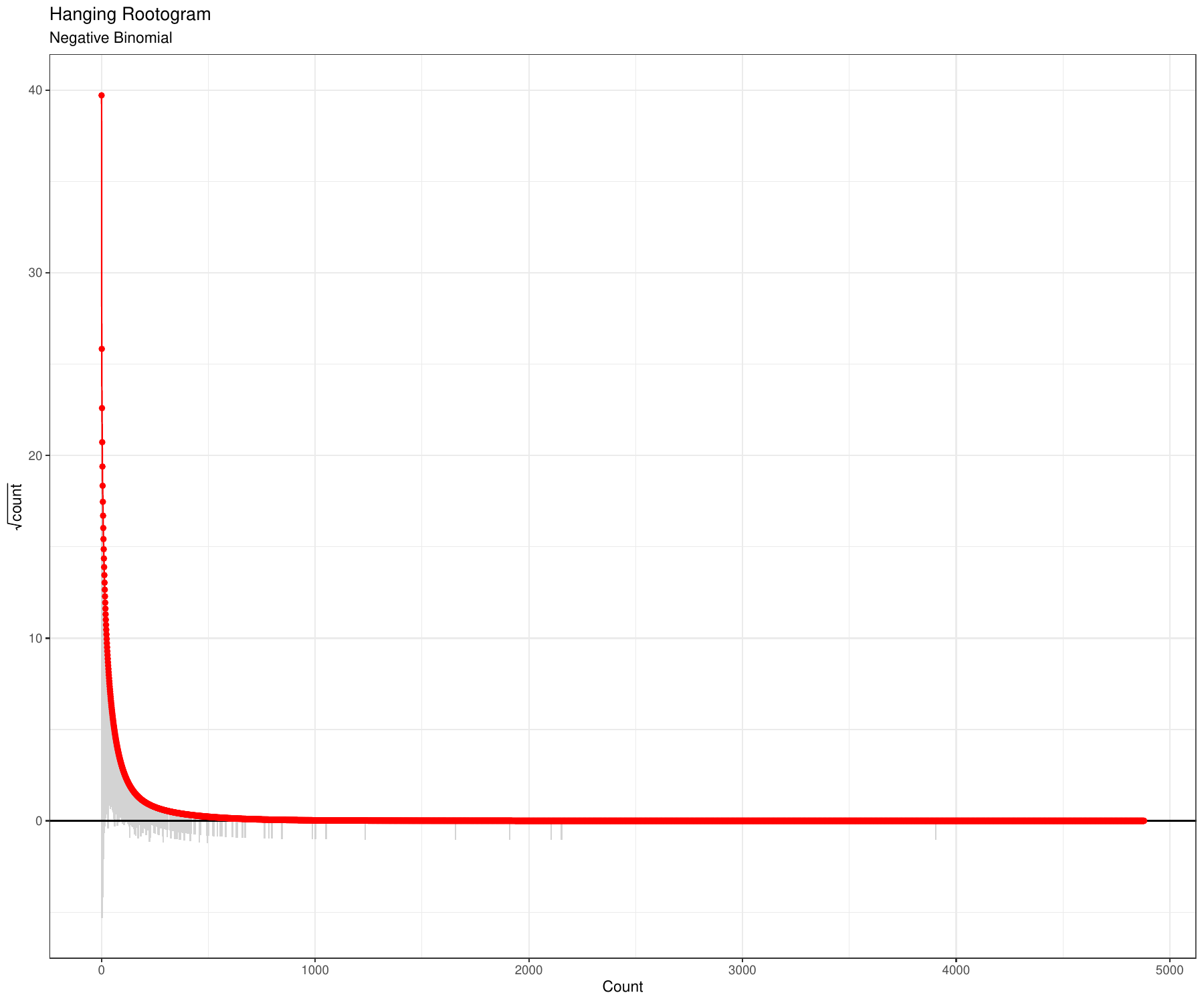}
    \caption{(Simulation 4) Overall rootogram diagnostics for Poisson (top) and negative binomial (bottom) models}
    \label{fig:sim4_root_all}
\end{figure}

\begin{figure}[H]
    \centering
    \includegraphics[width=\textwidth]{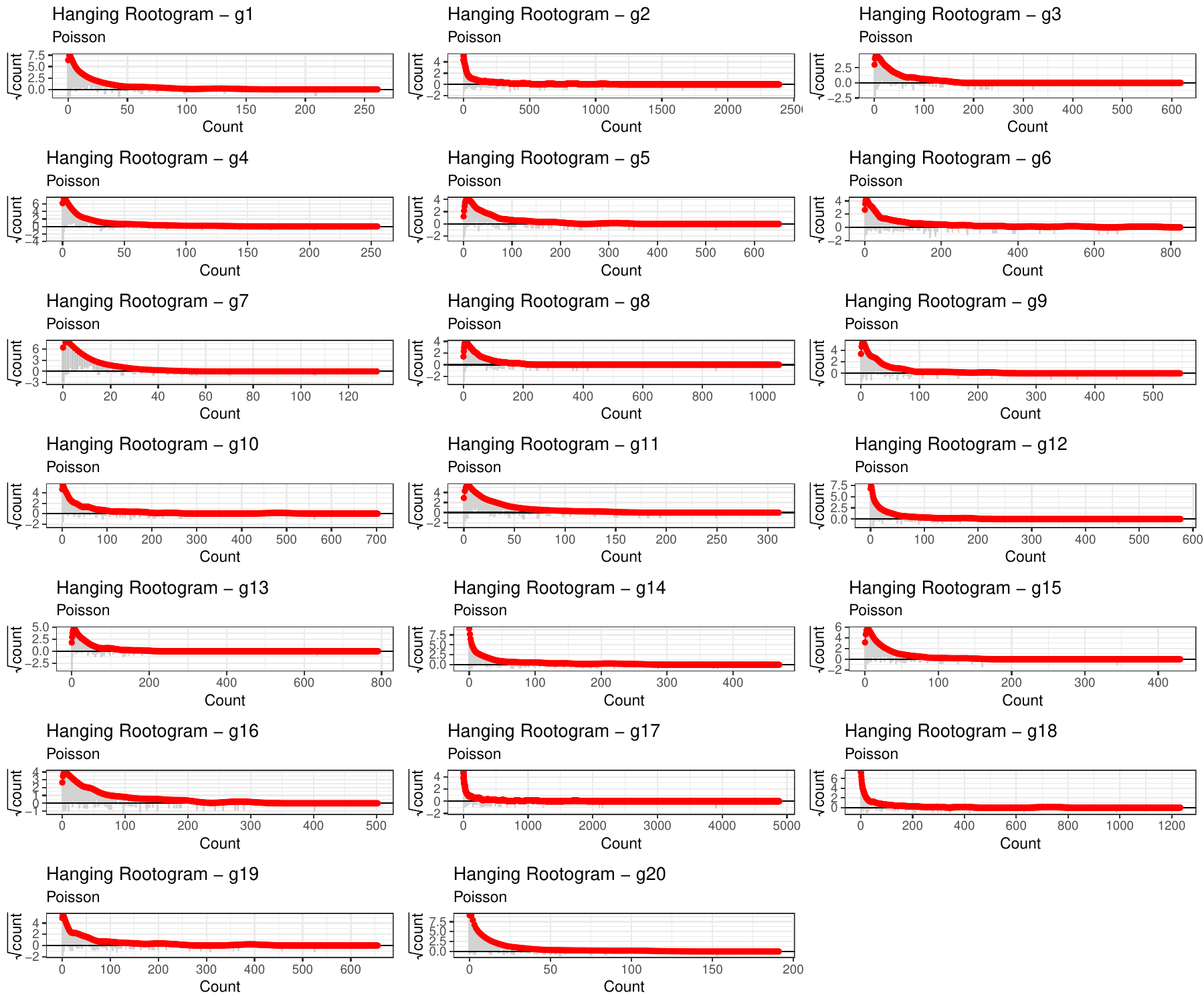}
    \caption{(Simulation 4) Poisson rootogram diagnostics by group}
    \label{fig:sim4_pois_root_group}
\end{figure}

\begin{figure}[H]
    \centering
    \includegraphics[width=\textwidth]{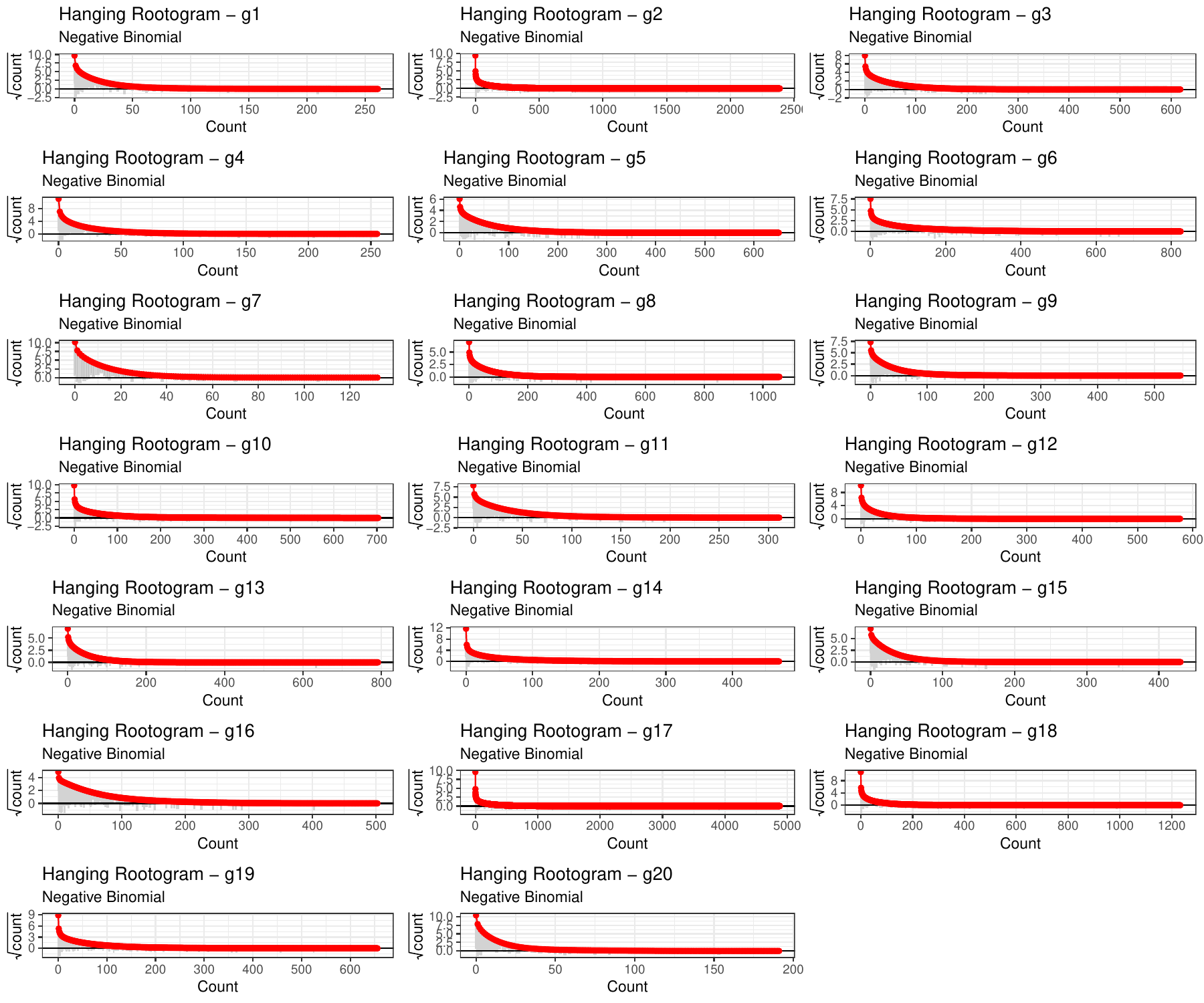}
    \caption{(Simulation 4) Negative binomial rootogram diagnostics by group}
    \label{fig:sim4_nb_root_group}
\end{figure}

\begin{figure}[H]
    \centering
    \includegraphics[width=0.7\textwidth]{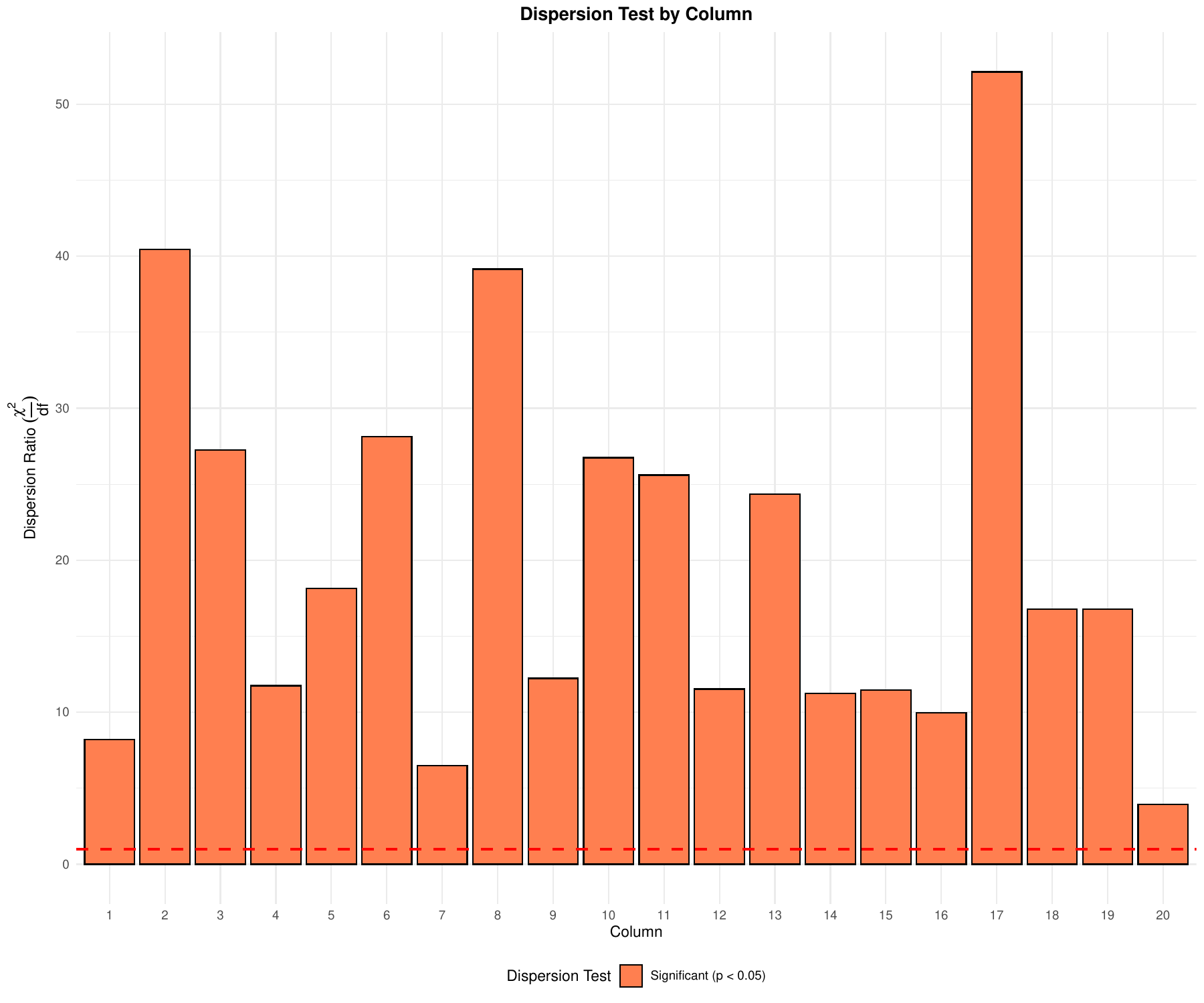}
    \caption{(Simulation 4) Dispersion diagnostics}
    \label{fig:sim4_disp}
\end{figure}

\noindent \textbf{Distribution:} The distributional fit is assessed through rootograms (Supplementary Figures~\ref{fig:sim4_root_all}, \ref{fig:sim4_pois_root_group}, and \ref{fig:sim4_nb_root_group}) and dispersion diagnostics (Supplementary Figure~\ref{fig:sim4_disp}). The overall rootograms, group rootograms, and dispersion metrics suggest that the neither the Poisson nor negative binomial distribution are sufficiently complex to explain the observed ARD, and therefore a full model would require a correlation structure based on the results above.

\begin{figure}[H]
    \centering
    \begin{subfigure}[b]{0.48\textwidth}
        \centering
        \includegraphics[width=\textwidth]{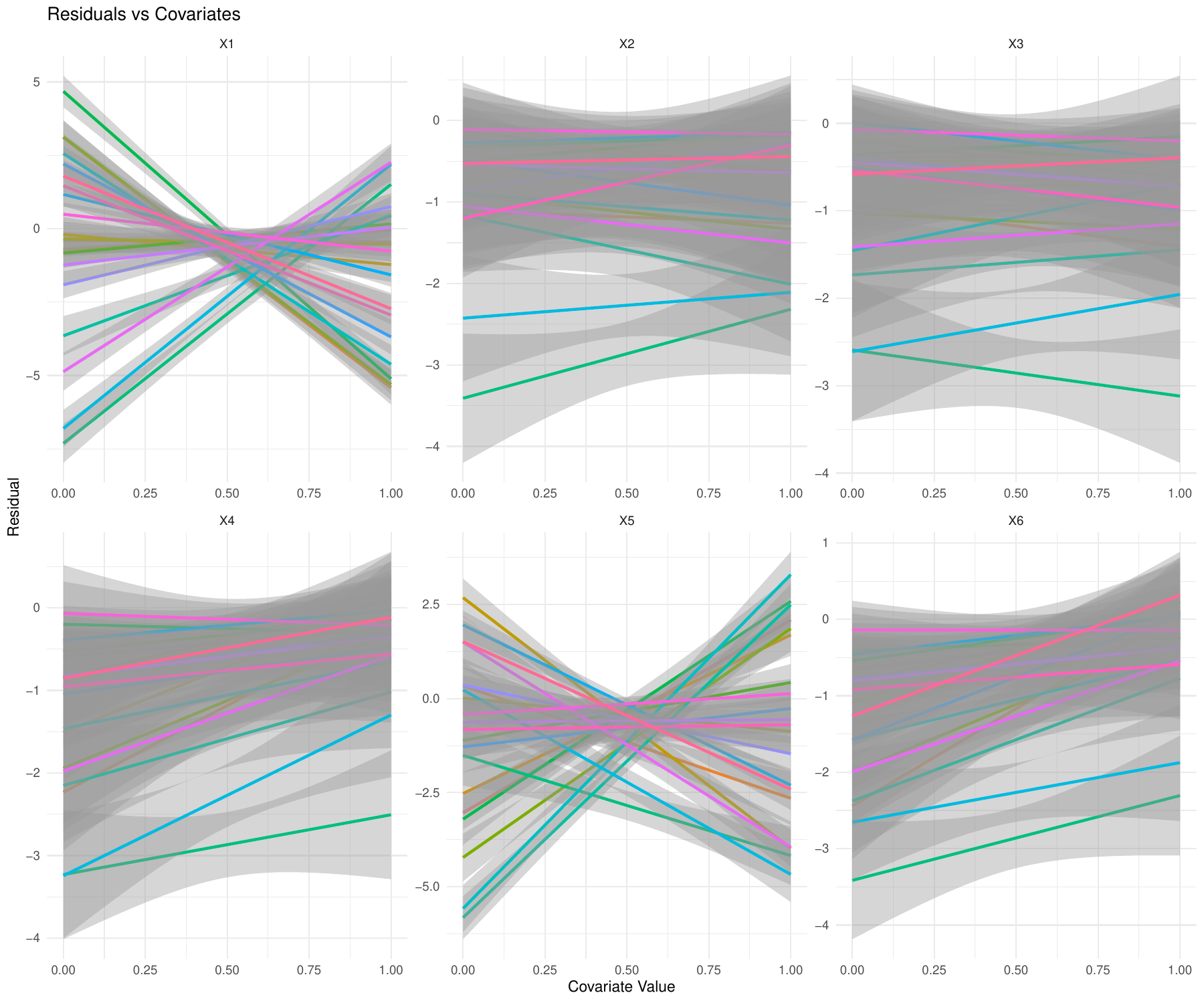}
        \caption{Local covariate effects}
        \label{fig:sim4_local_cov}
    \end{subfigure}
    \hfill
    \begin{subfigure}[b]{0.48\textwidth}
        \centering
        \includegraphics[width=\textwidth]{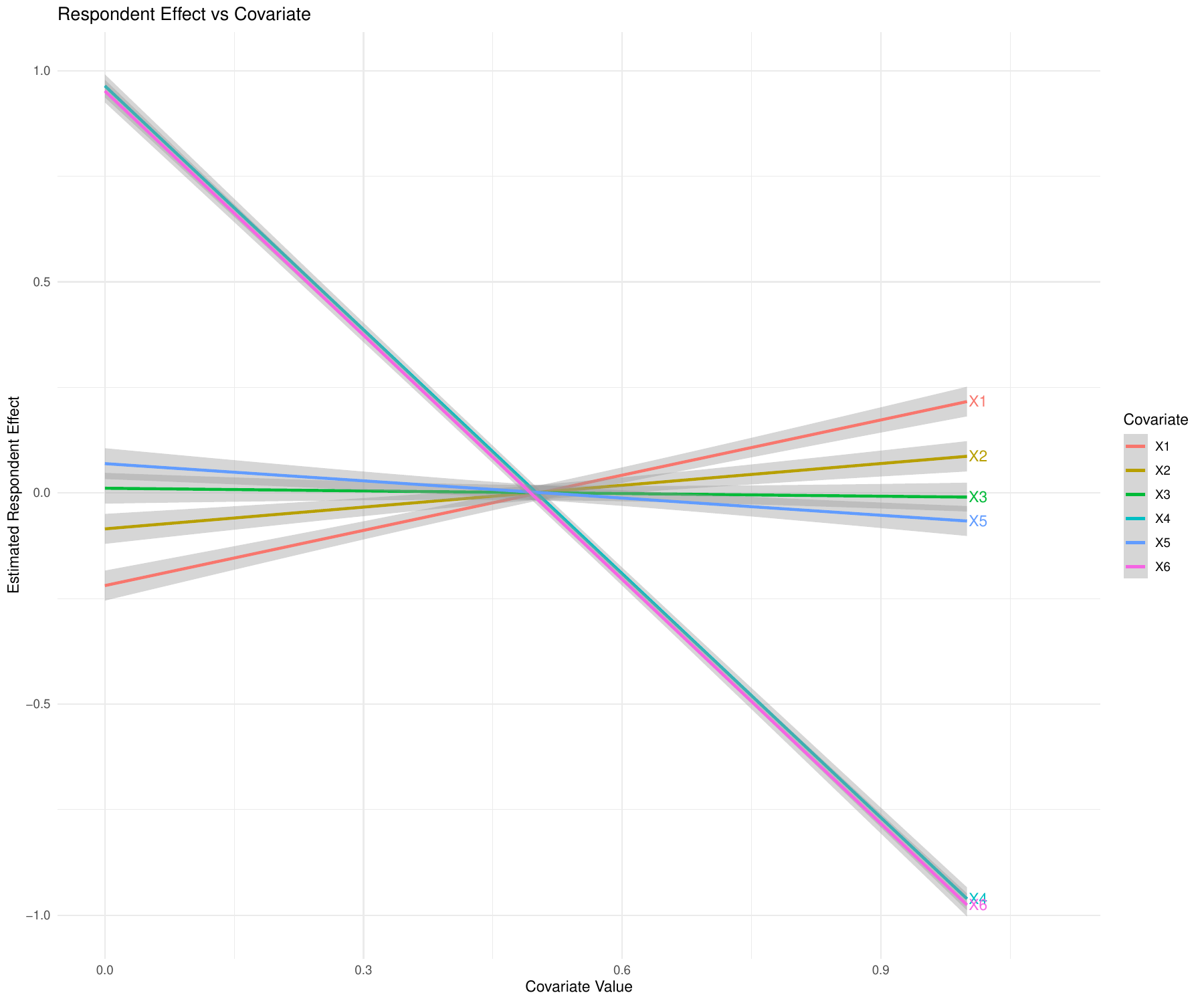}
        \caption{Global covariate effects}
        \label{fig:sim4_global_cov}
    \end{subfigure}
    \caption{(Simulation 4) Covariate selection diagnostics}
    \label{fig:sim4_cov}
\end{figure}

\begin{figure}[H]
    \centering
    \begin{subfigure}[b]{0.48\textwidth}
        \centering
        \includegraphics[width=\textwidth]{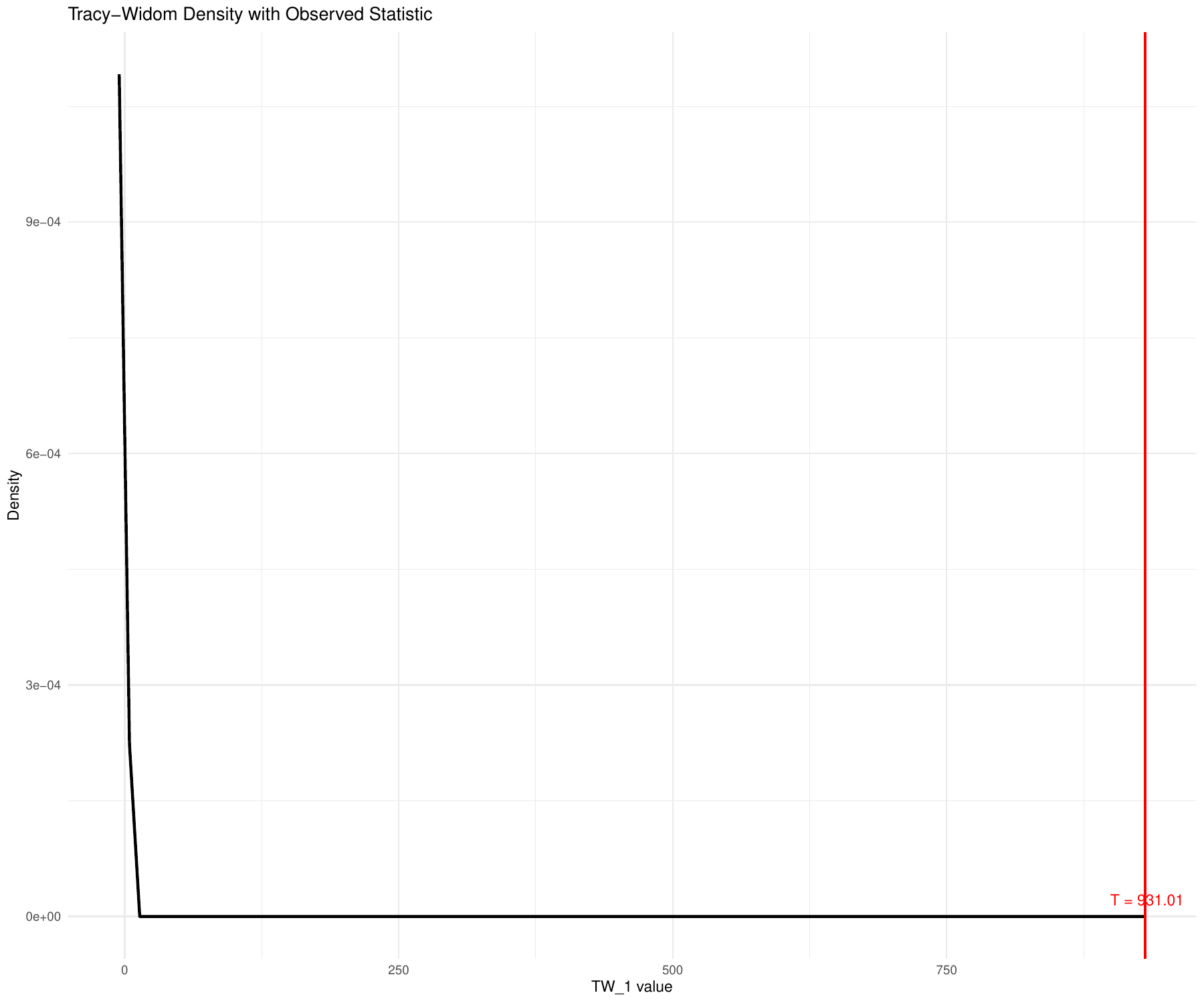}
        \caption{Poisson model}
        \label{fig:sim4_corr_pois}
    \end{subfigure}
    \hfill
    \begin{subfigure}[b]{0.48\textwidth}
        \centering
        \includegraphics[width=\textwidth]{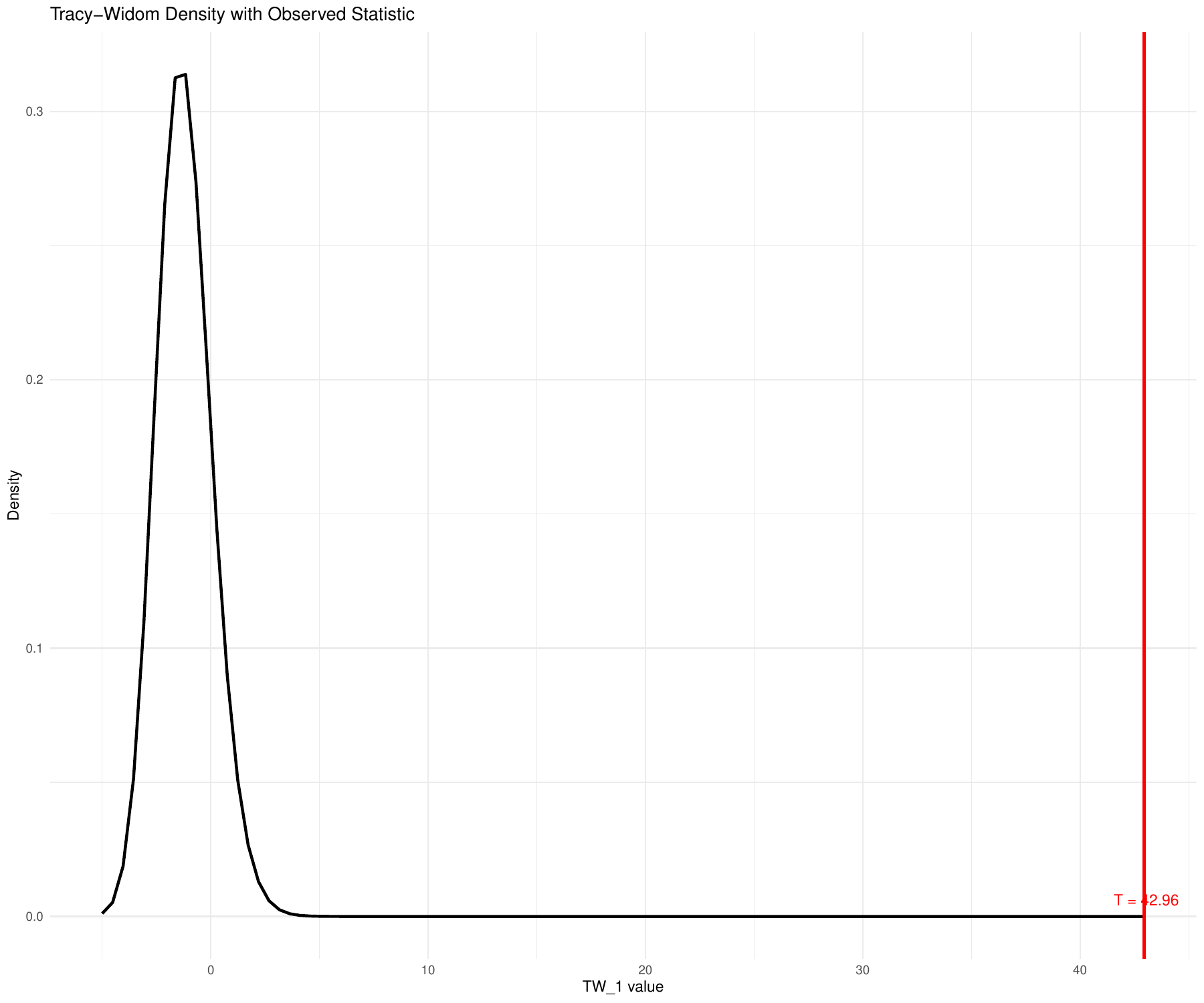}
        \caption{Negative binomial model}
        \label{fig:sim4_corr_nb}
    \end{subfigure}
    \caption{(Simulation 4) Group correlation diagnostics}
    \label{fig:sim4_corr}
\end{figure}

\clearpage

\section{Effect of Diagnostic Order on Model Selection}

Here we revisit Simulation 1 and demonstrate what happens when model diagnostics are performed in the wrong order. Specifically, checking for correlation and distribution issues before accounting for covariate effects can lead to incorrect model choice. Recall that the true model is Poisson distributed, with 3 covariates (2 local, 1 global), with no correlation structure.

\begin{figure}[H]
    \centering
    \begin{subfigure}[b]{0.5\textwidth}
        \centering
        \includegraphics[width=\textwidth]{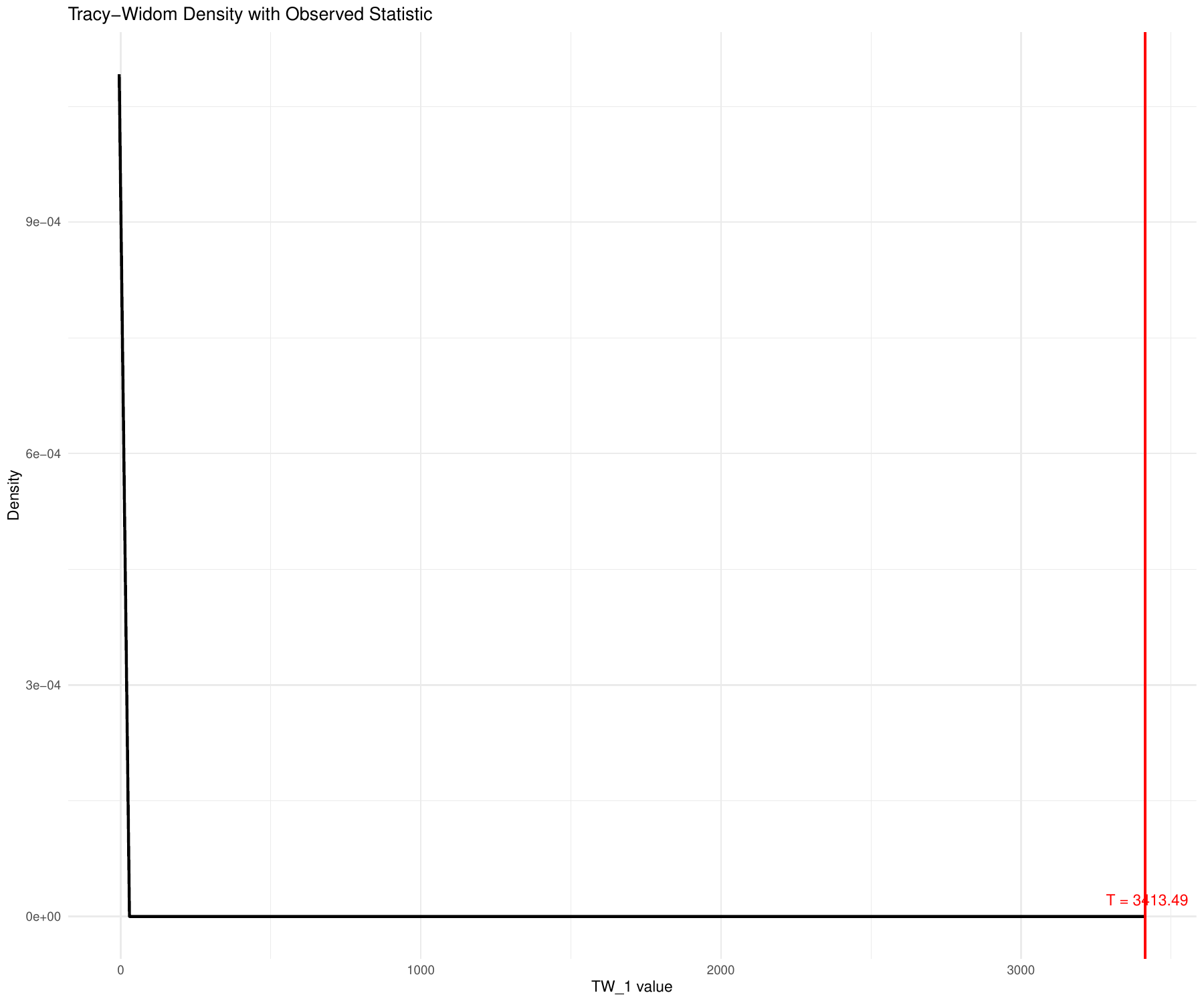}
        \caption{Poisson model}
        \label{fig:wrong_order_corr_pois}
    \end{subfigure}
    \hfill
    \begin{subfigure}[b]{0.5\textwidth}
        \centering
        \includegraphics[width=\textwidth]{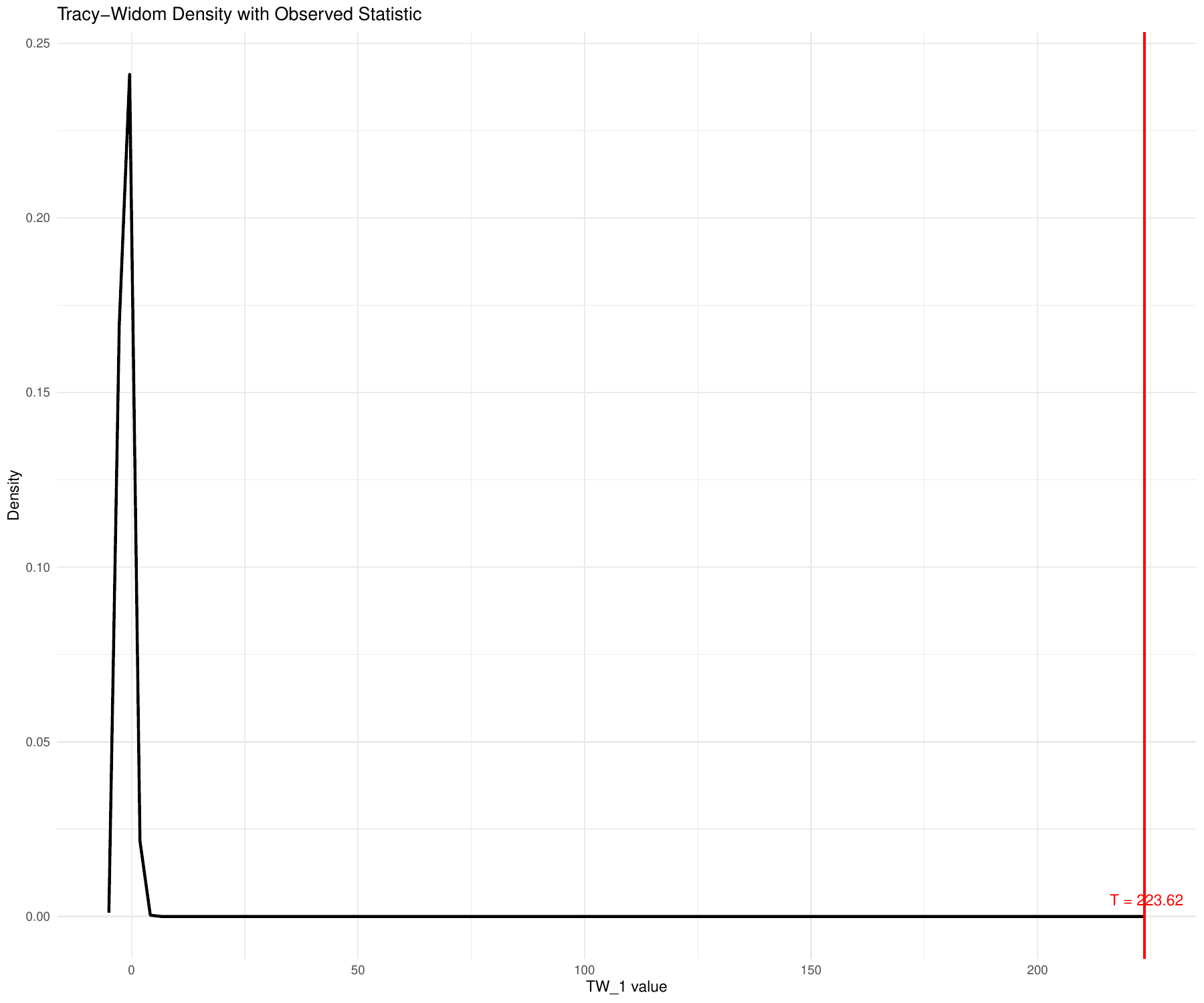}
        \caption{Negative binomial model}
        \label{fig:wrong_order_corr_nb}
    \end{subfigure}
    \caption{Group correlation diagnostics without covariate adjustment. Note the apparent correlation that may be confounded with missing covariates.}
    \label{fig:wrong_order_corr}
\end{figure}

Supplementary Figure~\ref{fig:wrong_order_corr} indicates that regardless of the fitted distribution, the Tracy-Widom test suggests that the residuals are highly correlated. While the missing covariates due induce correlation in the residuals, it is more appropriate to account for the covariates in the regression component of the model, and not in the residual covariance matrix.

\begin{figure}[H]
    \centering
    \begin{subfigure}[b]{0.5\textwidth}
        \centering
        \includegraphics[width=\textwidth]{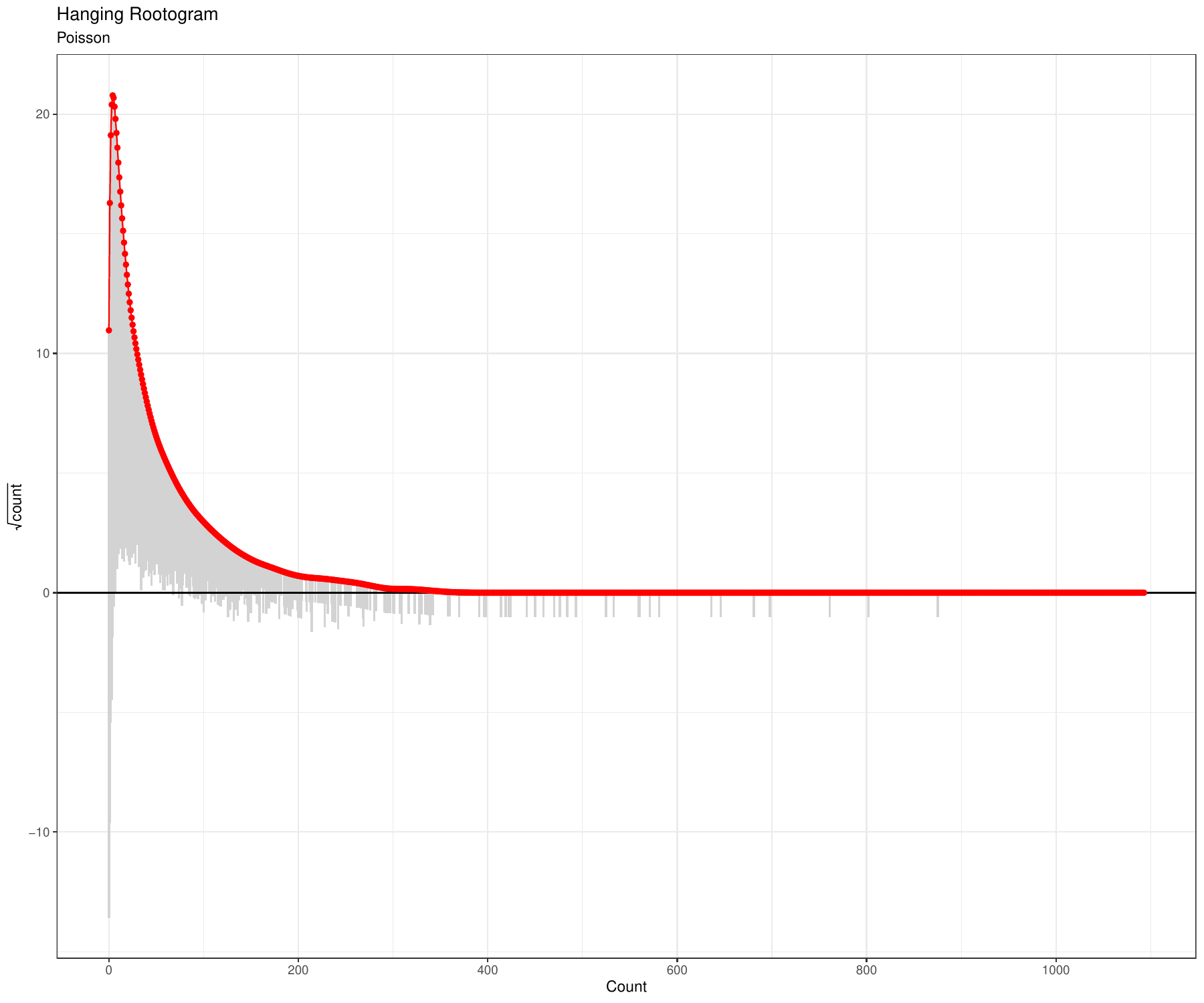}
        \caption{Poisson - overall}
        \label{fig:wrong_order_pois_all}
    \end{subfigure}
    \hfill
    
    \begin{subfigure}[b]{0.5\textwidth}
        \centering
        \includegraphics[width=\textwidth]{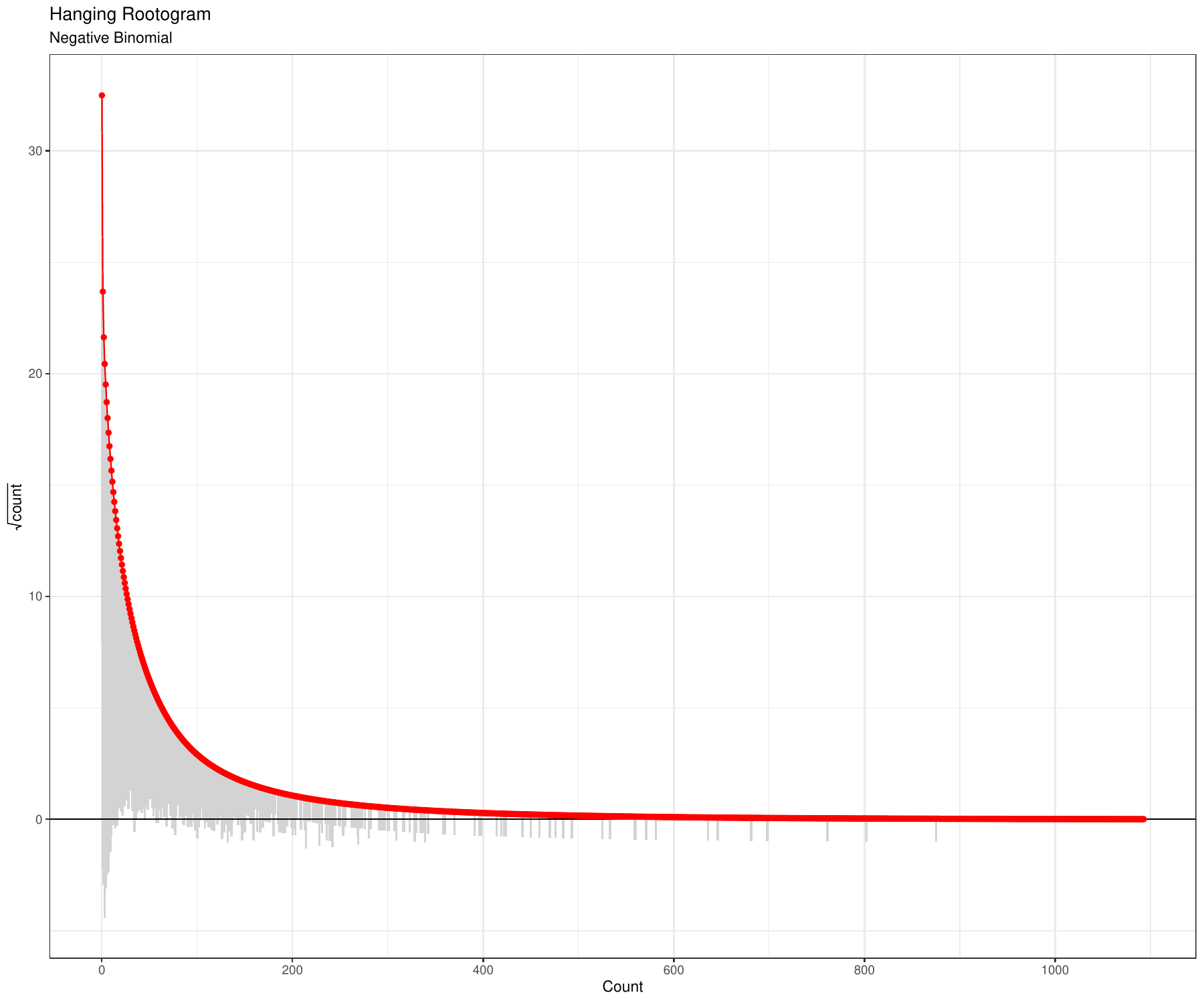}
        \caption{Negative binomial - overall}
        \label{fig:wrong_order_nb_all}
    \end{subfigure}
    \caption{Rootogram diagnostics without covariate adjustment. Poor fit may be due to missing covariates rather than distributional issues.}
    \label{fig:wrong_order_root}
\end{figure}

\begin{figure}[H]
    \centering
    \includegraphics[width=0.7\textwidth]{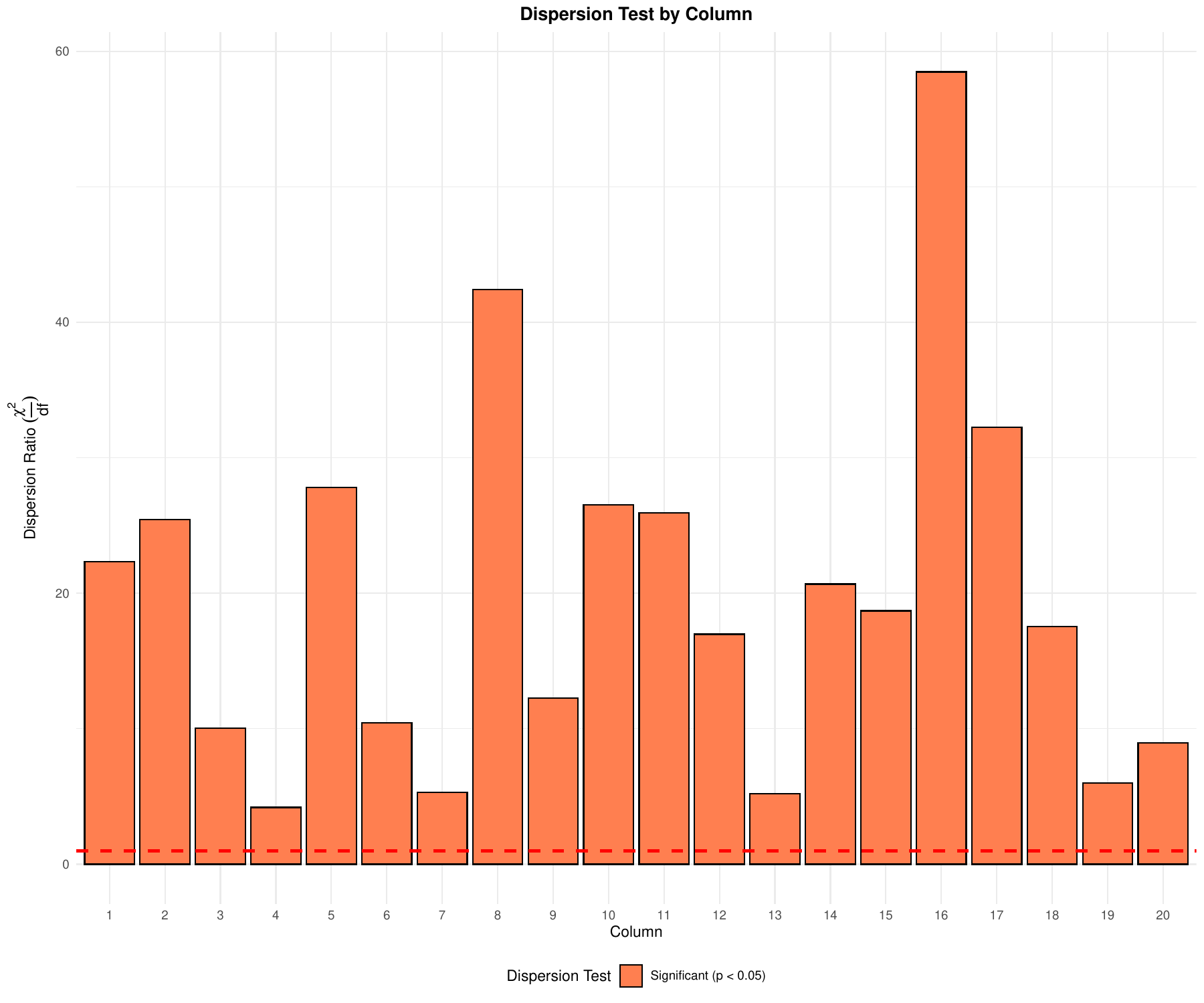}
    \caption{Dispersion diagnostics without covariate adjustment. Apparent overdispersion may be confounded with covariate effects.}
    \label{fig:wrong_order_disp}
\end{figure}

Similarly, Supplementary Figures~\ref{fig:wrong_order_root} and \ref{fig:wrong_order_disp} could incorrectly suggest that a negative binomial model is required. However, this overdispersion arises from omitted covariates rather than true conditional overdispersion. Once covariates are properly included, the Poisson distribution provides an adequate fit.

\clearpage

\end{document}